\documentclass[fleqn,usenatbib]{mnras}

\usepackage{newtxtext,newtxmath}

\usepackage[T1]{fontenc}

\DeclareRobustCommand{\VAN}[3]{#2}
\let\VANthebibliography\thebibliography
\def\thebibliography{\DeclareRobustCommand{\VAN}[3]{##3}\VANthebibliography}

\usepackage{xcolor}
\definecolor{steelblue}{rgb}{0.275, 0.510, 0.706}
\definecolor{scaryOrange}{rgb}{0.888, 0.597, 0.104}
\definecolor{scaryRed}{rgb}{0.888, 0.097, 0.104}
\definecolor{bottleGreen}{rgb}{0.0, 0.208, 0.153}
\definecolor{seagreen}{rgb}{0.190, 0.525, 0.361}
\definecolor{ceruleanCrayola}{rgb}{0.424, 0.763, 0.959}
\definecolor{anthracite}{rgb}{0.271, 0.270, 0.318}
\definecolor{refereebrown}{rgb}{0.11, 0.11, 0.0}
\definecolor{operamauve}{rgb}{0.718, 0.518, 0.655}

\newcommand{\REV}[1]{{}}

\usepackage{graphicx}	
\usepackage{amsmath}	
\usepackage{booktabs}
\graphicspath{{./}{figures/}}
\DeclareMathOperator\erf{erf}

\title[Growing Black Holes in Galactic Nuclei]{Growing Black Holes through Successive Mergers in Galactic Nuclei:\\ I. Methods and First Results}

\author[Atallah et al.]{
Dany Atallah,$^{1,2}$
Alessandro A. Trani,$^{3,4,5}$
Kyle Kremer,$^{6,7}$
Newlin C. Weatherford,$^{1,2}$
\newauthor Giacomo Fragione,$^{1,2}$
Mario Spera,$^{8}$
and Frederic A. Rasio,$^{1,2}$
\\
$^{1}$Department of Physics \& Astronomy, Northwestern University, Evanston IL 60208, USA\\
$^{2}$Center for Interdisciplinary Exploration \& Research in Astrophysics (CIERA), Evanston, IL \\
$^{3}$Niels Bohr International Academy, Niels Bohr Institute, Blegdamsvej 17, 2100 Copenhagen, Denmark\\
$^{4}$Research Center for the Early Universe, School of Science, The University of Tokyo, Tokyo 113-0033, Japan\\
$^{5}$Okinawa Institute of Science and Technology, 1919-1 Tancha, Onna-son, Okinawa 904-0495, Japan\\
$^{6}$TAPIR, California Institute of Technology, Pasadena, CA 91125, USA\\
$^{7}$The Observatories of the Carnegie Institution for Science, Pasadena, CA 91101, USA\\
$^{8}$SISSA, Via Bonomea 265, I-34136 Trieste, Italy
}

\pubyear{2022}
\begin{document}
\label{firstpage}
\pagerange{\pageref{firstpage}--\pageref{lastpage}}
\maketitle

\begin{abstract}
We present a novel, few-body computational framework designed to shed light on the likelihood of forming intermediate-mass (IM) and supermassive (SM) black holes (BHs) in nuclear star clusters (NSCs) through successive BH mergers, initiated with a single BH seed. Using observationally motivated NSC profiles, we find that the probability of a ${\sim}100 \, M_\odot$ BH to grow beyond ${\sim}1000 \, M_\odot$ through successive mergers ranges from ${\sim}0.1\%$ in low-density, low-mass clusters to nearly $90\%$ in high-mass, high-density clusters. However, in the most massive NSCs, the growth timescale can be very long ($\gtrsim 1\,$Gyr); vice versa, while growth is least likely in less massive NSCs, it is faster there, requiring as little as ${\sim}0.1\,$Gyr. The increased gravitational focusing in systems with lower velocity dispersions is the primary contributor to this behavior. We find that there is a simple ``7-strikes-and-you're-in'' rule governing the growth of BHs: our results suggest that if the seed survives 7~to~10 successive mergers without being ejected (primarily through gravitational wave recoil kicks), the growing BH will most likely remain in the cluster and will then undergo runaway, continuous growth all the way to the formation of an SMBH (under the simplifying assumption adopted here of a fixed background NSC). Furthermore, we find that rapid mergers enforce a dynamically-mediated ``mass gap'' between about ${50-300 \, M_\odot}$ in an NSC.
\end{abstract}



\section{Introduction} 
\label{sec:intro}
 Bridging the gap between stellar BHs $(\lesssim 50 \, M_\odot)$ and SMBHs $({\gtrsim 10^5 \, M_\odot})$ observed at the centers of most galaxies remains one of the major unsolved astrophysical problems \citep{Rees1984}. The incredible success in detecting stellar mass $\left(\lesssim 50 \, M_\odot \right)$ BHs as gravitational wave sources \citep[][]{LIGO_2019_population} in the local universe and SMBHs through dynamical measurements \citep{Ghez_review_SMBH_dyn_meas} and as engines of active galactic nuclei \citep{agn_review_2017} is overshadowed by the difficulty in observing IMBHs of $(10^2 - 10^5 \, M_\odot)$ through optical and GW searches. Though there have been many claimed IMBH candidates \citep{2002AJ....124.3270G, 2016A&A...588A.149K, 2017MNRAS.468.2114P}, most have been disproved \citep{2003ApJ...595..187M, 2011ApJ...732...67M, 2012A&A...542A..44K, 2018MNRAS.473.4832G, 2018ApJ...862...16T}; see \citet{2021RNAAS...5...47R} for a brief synopsis. Quantifying potential IMBH formation channels may provide critical context for directing future observations of these elusive objects \citep{JWST_IMBH_2022}.


Some of the most explored IMBH formation channels include runaway stellar collisions \citep[e.g.,][]{portegies_zwart_star_1999, zwart_runaway_2002, portegies_zwart_formation_2004, gurkan_massive_2006, freitag_runaway_2006,  mapelli_massive_2016,2020ApJ...903...45K, rizzuto_intermediate_2021, gonzalez_intermediate-mass_2021}, BH accretion \citep[e.g.,][]{Haehnelt_1998_AGNmass, 2014Sci...345.1330A, 2015ApJ...810...51M, 2016MNRAS.458.3047P, 2017ApJ...850L..42P, Ricarte_2018_bhseedaccretion, 2020MNRAS.493.3732D, ShiKremer2022} and repeated compact object mergers \citep{quinlan_dynamical_1989, mouri_runaway_2002,miller_production_2002,oleary_binary_2006, giersz_mocca_2015, antonini_merging_2016,banerjee_stellar-mass_2017, fischbach_are_2017, fragione_gravitational_2018,kovetz_limits_2018,2019arXiv190500902A, antonini_bhgrowth_2019, 2020ApJ...903...45K, 2019_Rodriguez_review, fragione_repeat_2020, FragioneLoeb2020, 2021ApJ...907L..25W, 2021MNRAS.505..339M, fragione_repeated_2022, 2022MNRAS.511.2631A}. In this work, we focus on the growth of massive BH seeds through repeated mergers with other stellar mass BHs. 

Several factors govern an environment's capacity for repeated or `multi-generation' BH mergers, where at least one component has already merged with another BH. Fundamentally, many stellar BHs must be concentrated in a sufficiently dense star cluster environment. In such environments, BH merger products may merge again via exchange encounters with existing binaries or GW capture \citep{fregeau_stellar_2004, downing_compact_2010,ziosi_dynamics_2014, antognini_dynamical_2016, rodriguez_dynamical_2016, zevin_eccentric_2019}. Critically, the recoil kicks caused by asymmetric emission of GW radiation during mergers commonly impart velocities ${\sim}10^2 - 10^3\,{\rm km}\,{\rm s}^{-1}$ (depending on mass ratio and spins), often large enough to eject the merger product from even the most massive stellar clusters \citep[e.g.,][see \citealt{Gerosa_hierarchical_2021} for a review]{campanelli_maximum_2007, HB2008,gerosa_escape_2019,FragioneLoeb2020,FragioneLoeb2022}. Recoil kick magnitudes peak at a moderate mass ratio ($m_1/m_2\approx0.38$ for low BH spins), remaining prominent until the BH is sufficiently massive relative to typical BHs in its local environment ($m_1/m_2 \lesssim 0.1$). 

The dense, high escape speed cores of nuclear star clusters (NSC) are likely to be viable factories for producing massive BHs via repeat BH-BH mergers. NSCs are broadly defined as dense stellar systems residing at the (dynamical) centers of most galaxies \citep{Neumayer_2011_NSCloc}. While early universe observations of NSCs are lacking, there is significant evidence suggesting their formation is heavily correlated with the formation of SMBHs/AGN \citep{Wehner_2006_NSCSMBH, 2006_Ferrarese_SMBHNSC, Seth_2008_AGN_NSC, kormendy_2013_smbhgalaxyclusterreview, MW_NSC_SMBH_correlation, NSC2020review}. Since AGN are observed in the early universe, it is reasonable to consider the possibility that NSCs may act as nurseries for the massive BHs we see today. Though the capabilities of the recently launched JWST are just being unveiled, it is becoming evident that galaxy structures are forming at higher redshifts than previously observed \citep{2022_JWST_earlyhighgalaxies}. If these galaxies are subject to similar dynamics as local galaxies, the prevalence of NSC nurseries in the early universe may become critical to SMBH growth models.  

The dynamical interactions within star clusters are traditionally explored using direct $N$-Body \citep{Aarseth_1999, 2015MNRAS.450.4070W} and Monte Carlo \citep{2013MNRAS.429.1221H, CMCcodepaper} calculations. Unfortunately, such codes reach intractable computational bottlenecks for clusters with $N \gtrsim 10^7$ bodies. In addition, the major Monte Carlo packages most capable of simulating large $N$ systems must forcibly break up any orbital hierarchies composed of three or more bodies because of the tendencies of such systems to dominate the computational demands of $N$-body calculations. Despite the undeniable power of these codes, the direct simulation and modeling of extremely massive NSCs remains a prohibitively expensive aspiration.

As an alternative to computationally expensive $N$-body simulations, the BH growth process may be modeled using carefully constructed BH scattering experiments that effectively reproduce the dynamical interactions a BH seed would likely undergo \citep[e.g.,][]{Gultekin2004, Miller2009}. While pioneering at the time, these previous calculations were limited in two ways: (i) they utilized only binary--single encounters without consideration of incoming binaries or growth of hierarchical multiples (triples, quadruples, etc.) and (ii) scattering experiments were terminated as soon as the seed BH experienced its first merger. These results were then used to extrapolate the growth rate of BH seeds.

Here we present a novel scattering infrastructure called \textsc{CuspBuilding} to go beyond these early efforts. As in previous models of this type \citep{oleary_binary_2006, Antonini2016,fragione_repeat_2020}, we compute dynamical interactions based on local properties of the background star cluster, such as the density, encounter rate, velocity distribution, and we assume that all interactions take place at the cluster center. The final product of each dynamical interaction becomes the target for the next interaction. We assume that the central regions of NSCs are dominated by BHs and we consider only interactions between BHs (no other objects). Crucially, our code evaluates all encounters in detail using a high-precision direct $N$-Body integrator while tracking and growing arbitrarily large hierarchical BH systems for the first time.


Our paper is organized as follows. Section~\ref{sec:methods} details our computational methodology and initial conditions. In Section~\ref{sec:results}, we present our numerical results, and, in particular, we characterize the probability of the rapid growth of seed BHs through successive mergers with other BHs, referenced from here out as a \textit{runaway}. In Section~\ref{sec:disc}, future considerations/improvements are discussed. We conclude by summarizing our key findings in Section~\ref{sec:conclusion}, emphasizing the most important limitations of our approach.

\begin{figure*}
\includegraphics[width=\textwidth]{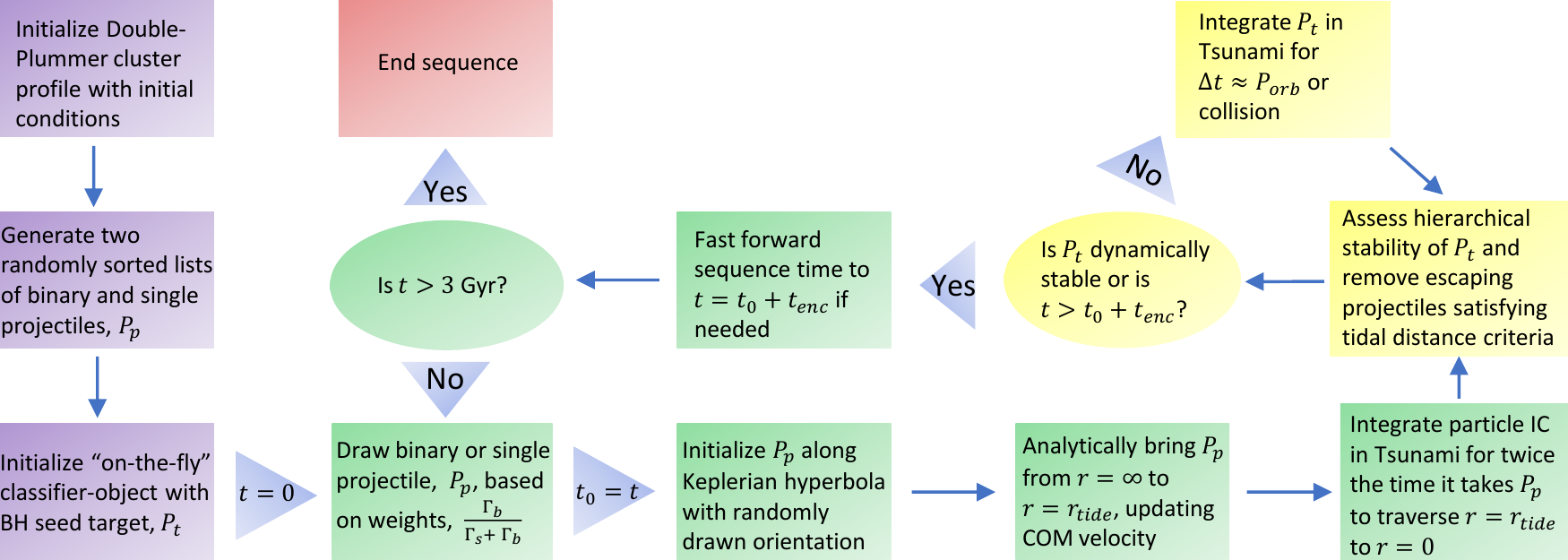}
\caption{An illustration of basic infrastructure logic. The parameters constraining our double-Plummer model is described in Sec. \ref{sec:dpmodel}, the process by which we initialize a scattering experiment in Sec. \ref{sec:scatinit}, the encounter rate and selection weighting for binary or single ``species'' in Sec. \ref{sec:ttne}, and the dynamical stability assessment in Sec.~\ref{sec:stability}.}
\label{fig:flowchart}
\end{figure*}

\section{Methodology}
\label{sec:methods}

\subsection{Tsunami}

We employ \textsc{tsunami}, a direct $N$-body integrator based on Mikkola's algorithmic regularization \citep{tsunami2022,trani2019a,trani2019b}, as the key building block for our Monte Carlo scattering framework, \textsc{CuspBuiliding}. \textsc{tsunami} handles all direct $N$-body calculations in \textsc{Cuspbuilding}, integrating the equation of motion, derived from a logarithmic Hamiltonian in extended phase space \citep{mikkola1999b}. Though not used here, \textsc{tsunami} includes the time-transformed regularization scheme \citep{mikkola1999a} and the unregularized leapfrog \citep[see][]{aarseth-nbody}.

First, round-off errors are commonly introduced when two particles are close to each other, but far from the center of a global coordinate system, such as the center-of-mass (COM) reference frame. These errors are minimized in \textsc{tsunami} because it does not integrate in a traditional reference frame. Instead, \textsc{tsunami} employs the chain-coordinate system introduced in \citet{kschain}. A ``chain'' of interparticle vectors is formed between all particles at every time-step. The first ``segment'' is selected to be the shortest interparticle distance in the system, with each successive segment connecting the nearest-neighbor particle closest to an end of the previous segment until all particles are included. All the particle coordinates are expressed relative to their nearest-neighbor on the chain. As the system evolves, \textsc{tsunami} updates the chain coordinates so that any chained vector is always shorter than adjacent non-chained vectors. New chain coordinates are always directly calculated using the previous chain coordinates without employing COM transformations.

Leapfrog is a second order algorithm, but \textsc{tsunami} takes advantage of Bulirsch-Stoer extrapolation to achieve higher accuracy \citep{stoer1980}. The primary mechanism behind Bulirsch–Stoer extrapolation is to consider the result of leapfrog integration as an analytic function of the stepsize $h$. The solution of a given time interval, $\Delta S$, is computed for smaller and smaller sub-steps, $h = \Delta S / N$, and is then extrapolated to $N\rightarrow \infty$ using polynomial functions. This scheme could be employed by any integration scheme, but is particularly advantageous to leapfrog integration because its error scales as $\delta E \propto h^2$ at the leading order.

These techniques allow \textsc{tsunami} to follow close encounters with extreme accuracy without reducing the integration time-step, unlike traditional integrators \citep[e.g.,][]{2004MNRAS.352....1F}. This makes \textsc{tsunami} an ideal code for integrating compact few-body systems, such as planetary systems \citep[e.g.][]{trani2020} and triple stellar systems \citep{manwadkar2021,hayashi2022,hellstrom2022}. 

\textsc{tsunami} implements non-Newtonian forces, such as dynamical and equilibrium tides for stars and planets \citep{hut81,samsing2018}, but are not useful in this work. Post-Newtonian corrections of order 1, 2, 2.5, 3, \& 3.5 \citep{2006MNRAS.372..219M,2008AJ....135.2398M,2014LRR....17....2B} are enabled during all encounters without exception.

\subsection{CuspBuilding}
\textsc{CuspBuiliding} is a Python program that allows a user to quickly set up scattering experiments to simulate the growth of a seed BH. \textsc{CuspBuiliding} dynamically interfaces with \textsc{tsunami}, automatically organizing particles into hierarchies and calculating all hierarchy properties, such as semi-major axes (SMA), eccentricities, and hierarchical organization. 

\textsc{CuspBuiliding} simulates the entire dynamical history of a hierarchy ``target,'' $P_\mathrm{t}$, by computing a series of back-to-back gravitational interactions (we will refer to each one as a ``sequence'') with ``projectile'' hierarchies, $P_{\rm p}$.  After each interaction is computed, the new $P_{\rm t}$ that is determined as the outcome becomes the target for the next interaction with another projectile (Fig.~\ref{fig:flowchart}). Here, $P_{\rm t}$ is always the hierarchy containing the massive BH that grew from the initial seed. In this study, the initial seed target is always a binary BH system (BBH) while the projectile is either a single BH or a BBH. Through successive interactions, we allow the target to grow into an arbitrarily large hierarchical system (triple, quadruple, etc.).  

\subsection{Host Cluster Environment} \label{sec:dpmodel}

\subsubsection{Double-Plummer Relations}\label{sec:dpr}

We model an NSC environment as a fixed background star cluster in gravothermal equilibrium, represented by a double-Plummer model. This model consists of two, concentric Plummer spheres where: (i) the larger sphere represents most of the mass and is composed only of ordinary stars and (ii) the smaller sphere represents an inner, stellar BH sub-cluster. Such a BH sub-cluster is expected to form generically in dense stellar systems through mass segregation \citep{2004ApJ...604..632G, 2011ApJ...741L..12B, 2013MNRAS.432.2779B, 2019ApJ...871...38K}.

An individual Plummer model is determined by only two parameters: the mass of the cluster, $M_{\rm cl}$, and the Plummer kernel (length scale), $b$. The equations for the gravitational potential, radial density profile, velocity dispersion, core radius, and half-mass radius are

\begin{equation}
\begin{aligned}
   &\Phi_{\rm cl}(r)  = -\frac{G M_{\rm cl}}{\sqrt{r^2 + b^2}} \\
 &\rho_{\rm cl}(r)  = \frac{3 M_{\rm cl}}{4 \pi b^3} \left(1 + \frac{r^2}{b^2}\right)^{-5/2} \\
 &\sigma_{\rm cl}^2 (r)  = -\frac{\Phi_{\rm cl}(r)}{6} \\
 &r_{\rm c}  = b \sqrt{\sqrt{2} - 1} \\
 &r_{\rm h}  = \frac{b}{\sqrt{2^{2/3} - 1}}.
\end{aligned}
\end{equation}
The smaller Plummer cluster is defined with its mass and radii scaled directly to $M_{\rm cl}$ and $\epsilon_{\rm b}$. It follows that the BH sub-cluster properties may be expressed
\begin{equation}\label{plumbh}
\begin{aligned}
 &M_{\rm bh}  = \epsilon_{\rm M} \, M_{\rm cl} \\
 &b_{\rm bh}  = \epsilon_{\rm b} \, b \\
 &\Phi_{\rm bh}(r)  = -\frac{G M_{\rm bh}}{\sqrt{r^2 + b_{\rm bh}^2}} \\
 &\rho_{\rm bh}(r)  = \frac{3 M_{\rm cl}}{4 \pi b_{\rm bh}^3} \left(1 + \frac{r^2}{b_{\rm bh}^2}\right)^{-5/2} \\
 &\sigma_{\rm bh}^2 (r)  = -\frac{\Phi_{\rm bh}(r)}{6}  \\
 &r_{\rm c,bh}  = b_{\rm bh} \sqrt{\sqrt{2} - 1} \\
 &r_{\rm h,bh}  = \frac{b_{\rm bh}}{\sqrt{2^{2/3} - 1}}, \\
\end{aligned}
\end{equation} 
where $\epsilon_{\rm M}$ and $\epsilon_{\rm b}$ are scaling parameters which relate the BH sub-cluster to the primary cluster. As a first guess, we employ the ``Late-Type" mass/radius scaling relations as observed in \citet{10.1093/mnras/stw093} such that
\begin{equation}\label{eq:georgiev}
    \log\left(\frac{r_{\rm \rm h}}{c_1}\right) = \alpha \log \left(\frac{M_{\rm cl}}{c_2}\right) + \beta
\end{equation}
where $\alpha$, $\beta$, $c_1$, and $c_2$ are fitting constants.

Thus, the global properties of our double-Plummer models may be fully explored by specifying a value for the primary cluster mass, $M_{\rm cl}$, and scaling parameters, $\{\epsilon_{\rm M}, \epsilon_{\rm b}\}$. For all models, ${\epsilon_{\rm M}=0.01}$ and we vary $\epsilon_{\rm b}$ such that ${\epsilon_{\rm b}=\{0.005, 0.01, 0.05\}}$. The scaling parameters are selected to mimic typical scaling observed between the total cluster and mass-segregated BH cores in \citet[][]{2020ApJ...903...45K}. For simplicity, we assume the two Plummer spheres are uncoupled, maintaining independent velocity profiles. Accordingly, the deviation from energy equipartition between the primary and BH sub-cluster at $r = 0$ is
\begin{equation}
\begin{aligned}
\eta & = \frac{\langle m_{\rm bh} \rangle \sigma_{\rm bh}^2 (0) } {\langle m_{\rm cl} \rangle \sigma_{\rm cl}^2 (0) }\\
		& = \frac{\langle m_{\rm bh}\rangle}{\langle m_{\rm cl} \rangle } \frac{ M_{\rm bh}}{M_{\rm cl}} \frac{b}{b_{\rm bh}}\\
       & = \frac{\epsilon_M \langle m_{\rm bh} \rangle}{\epsilon_b \langle m_{\rm cl} \rangle},
    \end{aligned}
\end{equation}
where $\langle m_{\rm bh} \rangle$ and $\langle m_{\rm cl} \rangle$ are the average particle masses in the primary and BH sub-cluster, respectively. For simplicity, the primary cluster is assumed to only be made up of ${\sim}1$$\, M_\odot$ objects across all models (i.e., ${\langle m_{\rm cl}\rangle=1\,M_\odot}$).

We treat each encounter between projectile, $P_{\rm p}$, and target, $P_{\rm t}$, as occurring within the core of the BH sub-cluster, $r < r_{\rm c,bh}$, and relate the hierarchies' relative velocity at infinity, $v_\infty$, to their velocity dispersions ($\sigma_{\rm p}$ and $\sigma_{\rm t}$, respectively). We approximate these dispersions by assuming energy equipartition between the target and projectile such that
\begin{equation}
\begin{aligned}
\sigma_{\rm p} &= \sigma_{\rm c,bh}\\
\sigma_{\rm t} &= \sqrt{\frac{\langle m_{\rm p} \rangle}{m_{\rm t}}}\sigma_{\rm c,bh} \\
\sigma_{\rm c,bh} &= \frac{1}{r_{\rm c,bh}} \int_{\rm 0}^{r_{\rm c,bh}}\sigma_{\rm bh}(r) \,dr.
\end{aligned}
\end{equation}
We then sample $v_{\rm \infty}$ from a Maxwellian velocity profile with the velocity dispersion dispersion defined in the relative motion frame of $P_{\rm t}$ \citep{2008gady.book.....B}
\begin{equation}\label{eq:vdisprel}
\sigma_{\rm rel} = \sqrt{\sigma_{\rm t}^2 + \sigma_{\rm p}^2}= \sigma_{\rm c,bh} \sqrt{1 + \frac{\langle m_{\rm p} \rangle}{m_{\rm t}}}.
\end{equation}
Considering that 

Escape velocities from the core of the BH sub-cluster, $v_{\rm e,bh}$, and the total double-Plummer environment, $v_{\rm e,cl}$, are calculated using
\begin{equation}\label{vesc}
    v_{\rm e}(r) = \sqrt{2[\Phi_{\rm tot}(r) - \Phi_{\rm tot}(0)]}
\end{equation}
where
\begin{equation}\label{eqn:phitot}
   \Phi_{\rm tot}(r) = \Phi_{\rm cl}(r) + \Phi_{\rm bh}(r).
\end{equation}
The escape velocities of the BH sub-cluster core, $v_{\rm e,cbh}$, and the primary cluster, $v_{\rm e,cl}$, may be found by setting $r=r_{\rm c,bh}$ and $r=\infty$, respectively. We halt all interactions $P_{\rm t}$ experiences if it is temporarily ejected into the halo of a cluster (i.e., $v_{\rm e,cl}>v_{\rm kick}>v_{\rm e,cbh}$) until $P_{\rm t}$ decays to the core of the BH sub-cluster (see Sec.~\ref{sec:dfd}). If $P_{\rm t}$ is ejected from the total cluster (i.e., $v_{\rm kick}>v_{\rm e,cl}$) the sequence ends.


\subsubsection{BH Mass Distribution, Spins, and Radii}\label{sec:msr}

We sample individual stellar BH masses in the BH sub-cluster from a Gaussian with a mean of $\langle m_{\rm bh} \rangle=20 \, M_\odot$ and a standard deviation of $5\, M_\odot$, truncated below $5\, M_\odot$ and above $40\, M_\odot$. Neutron stars are the only expected core collapse remnants below $5 \, M_\odot$ while pulsational pair instability likely suppresses the development of BH remnants between roughly $40~-~120 \, M_\odot$ \citep{Heger_Woosley_2002, 10.1093/mnras/stx1576, 2019ApJ...887...53F}. This distribution is found to be roughly comparable to the BH mass distributions of the largest cluster models in \citet{2020ApJ...903...45K}. In addition, BH binaries have their components drawn from the same Gaussian and thus have comparable masses. All BH radii are defined to be 10 times their Schwarzschild radius to accommodate for the limitations of the PN framework at small orbital separations.

Every BH in our set of sequences, both seeds and the background population, are initiated non-spinning with initial spin distributions to be explored in future work. Spin cannot be ignored though, because BBH merger products are spun up through merger \citep{campanelli_maximum_2007}. Including spin effects at merger is paramount to assessing the true probability of retaining BH seeds since the spins of merging BHs dramatically amplifies GW recoil kicks up to ${\sim} 10^3 - 10^4 \, $km/s. BH spin is defined with the dimensionless Kerr parameter, $\chi$, in units of $S/m^2$ and is stored as a vector quantity. BH spin vectors are rotated using Euler angles as appropriate; e.g., alongside the randomly generated orientations of target/projectile scatterings and when transforming to the orbital plane of a BBH merger.

\subsubsection{Collisions}\label{sec:collisions}

When the separation of any two BHs comes within their code radii, \textsc{Tsunami} stops integration and sends a collision flag to be handled by \textsc{CuspBuilding}. The BHs are then merged according to mass, spin, and gravitational-wave recoil in-plane prescriptions of \citet{2018PhRvD..97h4002H}, linearly summed to the \citet{PhysRevD.87.084027} out of plane contributions. The GW recoil kick velocity vector is then summed to the COM velocity of the merged BBH and the remaining system is integrated in \textsc{Tsunami} for a small buffer time (about $5$~yr) to allow the BHs to resettle into a new hierarchy (if any). 


\subsubsection{Binary BH Properties}\label{sec:bbhp}

All SMA are randomly drawn from a log-normal distribution with a most probable value of $10^{-2} a_{\rm hs}$, and a median at the $10^{-1} a_{\rm hs}$, where $a_{\rm hs}$ is the hard-soft boundary of the binary in question, defined to be \citep{2003gmbp.book.....H} 
\begin{equation}
\begin{aligned}
     a_{\rm hs} &= \frac{G m_1 m_2}{\left \langle m_{\rm bh} v_{\rm rel}^2 \right \rangle}\left(1+ \frac{\langle m_{\rm bh} \rangle}{m_1 + m_2}\right) \\
     &= \frac{G m_1 m_2}{3 \langle m_{\rm bh} \rangle \sigma_{\rm rel}^2}\left(1+ \frac{\langle m_{\rm bh} \rangle}{m_1 + m_2}\right)
\end{aligned}
\end{equation}

 where $\sigma_{\rm rel}=\sqrt{2}\sigma_{\rm c,bh}$ is expressed in the relative motion frame and $(m_1,m_2)$ are the masses of binary components. These values represent a first foray into the effects BBH distributions have on a seed BH's evolutionary history; a more robust exploration will be provided in future work. 
 

Eccentricities are randomly drawn from the thermal distribution \citep{jeans1919}
\begin{equation}
f(e) de  = 2 e de. 
\end{equation}
The distribution is truncated at $e = 0.95$ because highly eccentric binary BHs rapidly circularize and merge \citep{PhysRev.136.B1224}.

\subsection{The Encounter Rate}\label{sec:ttne}
The time between encounters, $t_{\rm enc}$, experienced by our target hierarchy, $P_{\rm t}$, in \textsc{CuspBuiliding} follows from the collisional time scale derivation expressed in \citet[][eq.~7.194]{2008gady.book.....B}, generalized by averaging over a Maxwellian distribution expressed in the relative motion frame. We compute separate encounter rates for each projectile ``species'' (i.e., single versus binary BHs) to obtain a total rate for the target. The total rate determines the time to the next encounter while the species-specific rates are used to randomly select whether the next projectile is a single BH or BBH. The general encounter rate between $P_{\rm t}$ and a species population is

\begin{equation}
\Gamma_{\rm p} = 2 \sqrt{2 \pi} n_{\rm p} \sigma_{\rm rel} r_{\rm p,max}^2 \left(1 + \frac{G \,(m_{\rm t} + \langle m_{\rm p} \rangle)}{ r_{\rm p,max} \, \sigma_{\rm rel}^2}\right)
\end{equation}
where $n_{\rm p}$ is the number density of the projectile species, $\langle m_{\rm p} \rangle$ is the average mass of the projectile species, and $ r_{\rm p,max}$ is the maximal distance of closest approach considered for each species (see eq.~\ref{eq:rpmax}).

A species specific $n_{\rm p}$ may be determined using the following relations:

\begin{equation}\label{eq:N_Ns_Nb}
\begin{split}
    &N_{\rm bh} \approx \frac{M_{\rm bh}}{\langle m_{\rm bh} \rangle}\\
    &N_{\rm bh} = N_{\rm s} + 2 N_{\rm b}\\
    &f_{\rm b} = \frac{N_{\rm b}}{N_{\rm s} + N_{\rm b}}\\
    &N_{\rm s} = \frac{1-f_{\rm b}}{1+f_{\rm b}} N_{\rm bh}\\
    &N_{\rm b} = \frac{f_{\rm b}}{1+f_{\rm b}} N_{\rm bh},\\
\end{split}
\end{equation}
where $f_{\rm b}$ is the binary fraction (defined to be 10\% for all cluster initial conditions), $M_{\rm bh}$ and $\langle m_{\rm bh} \rangle$ are the total mass of BH sub-cluster and average BH mass in the BH sub-cluster, respectively, $N_{\rm s}$ is the number of single BHs, and $N_{\rm b}$ is the number of BBH systems. The species specific number densities are then

\begin{equation}
\begin{split}
    n_{\rm s} = \frac{1-f_{\rm b}}{1+f_{\rm b}} n_{\rm c,bh}\\
    n_{\rm b} = \frac{f_{\rm b}}{1+f_{\rm b}} n_{\rm c,bh}
\end{split}
\end{equation}
where $n_{\rm c,bh}$ is the average number density within the core of the BH sub-cluster assuming all single BHs (e.g., $n_{\rm c,bh} \approx \rho_{\rm c,bh}/\langle m_{\rm bh} \rangle$). 

Finally, our effective encounter rate is defined to be 
\begin{equation}\label{eq:tenc}
\begin{split}
&\Gamma_{\rm tot} = \Gamma_{\rm s} + \Gamma_{\rm b}\\
&R_{\rm b} = \frac{\Gamma_{\rm b}}{\Gamma_{\rm tot}}\\
&t_\mathrm{enc} = \Gamma_{\rm tot}^{-1}
\end{split}
\end{equation}
where $R_{\rm b}$ is the likelihood a binary object will be selected from the projectile reservoir to be the next $P_{\rm p}$. This calculation is done frequently to account for the unpredictability in final mass and size that $P_{\rm t}$ may settle into following an encounter. 

\subsection{Interaction Initialization}
\label{sec:scatinit}

A sequence begins by defining a target particle group, $P_{\rm t}$, initiated with a massive seed BH ($m_{\rm s} \geq 40\, M_\odot$) and a BH companion with a mass drawn from the Gaussian mass distribution. It is straightforward to initiate $P_{\rm t}$ with any custom hierarchy beyond a binary object, but for the purpose of this paper, we always initiate $P_{\rm t}$ as a binary and allow its hierarchical rank to organically grow through successive interactions. The SMA of $P_{\rm t}$ is defined at the median of our log-normal SMA distribution, $10\%$ the binary's hard-soft boundary, and the eccentricity is defined by the median of the thermal distribution, $e_{\rm 0} = 1/\sqrt{2}$. The orientation of our initial $P_{\rm t}$ is always defined in the x-y plane with its phase initiated at apocenter. 
  
Next, a set of projectile groups $\{P_{\rm p}\}$ are compiled using the distributions discussed in Secs.~\ref{sec:dpr}, \ref{sec:msr}, and \ref{sec:bbhp}. From our set, we then select a $P_{\rm p}$ randomly to fire at $P_{\rm t}$ each round according to $R_{\rm b}$ (see eq.~\ref{eq:tenc}). The cross-sectional area explored may be constrained by defining a maximal distance of closest approach
     
\begin{equation}\label{eq:rpmax}
\begin{aligned}
r_{\rm p,max} &= k \, \max[a_{\rm t}+a_{\rm p}, \, a_{\rm GR}]\\
a_{\rm GR} &= \left(\frac{85 \sqrt{2} \pi G^{7/2} m_{\rm t} m_{\rm p} (m_{\rm t} + m_{\rm p})^{3/2}}{12 c^5 \sigma_{\rm rel}^2}\right)^{2/7}
\end{aligned}
\end{equation}
where $k$ is an arbitrary buffer constant, $a_{\rm p}$ is the radial size of $P_{\rm p}$, and $a_{\rm GR}$ is the maximal periastron distance capable of producing a GW binary capture (see \citet[][]{quinlan_dynamical_1989}). Increasing $k$ captures more of the parameter space at the cost of dramatically increasing the amount of interactions required to finish a sequence. To balance accuracy with computational time, we choose $k=3$ (in line with the findings of \citet[][]{Fregeau_2007}) because $k\gtrsim3$ overwhelmingly results in weak interactions between $P_{\rm t}$ and $P_{\rm p}$. Given $r_{\rm p,max}$, the maximal impact parameter is

\begin{equation}
b_\mathrm{max} = r_\mathrm{p,max} \sqrt{1 +  \frac{2 G (m_{\rm t} + m_{\rm p})}{ r_\mathrm{p,max} v_{\rm \infty}^2}}
\end{equation}
where $v_{\rm \infty}$ is the relative velocity of $P_{\rm p}$ at an infinite particle separation. If we consider $b_\mathrm{max}$ to be the radius of the maximal cross-sectional area between $P_{\rm t}$ and $P_{\rm p}$, the impact parameter distribution is

\begin{equation}
b = \sqrt{{\cal U}(0,1)} \, b_{\rm max}
\end{equation}
where ${\cal U}(0,1)$ is a random number on the uniform interval $\{0...1\}$.

The initial location and velocities of $P_{\rm p}$ is determined by analytically calculating its location along a Keplerian hyperbolic orbit until it reaches a minimal initial separation, $r_{\rm min}$. To determine $r_{\rm min}$, we first consider a tidal tolerance, $\delta$, such that

\begin{equation}\label{eq:delta}
\frac{|F_\mathrm{tide}|}{|F_\mathrm{rel}|} < \delta
\end{equation}
where $\delta = 10^{-4}$ is selected to be arbitrarily small, $F_\mathrm{rel}$ is the relative force between the two members of the outermost orbit in the hierarchy of $P_\mathrm{t}$, and $F_\mathrm{tide}$ is the tidal force exerted on the outermost orbit of $P_\mathrm{t}$ by $P_\mathrm{p}$. Following a similar procedure to that outlined in \citet{2016MNRAS.456.4219A}, a maximal possible tidal force on $P_\mathrm{t}$ may be expressed as

\begin{equation}
|F_\mathrm{tide}| = \frac{2 G \mu_{\rm t} m_p}{r^3} a_{\rm t} (1 + e_{\rm t})
\end{equation}
where $\mu_{\rm t}$ is the reduced mass of $P_{\rm t}$, $a_{\rm t}$ and $e_{\rm t}$ are the SMA and eccentricity, respectively, of the outermost orbit of $P_{\rm t}$, $r$ is the distance between the COM of $P_{\rm t}$ and the COM of $P_{\rm p}$, and ($m_{\rm t}$,~$m_{\rm p}$) are the total mass of ($P_{\rm t}$,~$P_{\rm p}$). We then define a minimal relative force between the constituents of $P_{\rm t}$

\begin{equation}\label{eqn:Frel}
|F_\mathrm{rel}| = \frac{G m_{\rm t0} m_{\rm t1}}{[a_{\rm t} (1 + e_{\rm t})]^2}
\end{equation}
where $m_{\rm t0}$ is the total mass of the masses contained within $a_{\rm t}$ and $m_{\rm t1}$ is the mass drawing the orbit of $a_{\rm t}$. Using eq.~\ref{eq:delta}, we may define our minimum initial separation to be

\begin{equation}\label{eqn:rmin}
r_\mathrm{min} = \max\left(10, \, \sqrt[3]{\frac{2 m_{\rm p}}{\delta \ m_{\rm t}}} \,\right)\, a_{\rm t} (1+e_{\rm t}).
\end{equation} 
With $v_{\rm \infty}$, $b$, and $r_{\rm min}$ in hand, the initial conditions of the interaction between $P_{\rm t}$ and $P_{\rm p}$ are handed to \textsc{tsunami}.

\subsection{Hierarchy Size Restraints}

A conservative limit on maximum hierarchy size, $r_{\rm max}$, is placed upon all hierarchies such that 

\begin{equation}\label{rmax}
\begin{split}
    r_{\rm max} &= \frac{1}{10} \left(\frac{4 \pi n_{\rm eff}}{3}\right)^{-1/3}\\
    &=0.1 \langle r_{\rm sep} \rangle\\
    n_{\rm eff} &= n_{\rm s} + n_{\rm b}
\end{split}
\end{equation}
No hierarchy extending beyond $10\%$ of the average inter-particle distance, $\langle r_{\rm sep} \rangle$, of the BH sub-cluster core is allowed to continue to the next interaction unmodified. \textsc{CuspBuilding} labels all members of a hierarchy with an SMA greater than this threshold to be ``unbound" alongside other BHs unbound to $P_{\rm t}$. \textsc{CuspBuilding} then removes unbound BHs when tidally appropriate (eq. \ref{eqn:rmin}).\footnote{Choosing $r_{\rm max} = 0.1 \langle r_{\rm sep} \rangle$ also allows us to ignore the tidal force enacted by the potential of the BH sub-cluster on a hierarchy, $F_{\rm tide,cl}$, since the ratio of $F_{\rm tide,cl}$ to $F_{\rm rel}$ (eq.~\ref{eqn:Frel}) at $r_{\rm max}$ is $\lesssim 10^{-4}$ across all models.}


\subsection{Orbital Stability}\label{sec:stability}
 We employ two stability criterion based on the tree of hierarchies contained in $P_{\rm t}$. The first criterion uses eqs.~(11), (12), and (13) in \citet{2018MNRAS.476..830M}, 
  \begin{equation}
  \begin{aligned}
      Q_{\rm st} &= A \left(\frac{\lambda \sqrt{N}}{1-e_{\rm out}}\right)^{1/6} (f g)^{1/3}  \\
      f &= f_1 - 0.3 \cos{i} \, f_2  \\
      f_1 & = 1 - \frac{2}{3} e_{\rm in} \left(1 - \frac{e_{\rm in}^2}{2}\right) \\
      f_2 & =  1 - \frac{e_{\rm in}}{2} + 2 \cos{i} \left(1-\frac{5}{2} e_{\rm in}^{3/2} - \cos{i} \right) \\
      g &= 1 + \frac{m_3}{m_1 + m_2},\\
  \end{aligned}
  \end{equation}
 where $Q_{\rm st} < \frac{a_{\rm out} \left(1-e_{\rm out}\right)}{a_{\rm in}}$ is the stability threshold, $\{a_{\rm in}, a_{\rm out}\}$ and $\{e_{\rm in}, e_{\rm out}\}$ are the inner/outer SMA and eccentricity of a hierarchical layer, $i$ is the orbital inclination relative to recursive layers, $m_1$ and $m_2$ are the masses of the inner binary, and $m_3$ is the mass of the outer tertiary. We also choose a conservative threshold for stability with $A=2.0$, $\lambda = 1$, and $N=10^4$. This criterion is applied recursively to each hierarchy layer of $P_{\rm t}$ to determine the absolute stability of the system.
 
 The second criterion employs the standard GW decay time, $T_{\rm decay}$, \citep{PhysRev.136.B1224} and is integrated for the inner-most binary. If $T_{\rm decay}<t_{\rm enc}$, then $P_{\rm t}$ is unconditionally labeled as unstable. If the rest of the hierarchical system is determined to be dynamically stable, the inner most binary is instantaneously merged within its respective orbital plane and a GW recoil kick is applied (see Sec.~\ref{sec:collisions}). 
 
Should $P_{\rm t}$ be labeled as stable following the above assessments, then it is simply ported to the next interaction without further integration in \textsc{Tsunami}. If it is unstable, $P_{\rm t}$ is integrated in \textsc{Tsunami} until stability is reached or until $t_{\rm f} = t_{\rm 0} + t_{\rm enc}$, where $t_0$ denotes the time at periapse of the last interaction. In the case that $P_{\rm t}$ is a binary, that binary is integrated using Peter's equation until $t_{\rm f}$. Outgoing $P_{\rm p}$ not bound to $P_{\rm t}$ are only extracted from the integrator once they have reached a tidal distance threshold $r_{\rm min}$ (eq.~\ref{eqn:rmin}).



\subsection{Dynamical Friction Delay} \label{sec:dfd}

To accurately mimic the shift in the $\Gamma_{\rm tot}$ (eq.~\ref{eq:tenc}) when $P_{\rm t}$ is kicked out of the BH sub-cluster, we increase $t_{\rm enc}$ by the dynamical friction decay timescale, $t_{\rm df}$ (i.e., $t_{\rm enc} \xrightarrow{} t_{\rm enc} + t_{\rm df}$). This increase is applied following each interaction when the final velocity of the new $P_{\rm t}$ in the COM frame of the most recent interaction, treated as the kick velocity for simplicity ($v_{\rm kick}$), exceeds the BH sub-cluster core escape velocity, $v_{\rm e,cbh}$.

There are two sources that may prompt kicks in our simulations: (i) dynamical kicks from strong COM velocity perturbations during close encounters with hierarchical systems and (ii) GW recoil during the asymmetric emission of GW radiation during merger. We note that dynamical kicks experienced by $P_{\rm t}$ across all cluster models are usually much less than $v_{\rm e,cbh}$, only rarely exceeding it during chaotic encounters. The overwhelming majority of sequence disrupting kicks are generated by GW recoil during merger. Our procedure for calculating $t_{\rm df}$ is as follows.

First, we calculate the apocenter distance at which the new $P_{\rm t}$ is ``deposited,'' $r_{\rm dep}$, assuming a purely radial orbit. This is computed numerically using conservation of energy in the cluster potential. Using $r_{\rm dep}$, we estimate $t_{\rm df}$ in the secular approximation, i.e., the decaying radial orbit is approximated by orbit averaging the energy dissipation due to dynamical friction.

Beginning with the velocity along a bound radial orbit, 

\begin{equation}\label{eq:rdotex}
    \dot r = \sqrt{2 [\Phi(r_{\rm a}) - \Phi(r)]},
\end{equation}
where $r_{\rm a}$ is the apocenter distance in the potential $\Phi(r)$. Solving for the period,
\begin{equation}\label{eq:tau_a}
    \tau(r_{\rm a}) = \int^\tau_0 dt = 4 \int^{r_{\rm a}}_0 (2[\Phi(r_{\rm a}) - \Phi(r)])^{-1/2} dr.
\end{equation}
In the secular approximation, the orbit-averaged energy loss per orbit may be expressed as

\begin{equation}\label{eq:delE}
\begin{aligned}
    \delta E(r_{\rm a}) &= \int^{\tau(r_{\rm a})}_0 F_{\rm df}(r,\dot r) \dot r dt\\
                &= 4 \int^{r_{\rm a}}_0 F_{\rm df}(r,\dot r) dr.\\
\end{aligned}
\end{equation}
Eqns.~(\ref{eq:tau_a}) and (\ref{eq:delE}) may be combined to obtain the orbit averaged rate of energy loss,

\begin{equation}\label{eq:dEdt}
\begin{aligned}
    \frac{\delta E (r_{\rm a})}{\tau(r_{\rm a})} = \frac{\int^{r_{\rm a}}_0 F_{\rm df}(r,\dot r) dr}{\int^{r_{\rm a}}_0 (2[\Phi(r_{\rm a}) - \Phi(r)])^{-1/2} dr}.\\
\end{aligned}
\end{equation}
Relating the change in orbital energy to the change in apocenter distance, $r_{\rm a}$,

\begin{equation}
    \Phi(r_{\rm a}) - \frac{1}{m_{\rm t}}\delta E = \Phi(r_{\rm a}-\delta r_{\rm a})
\end{equation}
and in the perturbation limit

\begin{equation}
\begin{aligned}
    \Phi(r_{\rm a}) - \frac{1}{m_{\rm t}}\delta E(r_{\rm a}) &\approx \Phi(r_{\rm a}) - \Phi'(r_{\rm a}) \delta r_{\rm a}\\
    \frac{\delta t}{\delta r_{\rm a}} &\approx \frac{m_{\rm t} \Phi'(r_{\rm a}) \delta t}{\delta E(r_{\rm a})}\\
    \delta t &\approx \frac{m_{\rm t} \Phi'(r_{\rm a}) }{\delta E(r_{\rm a})/\delta t}\delta r_{\rm a} , \\
\end{aligned}
\end{equation}
where $\delta t$ expresses a time-scale long compared to the orbital period but short compared to the orbital decay time. The total decay time time for the apocenter distance to decrease from its initial value, $r_{\rm dep}$, to the core of the BH sub-cluster, $r_{\rm c,bh}$, may then be evaluated as

\begin{equation}\label{eq:seculartdf}
 t_{\rm df} = \int^{t_{\rm df}}_0 dt \approx \int^{r_{\rm c,bh}}_{r_{\rm dep}} \frac{m_{\rm t} \Phi_{\rm tot}'(r_{\rm a}) }{\delta E(r_{\rm a})/\delta t} dr_{\rm a}
\end{equation}
with $\Phi_{\rm tot}(r)$ defined in eq.~\ref{eqn:phitot} and $\delta E(r_{\rm a})/\delta t$ defined above in eq.~\ref{eq:dEdt}.

Given an average mass in a cluster, $\langle m \rangle$, and the total mass of $P_{\rm t}$, $m_{\rm t}$, $F_{\rm df}$ is calculated with Chandrasekhar's dynamical friction formula as expressed for a Maxwellian velocity distribution in a spherically symmetric potential \citep[][Ch. 8.1]{2008gady.book.....B},

\begin{equation}\label{eq:fdyn}
\begin{aligned}
    &\vec F_{\rm df}(r, v) =-\frac{4 \pi G^2 m_{\rm t} (m_{\rm t}+\langle m\rangle)\rho(r) \, \alpha(r,v) \ln\Lambda(r,v)}{v^2}  \hat v \\
    & \alpha(r,v) = \erf(X) - \frac{2 X}{\sqrt{\pi}} e^{-X^2}\\ 
    & X = \frac{v}{\sqrt{2} \sigma(r)}\\
    &\Lambda(r,v) = \frac{\max[r,\, r_{\rm c,bh}]} {\max\left[a_{\rm t},\, \frac{G \left(m_t + \langle m\rangle\right)}{v^2 + 3\sigma(r)^2}\right]}
\end{aligned}
\end{equation}
where $a_{\rm t}$ is the SMA of $P_{\rm t}$, $\ln(\Lambda)$ is the Coulomb Logarithm, and $\vec{v} = v \hat{r}$ within the context of our radial orbit approximation. Assuming that $P_{\rm t}$ is subject to linearly independent dynamical friction forces from our double-Plummer cluster, one from the primary cluster and one from the BH sub-cluster, we define the dynamical friction force as
\begin{equation}
    F_{\rm df}(r, \dot r) = F_{\rm df,cl}(r,\dot r) + F_{\rm df,bh}(r,\dot r)
\end{equation}
with the density and velocity dispersion terms within eq.~\ref{eq:fdyn} defined by their respective clusters (Sec.~\ref{sec:dpmodel}).

\begin{figure}
\includegraphics[width=.475\textwidth]{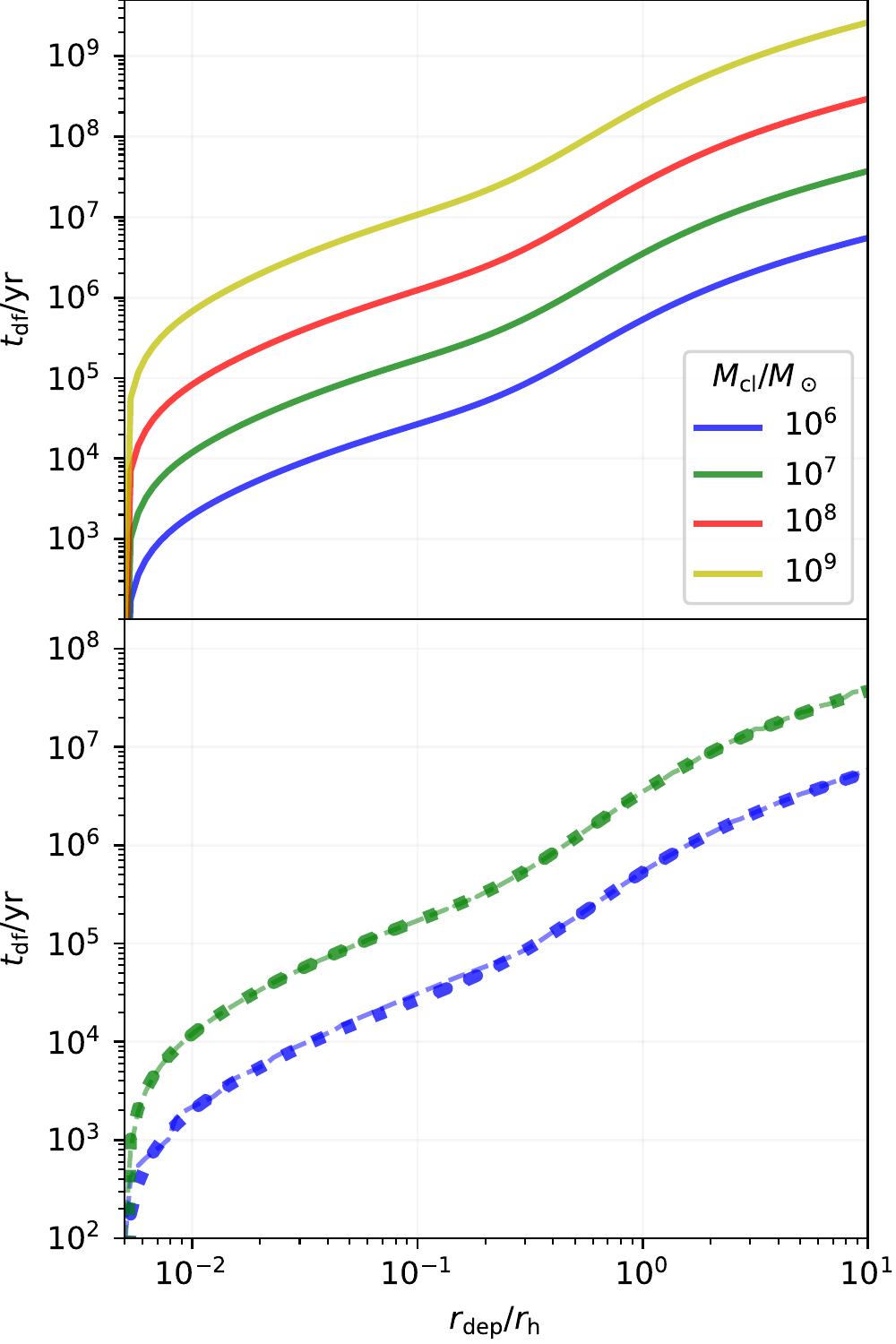}
\caption{Time to decay, $t_{\rm df}$, from a radial distance $r_{\rm dep}$ to $r_{\rm c,bh}$ for a $100 \, M_\odot$ BH via dynamical friction. Here we express the distance in units of the half-mass radius, $r_{\rm h}$. Color indicates $M_{\rm cl}$ and all curves are expressed for a double-Plummer model with scaling parameter $\epsilon_{\rm b} = 0.01$. The bottom panel displays the decay time resultant from integrating the full equation of motion (dashed curve) and our secular solution (eq.~\ref{eq:seculartdf}; dotted curve).}
\label{fig:tdfvsadep}
\end{figure}

An illustration of our prescription's $t_{\rm df}$ values as a function of $r_{\rm dep}$ is shown in the upper panel of Fig.~\ref{fig:tdfvsadep} and a comparison between the full numerical integration of the equation of motion and our secular approximation (eq.~\ref{eq:seculartdf}) are displayed in the lower panel.\footnote{While the Plummer distribution function yields an analytic solution for the Chandrasekhar dynamical friction force, we find minimal differences in the decay times between the Plummer and the Maxwellian friction forces.}

\subsection{Encounter Rate Dependence on Initial Conditions}

More encounters lead to more collisions. As such, it is crucial to document how the encounter rate, $\Gamma$ (eq.~\ref{eq:tenc}), reacts to changes in host cluster initial conditions. For convenience in the following calculation, BH sub-cluster quantities critical to $\Gamma$ may be approximated to be

\begin{equation}\label{eqn:bhclusterapprox}
    \begin{split}
    &n_{\rm c,bh} \approx \frac{\rho_{\rm c,bh}}{\langle m_{\rm p} \rangle} = \frac{3 M_{ \rm bh}}{4 \pi b_{\rm bh}^3 \langle m_{\rm p} \rangle}\\
    &r_{\rm p} \approx k a_{\rm hs} \approx \frac{k \, G m_1 m_2}{3 \langle m_{\rm p}\rangle \sigma_{ \rm c,bh}^2}\\
    &\sigma_{\rm c,bh} \approx \sqrt{\frac{G M_{\rm  bh}}{6 b_{\rm  bh}}}.
    \end{split}
\end{equation}
Substituting (\ref{eqn:bhclusterapprox}) into (\ref{eq:tenc}),

\begin{equation}
    \Gamma \approx \frac{ k^2 \sqrt{6 G} }{ \sqrt{\pi} \sqrt{M_{\rm \rm bh} b_{ \rm bh}^3} } \frac{ m_{\rm 1}^2 m_{\rm 2}^2 }{ m_{\rm p}^3} \left( 1 + \frac{3 m_{\rm p} \left(m_1 + m_2 + m_{\rm p} \right)}{2 k m_{\rm 1} m_{\rm 2}} \right)
\end{equation}
which scales as
\begin{equation}
    \Gamma \propto \left(M_{\rm \rm bh} b_{\rm \rm bh}^3 \right)^{-1/2}
\end{equation}
with changes in cluster initial conditions. Using our selected mass/radius scaling relations (eq.~\ref{eq:georgiev}), $\Gamma$ becomes
\begin{equation}\label{gamprop}
\begin{aligned}
    &b \equiv b(M_{\rm \rm cl})\\
    &b_{\rm \rm bh} = \epsilon_{\rm \rm b} \, b(M_{\rm \rm cl})\\
    &\Gamma \approx C_1 \left(\epsilon_{\rm \rm M} \, \epsilon_{\rm \rm b}^3 \, M_{\rm \rm cl} \, b\left(M_{\rm \rm cl}\right)^3 \right)^{-1/2}. 
\end{aligned}
\end{equation}
where $C_1$ is an arbitrary scaling constant.

\begin{figure}
\includegraphics[width=.47\textwidth]{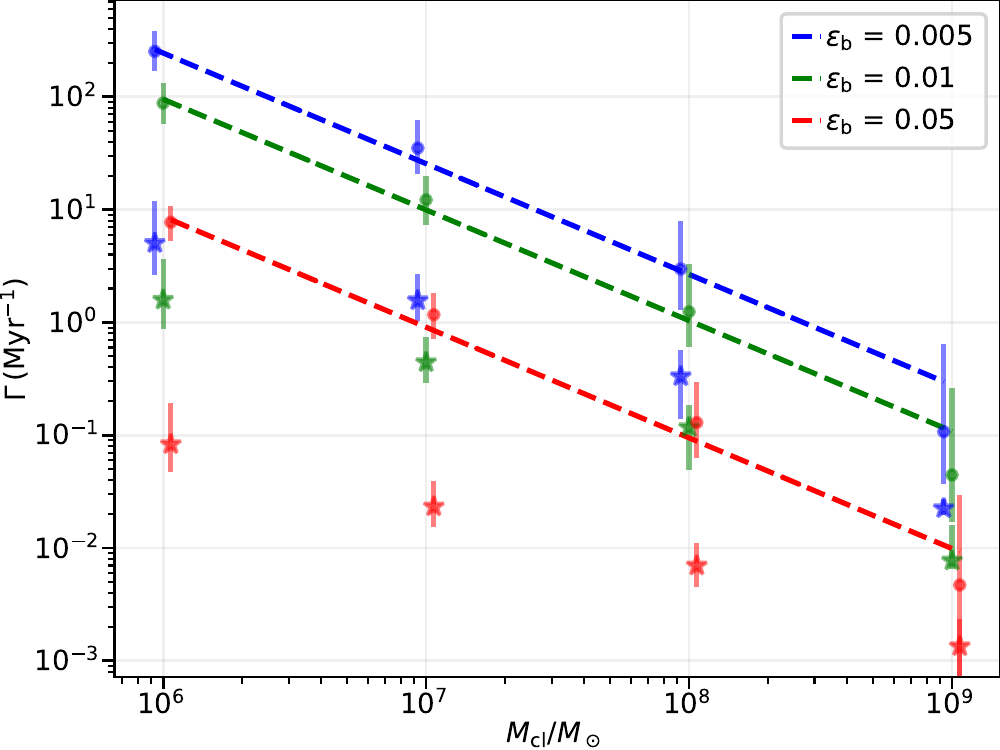}
\caption{Median encounter rate (circles) and merger rate (stars) experienced in the first ${\sim}100$ interactions in every sequence with respect to cluster initial condition. The dashed line is an analytic fit defined by eq.~\ref{gamprop}. Small horizontal offsets are applied for scaling parameter, $\epsilon_{\rm b}$, for easier visualization. Dynamical friction decay, $t_{\rm df}$, is not included here for consistency.}
\label{fig:encrate}
\end{figure}

Despite the massive dynamical variability in each sequence, eq.~\ref{gamprop} produces a reasonable tracing of the median encounter rate with respect to cluster mass, $M_{\rm cl}$, as can be seen in Fig.~\ref{fig:encrate}. A key takeaway is that encounter and merger rates are inversely proportional to the cluster mass, $M_{\rm cl}$, across all mass and density scales.

We note that the cluster is implicitly assumed to be in a state of balanced evolution (i.e. gravothermal equilibrium; see \citep{2013MNRAS.432.2779B, antonini_bhgrowth_2019}), but we only compute our encounter rate for the growing seed. The self-consistent treatment of binary interactions is represented by our simple distribution of binary properties and binary fraction. Our above calculation only concerns the rate of interactions of the growing seed hierarchy and is not meant to apply to binary interactions in general.

\begin{table}%
\centering%
\caption{Initial condition table describing all double-Plummer models. In order from left to right, the columns are the total cluster mass, $M_{\rm cl}$, the sub-cluster radial scaling parameter, $\epsilon_{\rm b}$, the half-mass radius of the cluster, $r_{\rm h}$, the escape velocity of the double-Plummer model, $v_{\rm e,cl}$, the escape velocity the BH sub-cluster core, $v_{\rm e,cbh}$, the average mass density of the BH sub-cluster core, $\rho_{\rm c,bh}$, the average velocity dispersion of the sub-cluster core, $\sigma_{\rm c,bh}$, and the deviation from energy equipartition between the BH sub-cluster and the primary cluster, $\eta$.}
\fontsize{7.9pt}{8}\selectfont
\addtolength{\tabcolsep}{-3.5pt} 
\begin{tabular}{l || llrrrrrr}
\toprule
{} &      $M_{\rm cl}$ &        $\epsilon_{\rm b}$ & $r_{\rm h}$ & $v_{\rm e,cl}$ & $v_{\rm e,cbh}$ & $\rho_{\rm c,bh}$ & $\sigma_{\rm c,bh}$ &                                                                              $\eta$ \\
{} & $\left(M_{\odot}\right)$ & $r_{\rm h,bh}/r_{\rm h}$ &  $(pc)$ &     $(km/s)$ &     $(km/s)$ &       $\left(M_\odot \, pc^{-3}\right)$ &                       $(km/s)$ & $\frac{ T_{\rm bh}}{ T_{\rm cl}}$ \\
\midrule
0  &      $10^{6}$ &                0.05 &     2.1 &         79 &         27 &              ${3.3}\times 10^{6}$ &                            9 &                                                4 \\
1  &      $10^{6}$ &                0.01 &     2.1 &        103 &         58 &              ${4.1}\times 10^{8}$ &                           20 &                                               20 \\
2  &      $10^{6}$ &               0.005 &     2.1 &        126 &         82 &              ${3.3}\times 10^{9}$ &                           29 &                                               40 \\
\hline
3  &      $10^{7}$ &                0.05 &     4.5 &        174 &         59 &              ${3.6}\times 10^{6}$ &                           20 &                                                4 \\
4  &      $10^{7}$ &                0.01 &     4.5 &        224 &        127 &              ${4.5}\times 10^{8}$ &                           44 &                                               20 \\
5  &      $10^{7}$ &               0.005 &     4.5 &        274 &        180 &              ${3.6}\times 10^{9}$ &                           63 &                                               40 \\
\hline
6  &      $10^{8}$ &                0.05 &     9.4 &        379 &        128 &              ${3.9}\times 10^{6}$ &                           43 &                                                4 \\
7  &      $10^{8}$ &                0.01 &     9.4 &        490 &        278 &              ${4.9}\times 10^{8}$ &                           97 &                                               20 \\
8  &      $10^{8}$ &               0.005 &     9.4 &        600 &        392 &              ${3.9}\times 10^{9}$ &                          137 &                                               40 \\
\hline
9  &      $10^{9}$ &                0.05 &    19.6 &        829 &        280 &              ${4.2}\times 10^{6}$ &                           95 &                                                4 \\
10 &      $10^{9}$ &                0.01 &    19.6 &       1070 &        606 &              ${5.3}\times 10^{8}$ &                          212 &                                               20 \\
11 &      $10^{9}$ &               0.005 &    19.6 &       1310 &        857 &              ${4.2}\times 10^{9}$ &                          300 &                                               40 \\
\bottomrule
\end{tabular}

\label{table:IC}
\end{table}
\section{Results} \label{sec:results}
In this paper we present initial results from
\begin{enumerate}
    \item calculations of $\sim$17 million interactions using a 3-by-4 grid of 12 host cluster initial conditions, with 1500 realizations per cluster initial condition,
    \begin{equation}
        \begin{aligned}
        M_{\rm cl}/M_\odot & = \{10^6, 10^7, 10^8, 10^9\} \\
        \epsilon_{\rm b} &= \{0.05, 0.01, 0.005\}, \\
        \end{aligned}
    \end{equation}
    performed for two seed masses, ${m_{\rm s0}/M_\odot=\{50, 100\}}$.
    \item calculations of $\sim$18 million interactions using a 3-by-1 grid of 3 host cluster initial conditions, with 1500 realizations per cluster initial condition,
    \begin{equation}
    \begin{aligned}
    M_{\rm cl}/M_\odot & = \{10^6\} \\
    \epsilon_{\rm b} &= \{0.05, 0.01, 0.005\}, \\
    \end{aligned}
    \end{equation}
    performed for six different seed masses, \\${m_{\rm s0}/M_\odot=\{50, 100, 150, 200, 250, 300\}}$.
\end{enumerate}

The selected seed masses are motivated by CMC simulations of young, massive clusters \citep{Kremer_2020_IMBH, gonzalez_intermediate-mass_2021}, though they may be more massive in principle. All initial conditions pertaining to variations of our double-Plummer models are displayed in Table~\ref{table:IC}.

Only a single computationally expensive interaction is needed to completely prevent the completion of a sequence. Despite this, the majority of models reach completion with each set of sequences having a completion rate of ${>99\%}$.

\subsection{Sequence End-States}

A BH seed sequence loops until one of three critical junctures is reached:

\begin{enumerate}

 \item The seed is ejected from the cluster due to a dynamical or GW recoil kick (i.e., ${v_{\rm kick}>v_{\rm e,cl}}$).

 \item The seed is experiencing runaway growth. The seed is labeled as a runaway if it reaches a mass ${m_{\rm s} \geq 1000 \, M_\odot}$ within the {3~Gyr} sequence duration.  

 \item The seed has survived encounters for a time ${t_{\rm f}=3 \,\rm{Gyr}}$ without escaping or runaway.
\end{enumerate}

Escaping BHs are labeled as ``escapees" and form the most common end-state in most double-Plummer models. The ``runaway", is a BH which is experiencing an exponential growth due to an exponentially increasing rate of mergers. We find in our simulations that 100\% of models that reach $1000\, M_\odot$ grow indefinitely, thus we adopt $m_{\rm run}=1000\, M_\odot$ as a our threshold for runaway and stop all simulations when a BH reaches this mass. We also adopt an \textit{effective runaway fraction}, 
\begin{equation}\label{eq:effrunfrac}
    f_{\rm run} = \frac{m_{\rm s} - m_{\rm s0}}{m_{\rm run} - m_{\rm s0}}
\end{equation}
where $m_{\rm s}$ is the BH seed mass at the point of evaluation and $m_{\rm s0}$ is the initial BH seed mass. This quantity is useful to draw a consistent comparison of mass growth across different initial seed masses. Finally, if $P_{\rm t}$ reaches 3~Gyr of evolution time without escaping or reaching ${m_{\rm run}=1000\, M_\odot}$, we label the seed as ``stalled".

\subsection{Final Mass Distributions}

\begin{figure*}
\begin{minipage}[b]{.49\textwidth}
\centering
\includegraphics[width=.965\textwidth]{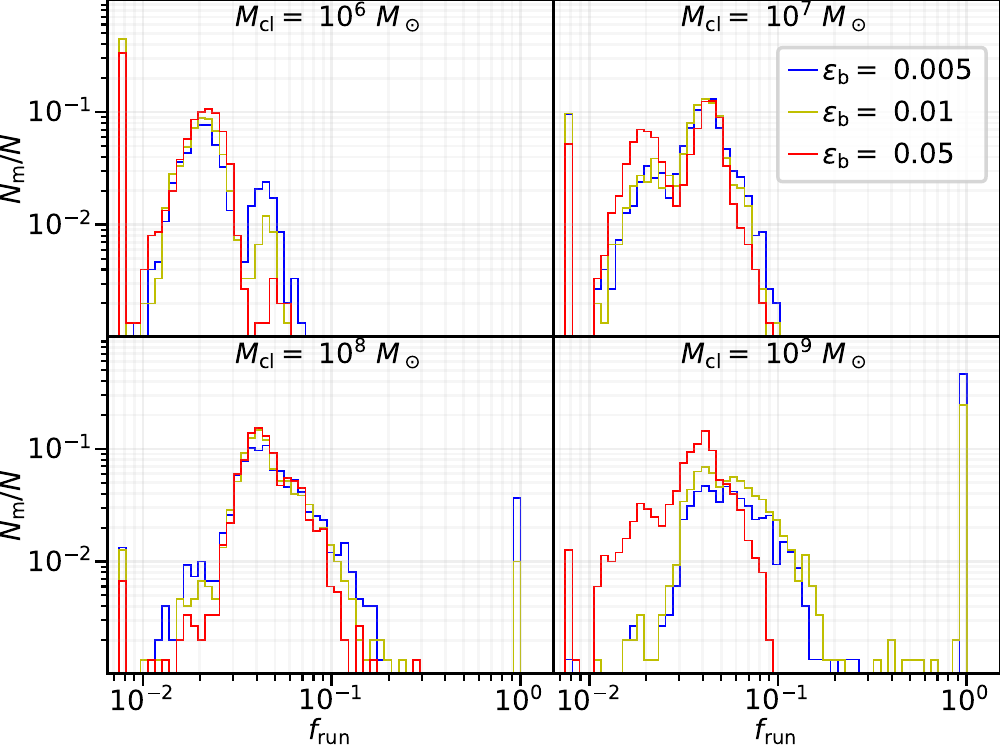}{\\(a) $m_{\rm s0} = 50 \,M_\odot$}
\end{minipage}
\begin{minipage}[b]{.49\textwidth}
\centering
\includegraphics[width=.965\textwidth]{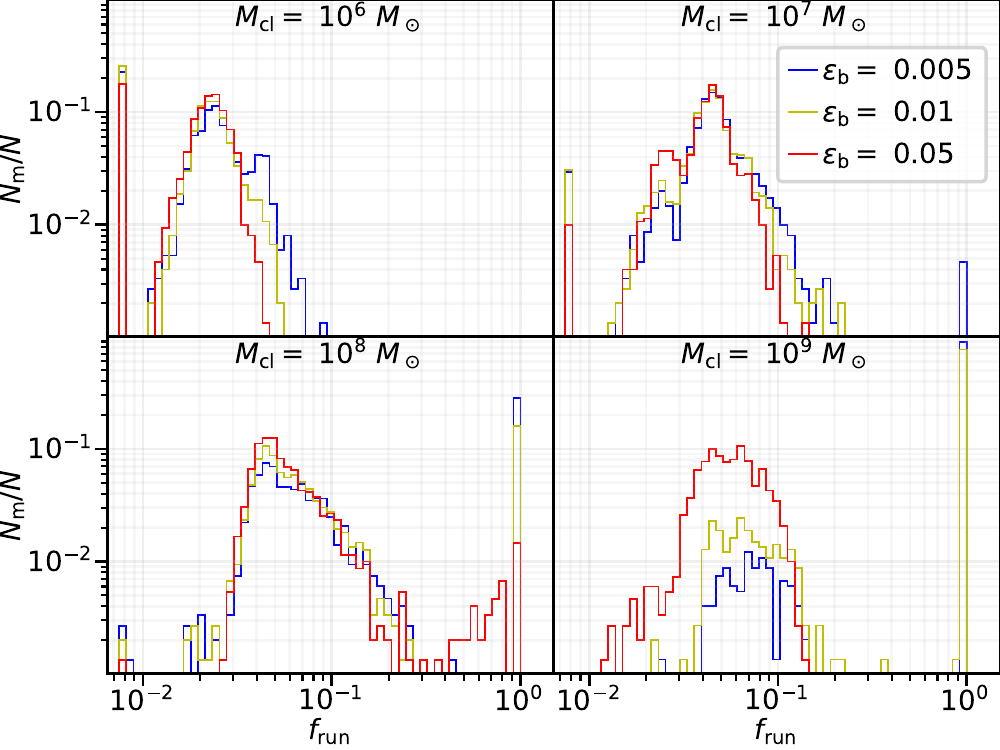}{\\(b) $m_{\rm s0} = 100 \,M_\odot$}
\end{minipage}

\caption{Distribution of final seed masses expressed in terms of the \textit{effective runaway fraction}, $f_{\rm run}$ (see eq.~\ref{eq:effrunfrac}), with each grid member showing the final seed mass distribution with cluster mass, $M_{\rm cl}$. BH seeds which do not undergo any mergers are deposited into the leftmost bin of each grid ($f_{\rm run}=8 \times 10^{-3}$).}
\label{fig:mfdist}
\end{figure*}

In Fig.~\ref{fig:mfdist}, we show the final mass distribution of all sequence end-states. The final seed mass, $m_{\rm f}$, distributions follow intuitive trends, favoring larger $m_{\rm f}$ as $\rho_{\rm c,bh}$ and $m_{\rm s}$ increase. These terms increase the escape velocity and reduce the magnitude of GW recoil kicks, mitigating the tendency for merger products to escape their host cluster and encouraging follow-up mergers.

Notably, there are four distinct peaks each panel may exhibit: at $f_{\rm run} = 0$ (shifted to $7 \times 10^{-3}$ for clarity), the fraction of BH seeds which do not undergo a merger is displayed; $>90\%$ also being escapees through dynamical kicks. The peaks lying at $f_{\rm run} \approx 2 \times 10^{-2}$ and $4 \times 10^{-2}$ correspond to escapees following one and two mergers, respectively. The peak at $\approx 2 \times 10^{-2}$ tends to shrink and disappear with increasing $\rho_{\rm c,bh}$ and $m_{\rm s}$. In such models, only a GW recoil kick amplified by a spinning seed is powerful enough to eject the merger product. The final peak lies at the runaway threshold, $m_{\rm s} = m_{\rm run}$, corresponding to $f_{\rm run} = 1$. The runaway probability tends to increase by a factor of $10-50$ with an order of magnitude increase in $M_{\rm cl}$, while a factor of 2 decrease in $\epsilon_{\rm b}$ corresponds to a factor of $\sim$2 increase in runaway probability.

\begin{figure*}
\begin{minipage}[b]{.39\textwidth}
\centering
\includegraphics[width=\textwidth]{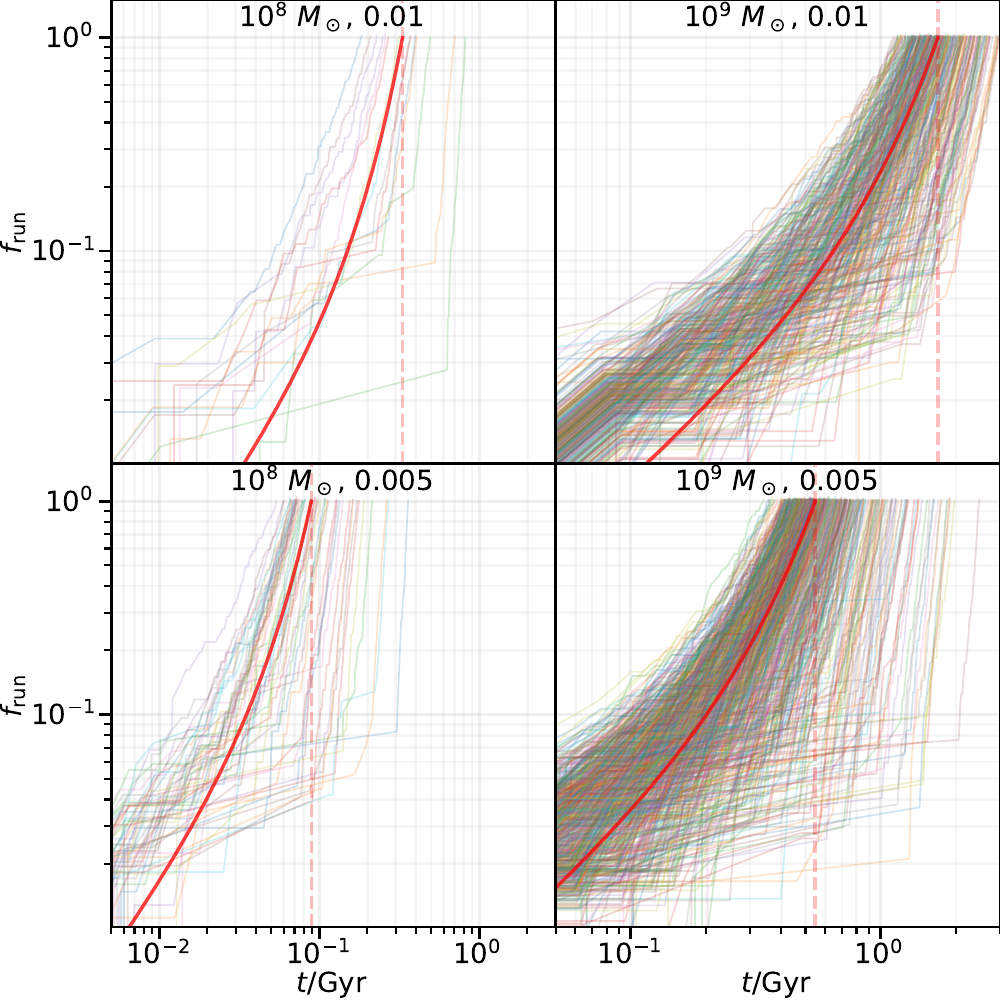}{\\(a) $m_{\rm s0} = 50 \,M_\odot$}
\end{minipage}
\begin{minipage}[b]{.59\textwidth}
\centering
\includegraphics[width=\textwidth]{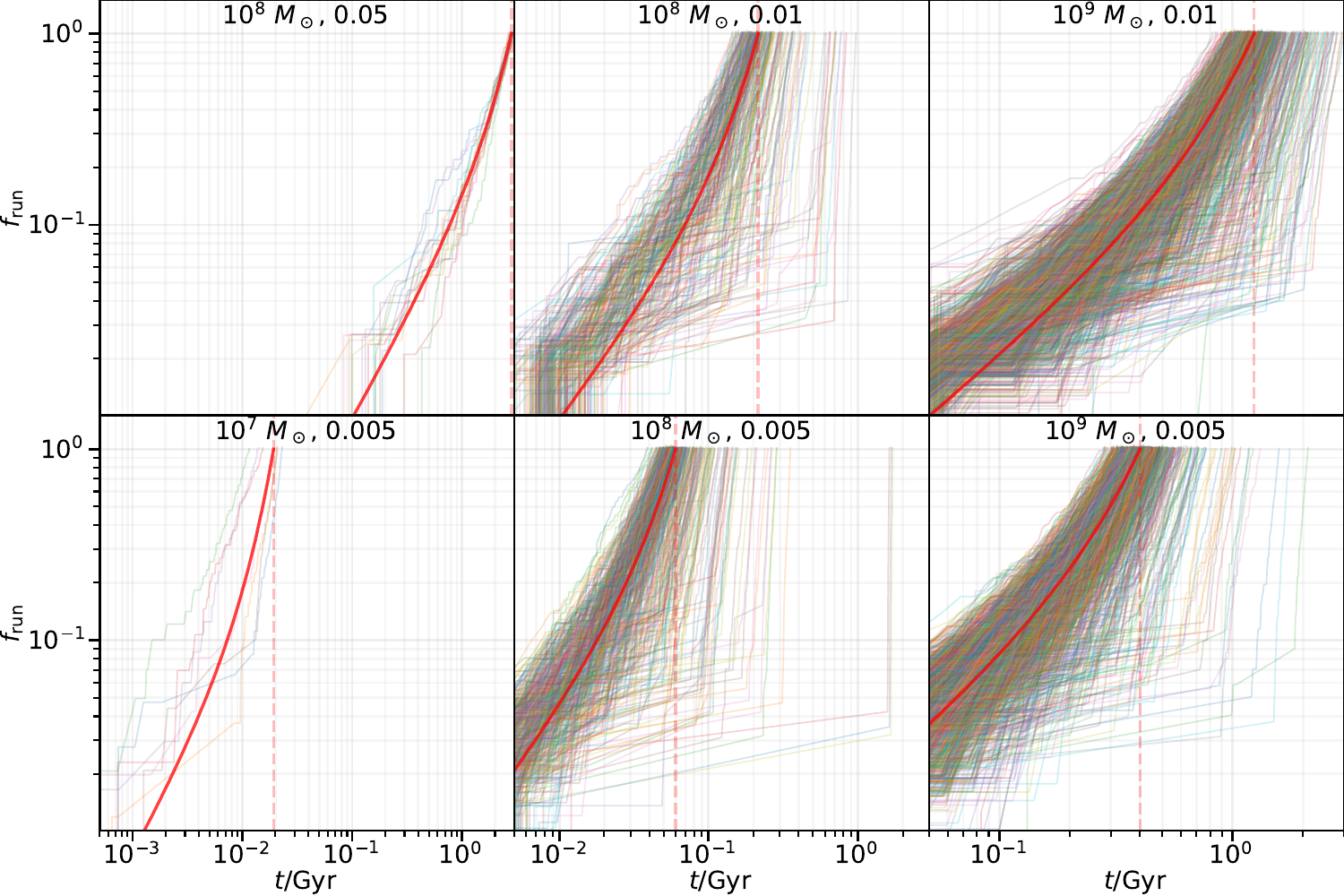}{\\(b) $m_{\rm s0} = 100 \,M_\odot$}
\end{minipage}

\caption{All runaway BH seed masses with respect to time, organized by double-Plummer model $(M_{\rm cl}, \epsilon_{\rm b})$. Each lightly shaded line represents an individual seed's effective runaway fraction, $f_{\rm run}$, with respect to time. The red line is a simple exponential fit to the median $f_{\rm run}$ across all runaway sequences. Thus, the red line shows the mass growth of a ``typical" runaway BH. The dashed vertical line denotes the time the median growth curve reaches our runaway threshold, $m_{\rm run} = 1000\,M_\odot$. We do not include $M_{\rm cl} = 10^6 \,M_\odot$ models because no runaways occur for $m_{\rm s0}/M_\odot = \{50, 100\}$ and $M_{\rm cl} = 10^7\, M_\odot$ panels are omitted for $m_{\rm s0} = 50 \, M_\odot$ due to a lack of runaways. The $m_{\rm s0} = 100 M_\odot$ seed includes the only $M_{\rm cl} \geq 10^8 M_\odot$, $\epsilon_{\rm b} = 0.05$ model hosting runaways within the $3 \, \text{Gyr}$ time limit and has have been included above the $(10^7 M_\odot, 0.005)$ panel.}
\label{fig:masshistrem}
\end{figure*}

In Fig.~\ref{fig:masshistrem}, we display the mass growth rate versus time for the set of all runaway sequences in each double-Plummer model. The figure shows that a decrease in $\epsilon_{\rm b}$ by a factor of ${\sim}2$ corresponds to a factor of ${\sim}2-3$ reduction in the time it takes to reach the runaway mass, $t_{\rm run}$, for a constant $M_{\rm cl}$. While runaways are less frequent in lower mass clusters, BH seeds reach runaway ${\sim}10$ times faster when decreasing $M_{\rm cl}$ by an order of magnitude; i.e., ${t_{\rm run}' \approx (M_{\rm cl}'/M_{\rm cl})(\epsilon_{\rm b}'/\epsilon_{\rm b}) t_{\rm run}}$. These trends follow directly from the fits which produce Fig.~\ref{fig:encrate}, expressing that encounter rates of individual BHs tend to decrease with a larger cluster mass, but increase with central density with \citet[][]{10.1093/mnras/stw093} NSC models. Notably, our fixed double-Plummer models predict the runaway seed BH will reach $10^5\, M_\odot$ within ${{\sim}3 t_{\rm run}}$ across all models. This means our seeds may reach SMBH status within roughly $0.2-1.0$~Gyr and $1.2-5.4$~Gyr in $10^8\, M_\odot$ and $10^9\, M_\odot$ clusters, respectively (excluding $\epsilon_{\rm b} = 0.05$ models).

\subsection{Critical Mass}

In rudimentary analysis, once the seed BH reaches a ``critical mass", $m_{\rm crit}$, it becomes unlikely that any strong encounter may prevent runaway growth. To find a reasonable boundary on $m_{\rm crit}$, we define it to be the boundary at which $>$$50\%$ of seeds which reach $m_{\rm crit}$ will also reach $m_{\rm run}$. We will also limit our calculations to models where more than one seed achieves runaway status. In any other circumstance, there is no way to calculate a robust probability of a runaway occurring with remaining, ``stalled" BHs since the assumption of a fixed background becomes increasingly undependable as evolution time is increased.

\begin{figure*}
\begin{minipage}[b]{\textwidth}
\centering
\includegraphics[width=\textwidth]{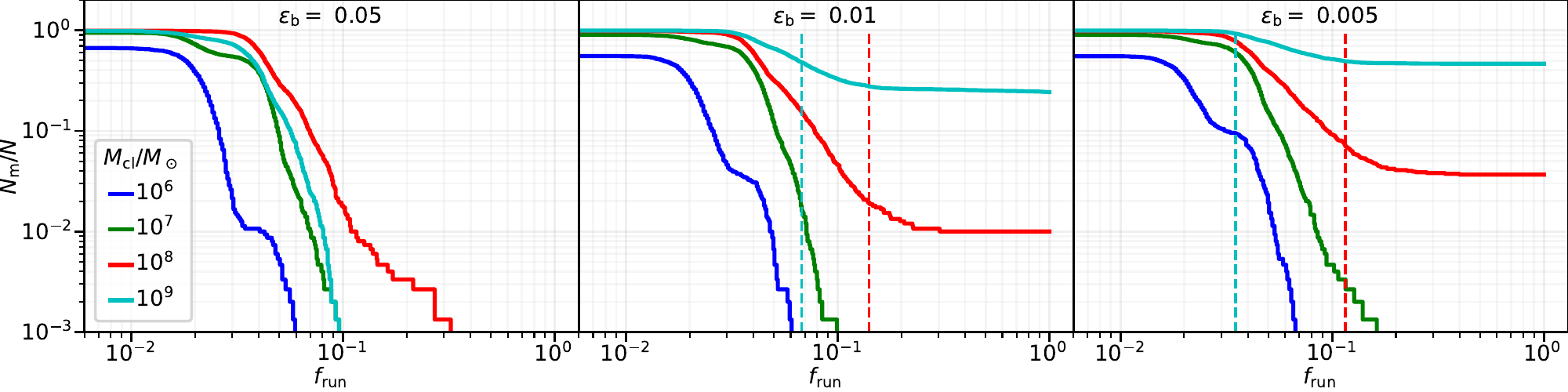}{\\(a) $m_{\rm s0} = 50 \,M_\odot$}
\end{minipage}
\par\bigskip
\begin{minipage}[b]{\textwidth}
\centering
\includegraphics[width=\textwidth]{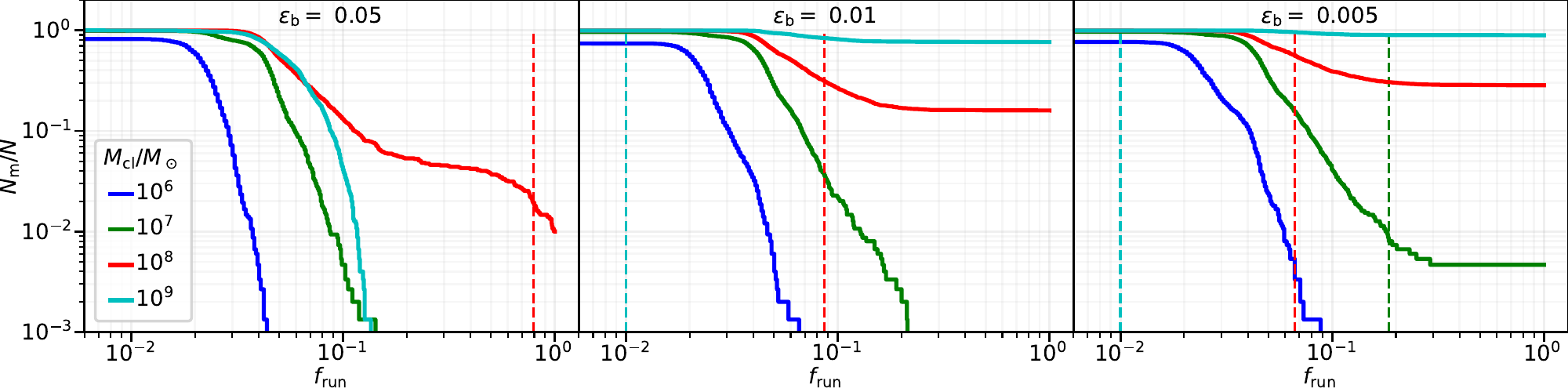}{\\(b) $m_{\rm s0} = 100 \,M_\odot$}
\end{minipage}
\caption{The frequency of BH seeds, $N_{\rm m}/N$, exceeding our \textit{effective runaway fraction}, $f_{\rm run}$ (see eq.~\ref{eq:effrunfrac}). The colored, dashed lines mark $m_{\rm crit}$, defined as the point where 50\% of seeds will also reach $m_{\rm run}$ within the 3~Gyr lifetime allowance. In the left-most panels, with the exception of $(m_{\rm s0}, M_{\rm cl}, \epsilon_{\rm b}) = (100 M_\odot, 10^8 M_\odot, 0.05)$, there are no dashed lines because there are no runaways.}
\label{fig:fvsmf}
\end{figure*}

In this spirit, Fig.~\ref{fig:fvsmf} details $m_{\rm crit}$ for each model with a dashed vertical line and conveys several key ideas. First, increasing $M_{\rm cl}$ dramatically increases the fraction of BHs reaching larger final masses. While increasing the BH sub-cluster density by an order of magnitude (by decreasing $\epsilon_{\rm b}$ from 0.01 to 0.005) is effective at increasing runaway probability (by ${\sim}2-5$ times with $M_{\rm cl}$ and $m_{\rm s}$ held constant), increasing $M_{\rm cl}$ by an order of magnitude is always more effective, increasing runaway probability by ${\sim}10-20$ times with $\epsilon_{\rm b}$ and $m_{\rm s}$ held constant). In addition, for all models producing a runaway in $M_{\rm cl}\geq 10^7 M_\odot$, $\epsilon_{\rm b} = \{0.01, 0.005\}$ clusters, $m_{\rm crit} - m_{\rm s0} \lesssim 180 \,M_\odot$.

\begin{figure*}
\begin{minipage}[b]{\textwidth}
\centering
\includegraphics[width=\textwidth]{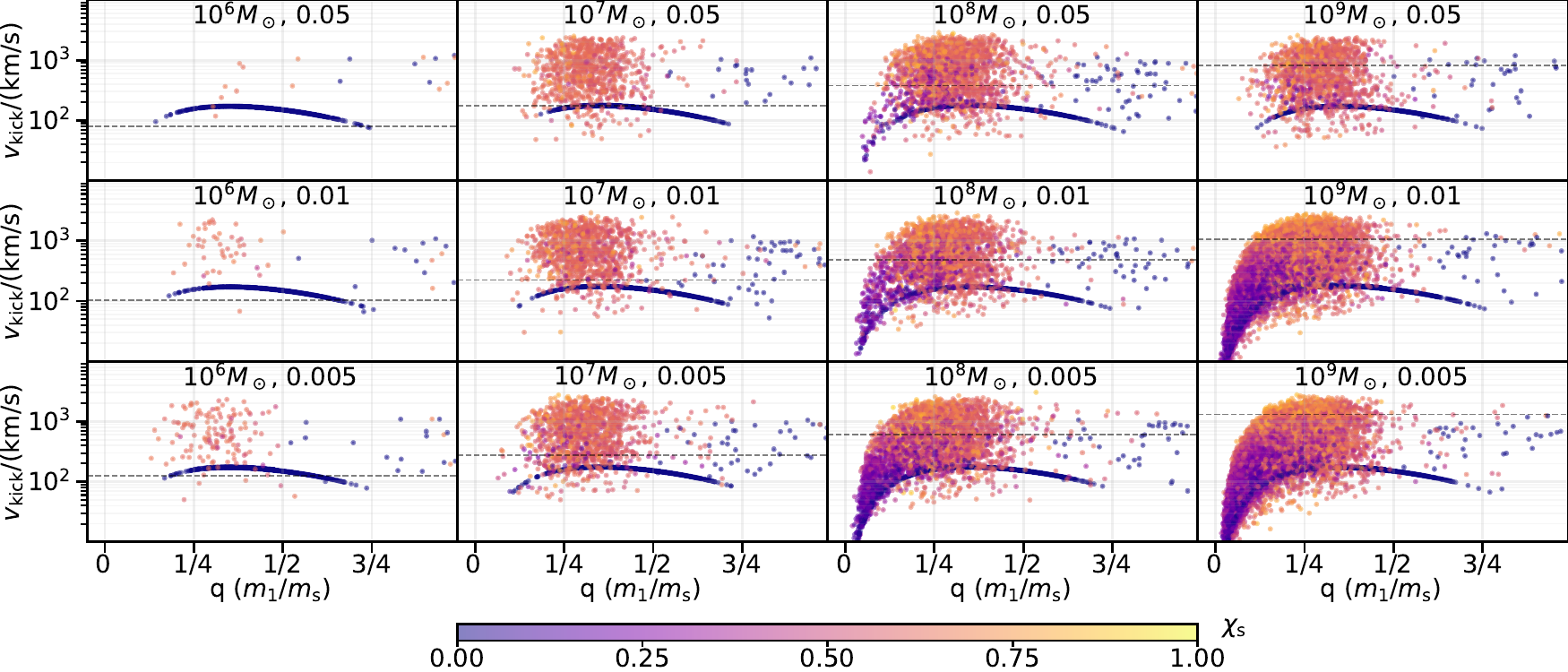}{\\(a) $m_{\rm s0} = 50 \,M_\odot$\\}
\end{minipage}
\par\bigskip
\begin{minipage}[b]{\textwidth}
\centering
\includegraphics[width=\textwidth]{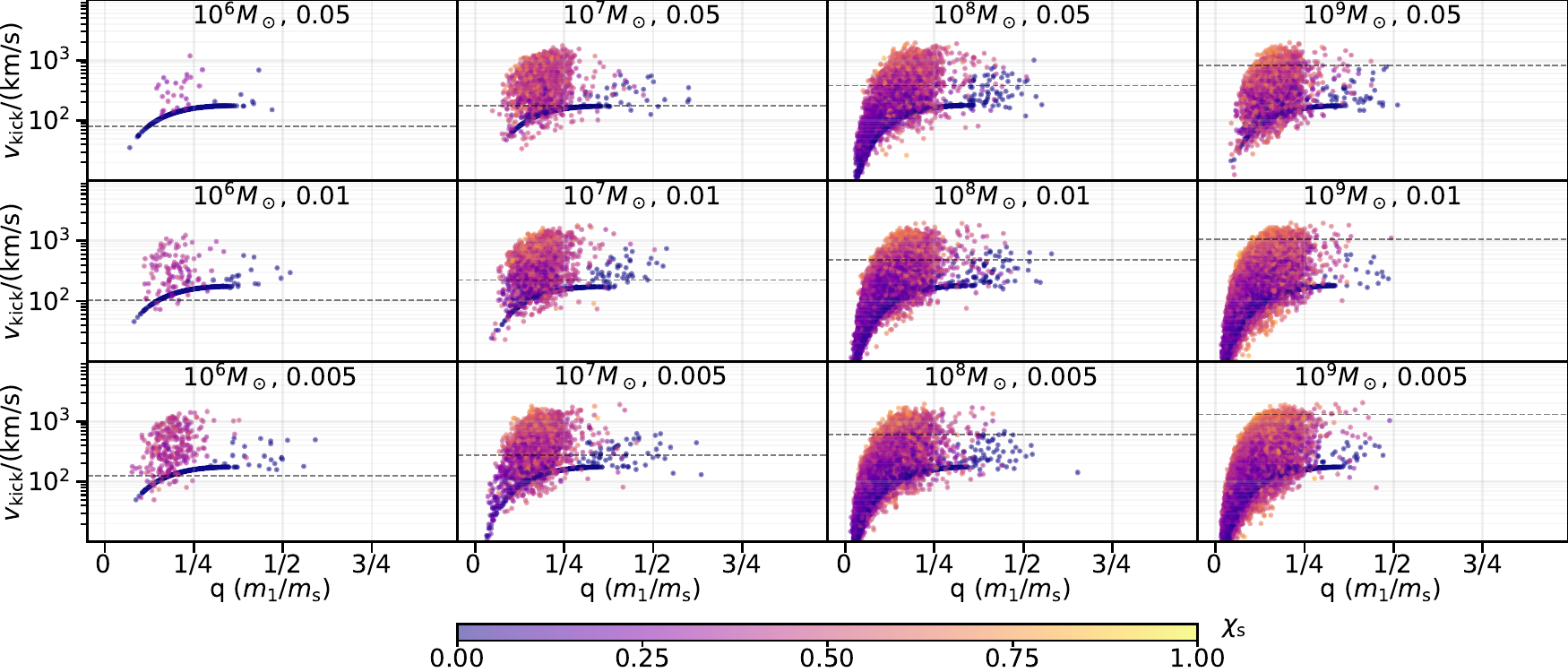}{\\(b) $m_{\rm s0} = 100 \,M_\odot$}
\end{minipage}
\caption{Scatter plot of every GW recoil kick experienced by a seed in all sequences. The kick velocity, $v_{\rm kick}$, is displayed along the y-axis and the mass ratio, in terms of the mass of the secondary BH, $m_1$, merging with the seed BH, $m_{\rm s}$, is displayed along the x-axis. The dashed, gray line is the central escape velocity of the double-Plummer cluster. Color represents the spin of the seed BH, $\chi_{\rm s}$, at merger. As expected, GW recoil magnitude decreases as the mass ratio, $q$, drifts from the zero-spin peak of $q\approx0.38$, but tends to be dramatically amplified by spin.}
\label{fig:qvsvfk}
\end{figure*}

When comparing runaway probability between $50 \, M_\odot$ and $100 \, M_\odot$ seeds, runaways sourced from $100 \, M_\odot$ seeds are roughly $1.5-10$ times more frequent. Weaker GW recoil kicks follow immediately from an increased initial seed mass, $m_{\rm s0}$, beyond the typical BH mass, $\langle m_{\rm bh}\rangle$ (i.e., $q_{100}<q_{50}<0.38$, where $q=0.38$ is the peak of the GW recoil kick distribution). The smaller $q$ is for non-spinning merging BHs, the smaller the spin of the merger product, helping reduce the magnitude of GW recoil from additional mergers; see Fig.~\ref{fig:qvsvfk} for a comparison of GW recoil kicks to $v_{\rm e,cl}$ for various cluster models. An $m_{\rm s0}=100 \, M_\odot$ seed initiated in $\left(M_{\rm cl}, \epsilon_{\rm b} \right) = \left(10^9 \,M_\odot, \left\{0.01, 0.005\right\} \right)$ clusters have $m_{\rm crit} = m_{\rm s0}$ (binned at $f_{\rm run}=10^{-2}$ for convenience), with about $76-89\%$ of seeds reaching runaway. It is difficult to not produce a runaway from a $100 \, M_\odot$ seed in these clusters given the extremely large central escape velocities ($>$$1000\,$km/s).

\begin{figure*}
\begin{minipage}[b]{\textwidth}
\centering
\includegraphics[width=.95\textwidth]{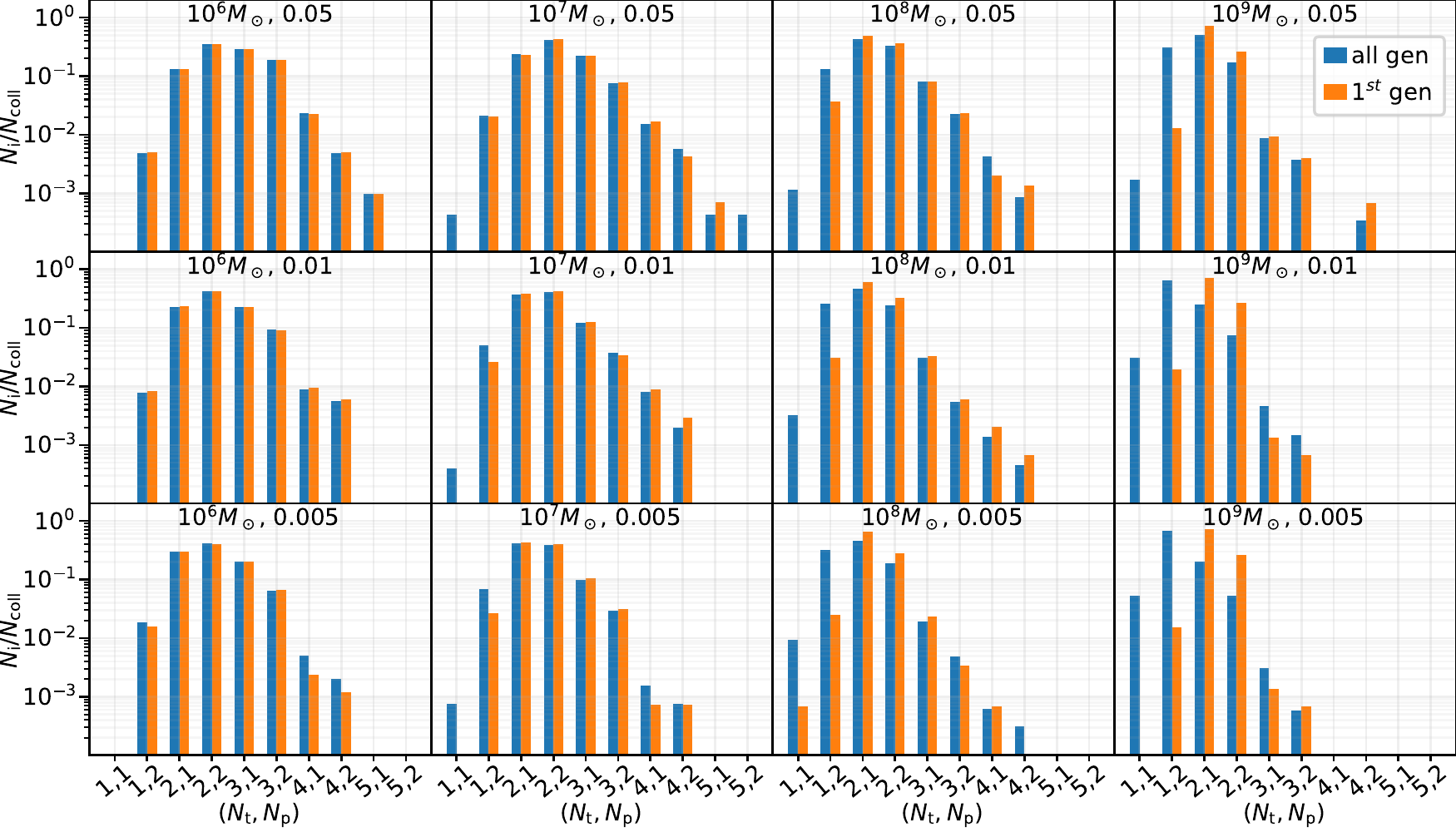}{\\(a) $m_{\rm s0} = 50 \,M_\odot$}
\end{minipage}
\par\bigskip
\begin{minipage}[b]{\textwidth}
\centering
\includegraphics[width=.95\textwidth]{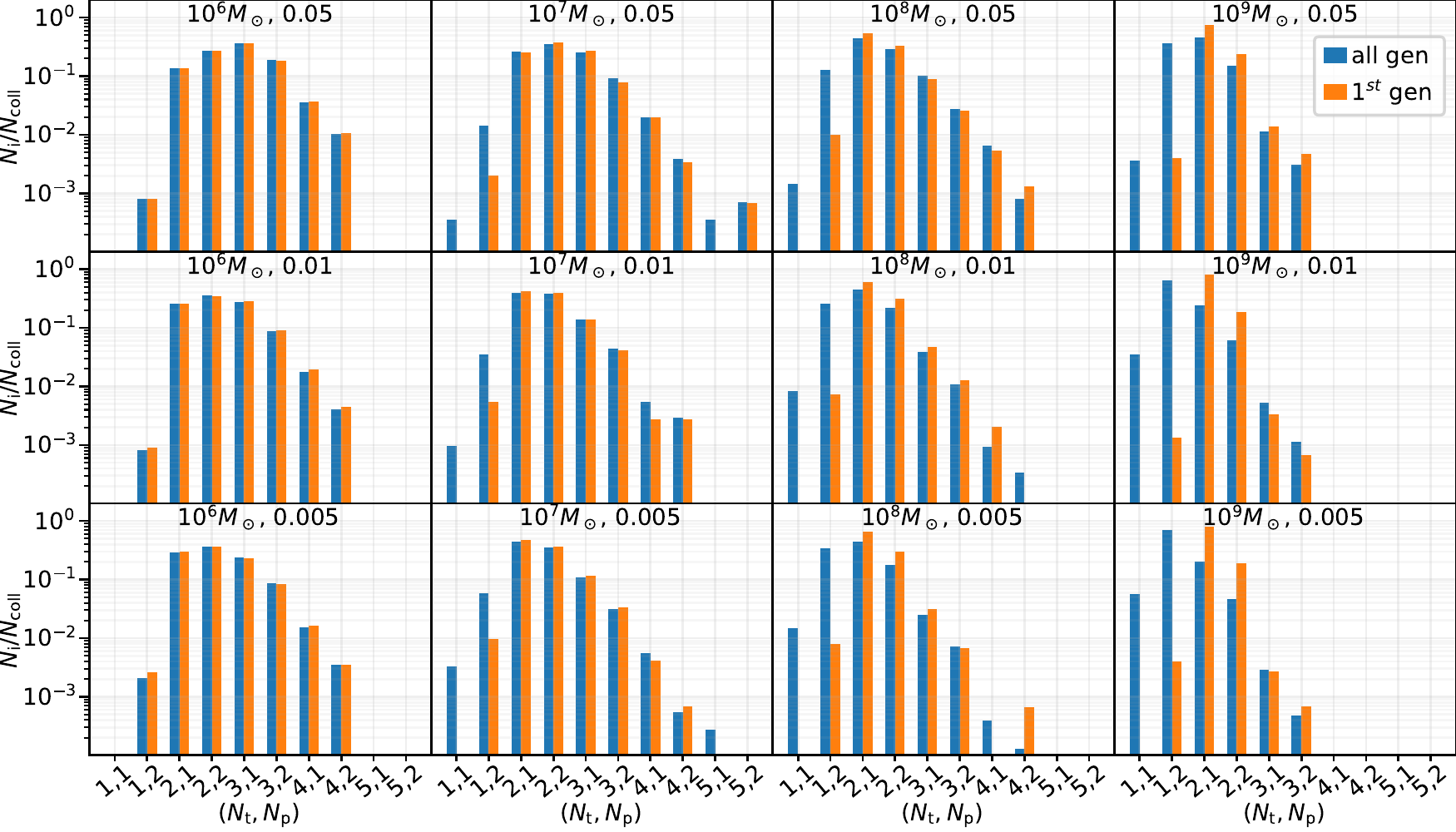}{\\(b) $m_{\rm s0} = 100 \,M_\odot$}
\end{minipage}
\caption{For each cluster initial condition, the above set shows the rate of $(N_{\rm t}, N_{\rm p})$ encounters in which the seed mass, $m_{\rm s}$, underwent a merger across all sequences; distinguishing between $1^{\rm st}$ generation mergers and across all generations with bar color. $N_{\rm t}$ and $N_{\rm p}$ represent the number of bodies in the target hierarchy and projectile system, respectively. For example, (2,1) and (1,2) are binary-single encounters, (3,2) is a triple-binary, etc. The top panel (a) adopts an initial seed mass $m_{\rm s0} = 50 \, M_\odot$ and the bottom panel (b) adopts an initial seed mass $m_{\rm s0} = 100 \, M_\odot$.}
\label{fig:scatcollhist}
\end{figure*}
\subsection{The Importance of Hierarchies}

While our models produce simple exponential growth rates over the lifetime of a runaway BH seed, the hierarchical interactions fueling them are rich. Other than via explicit integration in direct $N$-body simulations--which are too computationally expensive to study the dense regimes likely to form massive BHs--cluster models often neglect to account for the influence of large hierarchies in BH dynamics. For example, Monte Carlo codes like \texttt{CMC} and \texttt{MOCCA} \citep[see][for the respective and most recent comprehensive overviews]{2013MNRAS.431.2184G, CMCcodepaper} incorporate \texttt{fewbody}-based scattering only for binary-single and binary-binary interactions. By enabling our BH seed to occupy any sized hierarchy allowed by the physical constraints defined in Sec.~\ref{sec:dpmodel}, we find a $53-95\%$ chance that one or more interactions where $P_{\rm t}$ contains three or more (3+) bound BHs will occur over a single sequence in $10^6 \,M_\odot$ and $10^7 \, M_\odot$  double-Plummer models (Table~\ref{table:cusptable}). For the higher mass $10^8 \,M_\odot$ and $10^9 \, M_\odot$ models, the probabilities range from $4-63\%$.

The $M_{\rm cl}/M_\odot = \{10^6, 10^7\}$ double-Plummer models constitute the most massive cluster models explorable by common $N$-body stellar dynamics infrastructures (CMC, MOCCA, NBODY6++, etc). As can be seen in Fig.~\ref{fig:scatcollhist}, single-single, binary-single, and binary-binary interactions constitute roughly $65\%$ and $80\%$ of all ``merger-producing'' interactions in $10^6 \,M_\odot$ and $10^7 \, M_\odot$ clusters, respectively. The approximation that binary-single and binary-binary interactions may encompass $>$$90\%$ of a BH seed's dynamical history is valid only in clusters with $M_{\rm cl}\geq10^8\, M_\odot$; regimes the previously mentioned numerical infrastructures are incapable of probing. In fact, if the dynamical history of a massive BH seed ($\gtrsim 100\, M_\odot$) is considered within globular cluster-like (GC) cluster masses ($\lesssim$$10^6 \,M_\odot$), about $95\%$ of $P_{\rm t}$ will experience at least one 3+ hierarchical interaction in a sequence, and $\gtrsim50\%$ will experience at least one 4+ hierarchical interaction (Table~\ref{table:cusptable2}).

The rate of single-single GW capture is inversely proportional to $a_{\rm hs}$; occurring in about $0.04\%$, $0.3\%$, $3\%$ of interactions in $10^7 \, M_\odot$, $10^8 \, M_\odot$, and $10^9 \, M_\odot$ clusters, respectively, for $m_{\rm s0} = 50 M_\odot$ seeds. Since increasing the density and velocity dispersion of bodies in the BH sub-cluster reduce the lifetime (GW decay time) and physical cross-section of BBHs relative to $t_{\rm enc}$ and $a_{\rm GW}$ respectively, it becomes more likely that the set of strong encounters experienced by the seed will include strong single-single interactions.

\begin{figure*}
\begin{minipage}[b]{\textwidth}
\centering
\includegraphics[width=.95\textwidth]{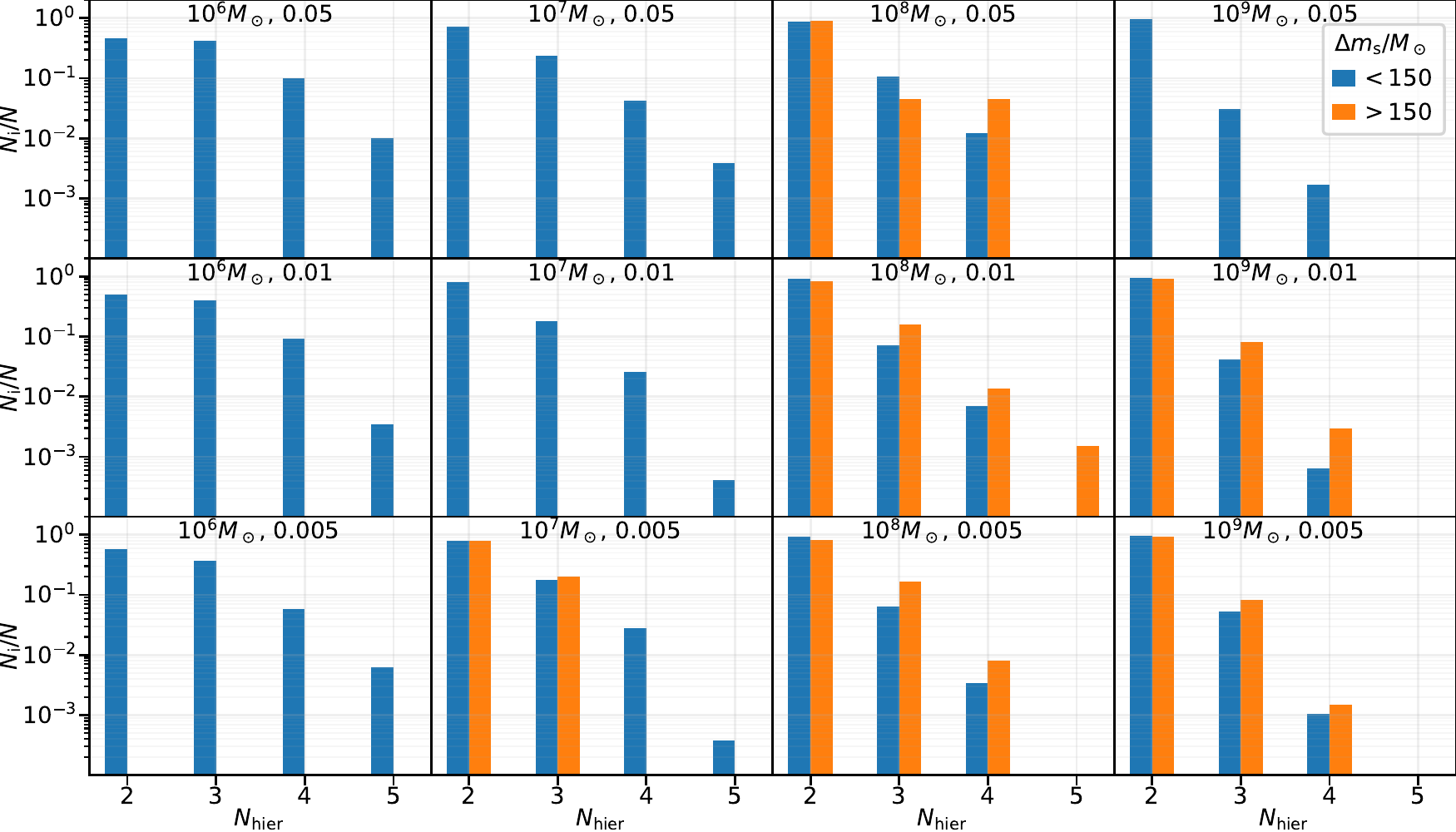}{\\(a) $m_{\rm s0} = 50 \,M_\odot$}
\end{minipage}
\par\bigskip
\begin{minipage}[b]{\textwidth}
\centering
\includegraphics[width=.95\textwidth]{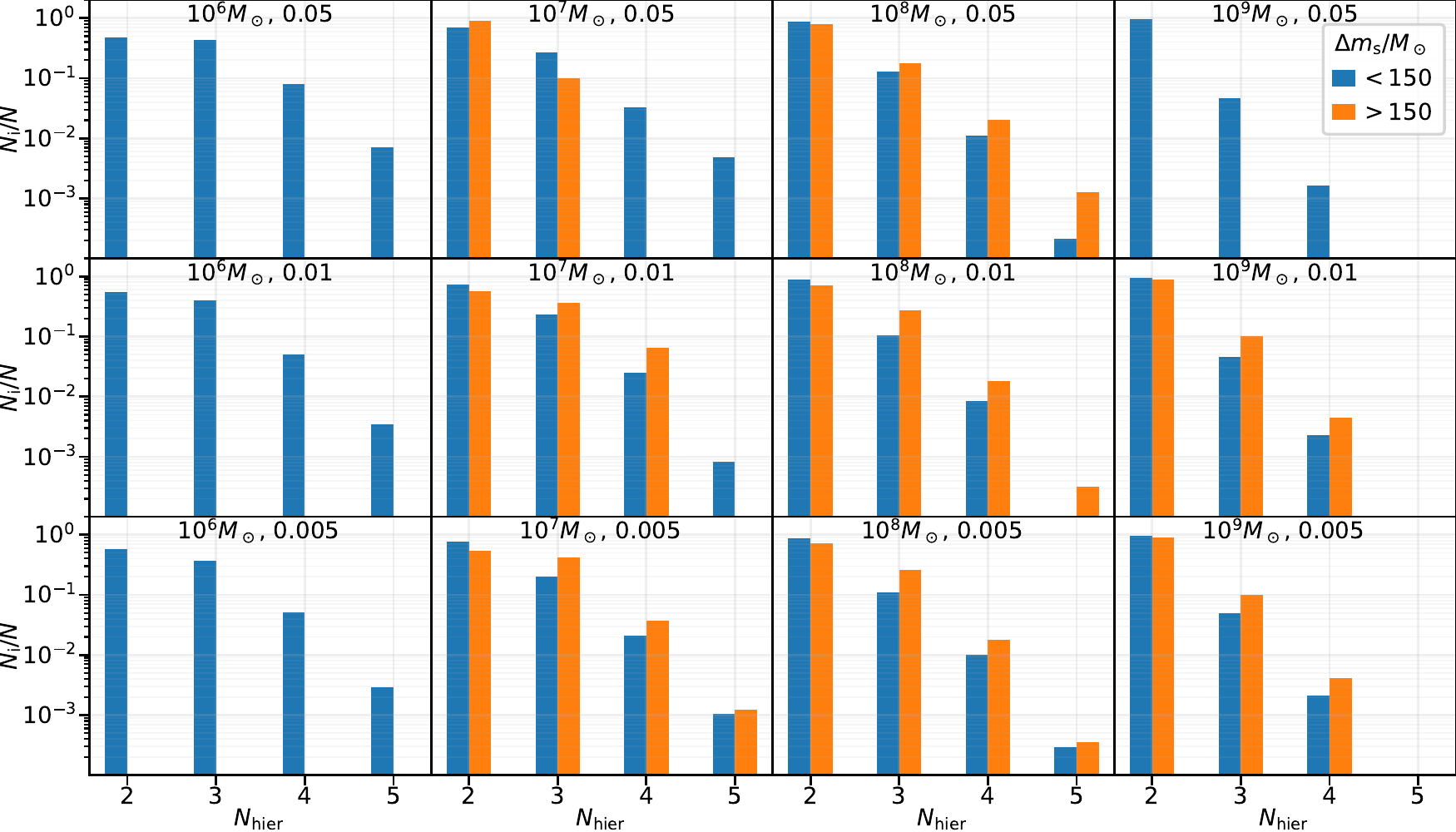}{\\(b) $m_{\rm s0} = 100 \,M_\odot$}
\end{minipage}
\caption{The probability, $N_{\rm i}/N$, of a BH seed merger to occur within hierarchy rank, $N_{\rm hier}$. Color bars are organized by $\Delta m_{\rm s} = m_{\rm s} - m_{\rm s0}$, where $m_{\rm s}$ is the seed mass post merger and $m_{\rm s0}$ is the initial seed mass. The top panel (a) adopts an initial seed mass $m_{\rm s0} = 50 M_\odot$ and the bottom panel (b) adopts an initial seed mass $m_{\rm s0} = 100 M_\odot$. Lower-density clusters feature mergers in a broader range of hierarchy types.}
\label{fig:collhierhist}
\end{figure*}

\begin{figure*}
\centering
\includegraphics[width=.9\textwidth]{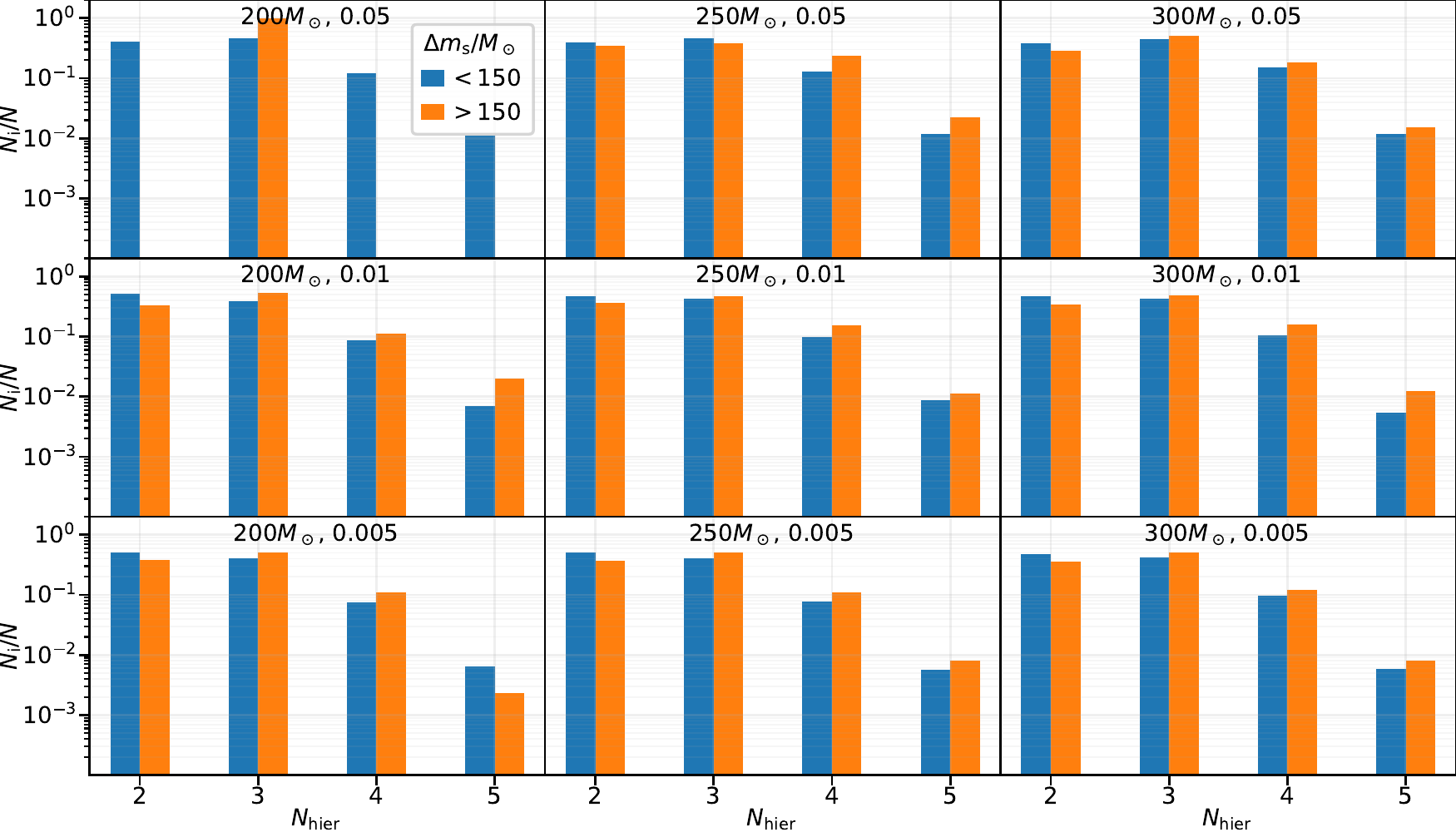}{\\$M_{\rm cl} = 10^6\, M_\odot$}
\caption{The probability, $N_{\rm i}/N$, of a BH seed merger to occur within hierarchy rank, $N_{\rm hier}$. Color bars are organized by $\Delta m_{\rm s} = m_{\rm s} - m_{\rm s0}$, where $m_{\rm s}$ is the seed mass post merger and $m_{\rm s0}$ is the initial seed mass. Sub-plots are labeled with initial seed mass and scale parameter $(m_{\rm s0}, \epsilon_{\rm b})$. As seed mass increases, hierarchical mergers become more prevalent than isolated binary mergers.}
\label{fig:collhierhist1e+06}
\end{figure*}

Fig.~\ref{fig:collhierhist} displays the hierarchical configuration of the BH seed and it's bound companions at the time of merger. Hierarchical triples, quadruples, and quintuples constitute about $40\%$, $9\%$, and $0.3\%$ of configurations, respectively, at the time of merger in an $m_{\rm s0}=50 M_\odot$, $M_{\rm cl} = 10^6 M_\odot$ model; none of which are sourced from a seed BH on the runaway track. While rarer in an $M_{\rm cl} = 10^9 M_\odot$ model, triples and quadruples constitute roughly $8\%$ and $0.3\%$ hierarchical configurations at the time of merger, respectively, for a BH seed on the runaway track. Within a GC-like cluster profile, the 3+ hierarchical mergers are more probable ($\gtrsim50\%$) than isolated binary mergers for massive seeds which do not escape the cluster following $5-7$ mergers (Fig.~\ref{fig:collhierhist1e+06}).

\begin{figure*}
\centering
\begin{minipage}[b]{\textwidth}
\centering
\includegraphics[width=.99\textwidth]{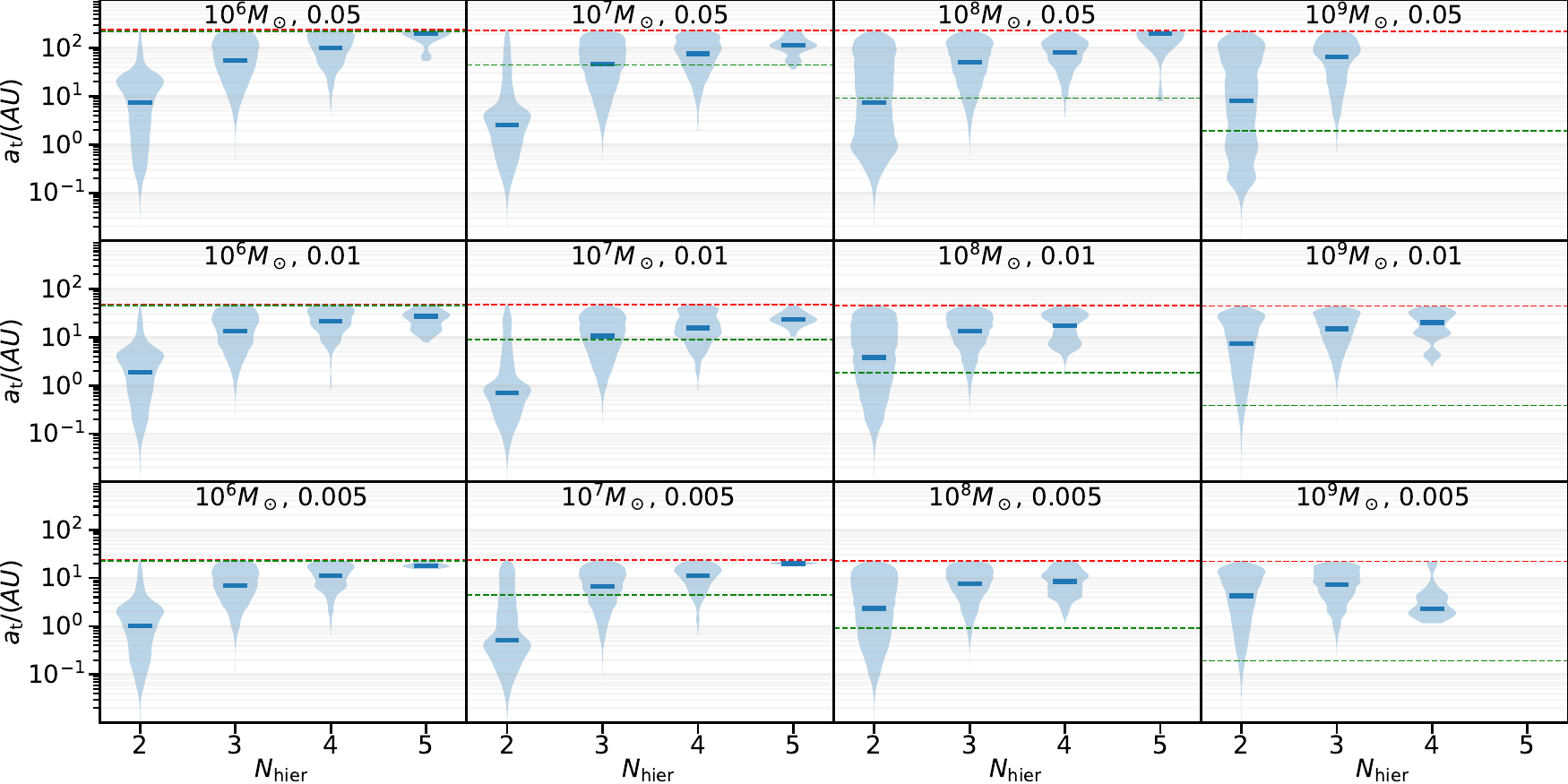}{\\$m_{\rm s0} = 100 \,M_\odot$}
\end{minipage}
\caption{Violin plot displaying the spread of target hierarchy orbital radii, $a_{\rm t}$, which survive to the next interaction. Internal blue lines are the median of each hierarchy violin plot, the dashed green line is a reference hard-soft boundary for a binary consisting of a 100$\,M_\odot$ seed and $20 \, M_\odot$ companion, and the dashed red line is the orbital radius limit defined at $10\%$ of the average interparticle distance of each model; beyond which bound orbits are broken up.}
\label{fig:hiersummary}
\end{figure*}

Fig.~\ref{fig:hiersummary} suggests that binaries live most comfortably with an orbital radius an order of magnitude smaller than the absolute orbital radius threshold of the respective cluster model (see eq.~\ref{rmax}). Unsurprisingly, the set of orbital radii of higher order hierarchies congregate at the enforced threshold. This effect is expected because hierarchical stability is most strongly correlated with orbital separation between recursive layers (see Sec.~\ref{sec:stability}). In our models, quintuples are the largest hierarchical configurations we can form frequently enough to provide sample statistics, though sextets and septets form infrequently ($\lesssim 0.1\%$ of interactions). Fig.~\ref{fig:collhierhist} suggests that, given a fixed orbital radius threshold, a BH seed becomes more likely to support larger hierarchies at merger as it accumulates mass. Thus, a potential signature of a BH seed which has undergone runaway growth through dynamical encounters may be a large, dynamically stable, (5+) hierarchical cusp. 

To explore the effect that rank 3+ hierarchies have in a BH seed sequence, a set of $m_{\rm s0} = 50 \,M_\odot$ models was calculated in \textsc{cuspbuilding} while forcefully separating the outermost orbit(s) of 3+ hierarchical systems at the end of an interaction--limiting interactions to 1+1, 2+1, or 2+2 configurations. We find nominal increases in the runaway probability when hierarchies are allowed to organically form and grow; constituting a factor of ${1.03 - 1.30}$ increase when compared to the models where separation of 3+ hierarchies is enforced post-interaction. The typical time required to runaway remains unchanged. While hierarchical assembly is likely to play a significant role in a BH seed's dynamical history, it may only have a nominal effect on runaway frequency or evolution time of any individual seed BH in clusters well described by our fixed double-Plummer models.

\subsection{Escapees}

Seeds are unlikely to stall and will either escape through strong velocity perturbations or cross the runaway threshold, $m_{\rm run}$; the exceptions are models where $t_{\rm enc}$ is a significant fraction of our $3 \,$Gyr evolution time, namely $\left(M_{\rm cl}, \epsilon_{\rm b} \right) = \left(10^9 M_\odot, \{0.05, 0.01\} \right)$. While dynamical kicks play a small role in ejecting seeds in lower mass star clusters, constituting up to $45\%$ of $50 M_\odot$ BH seed ejections in $10^6 M_\odot$ cluster models, GW recoil is most responsible for abruptly halting the growth of a BH seed in every double-Plummer model.

\begin{figure*}
\centering
\begin{minipage}[b]{\textwidth}
\centering
\includegraphics[width=.98\textwidth]{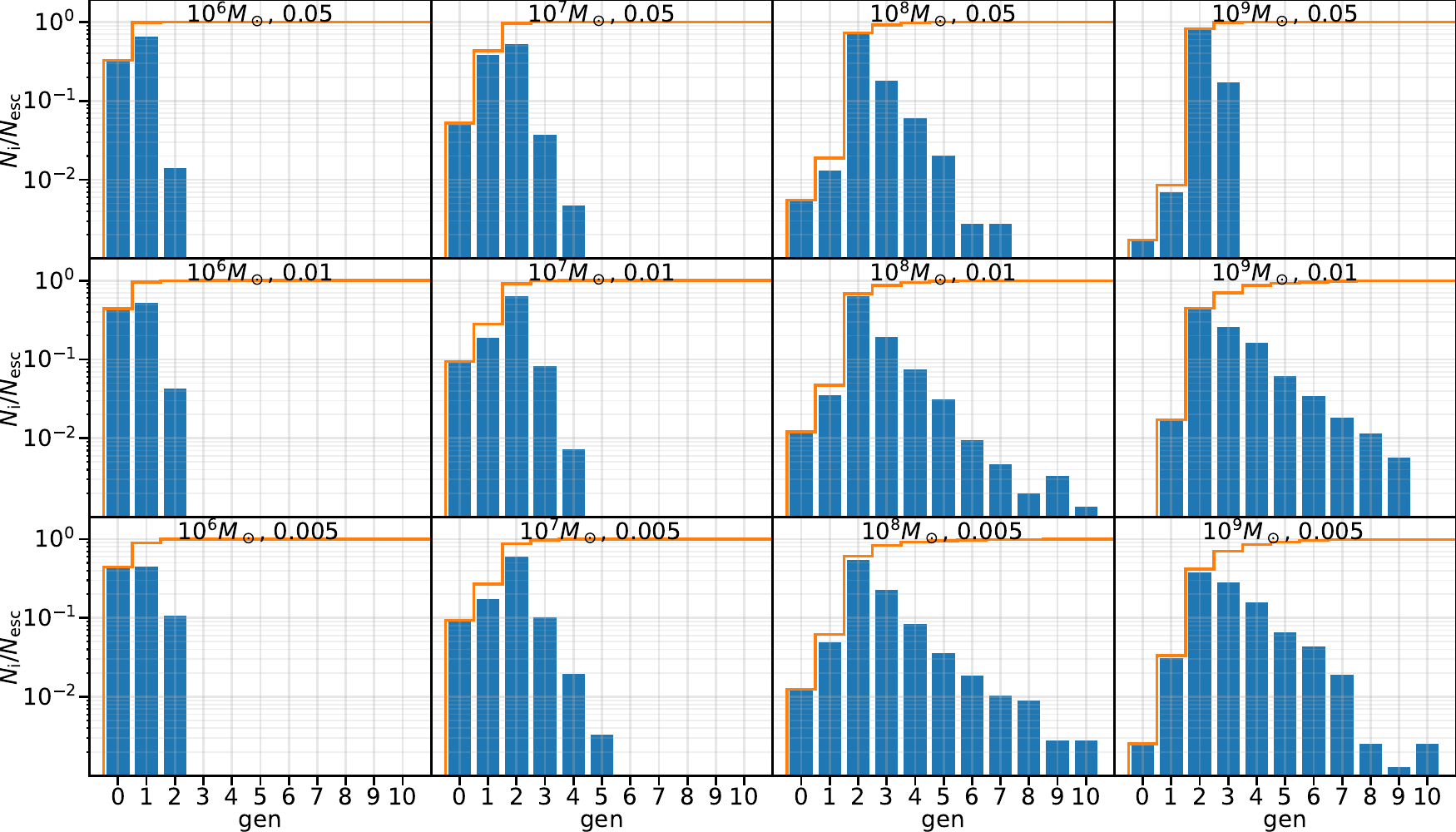}{\\(a) $m_{\rm s0} = 50 \,M_\odot$}
\end{minipage}
\par\bigskip
\begin{minipage}[b]{\textwidth}
\centering
\includegraphics[width=.98\textwidth]{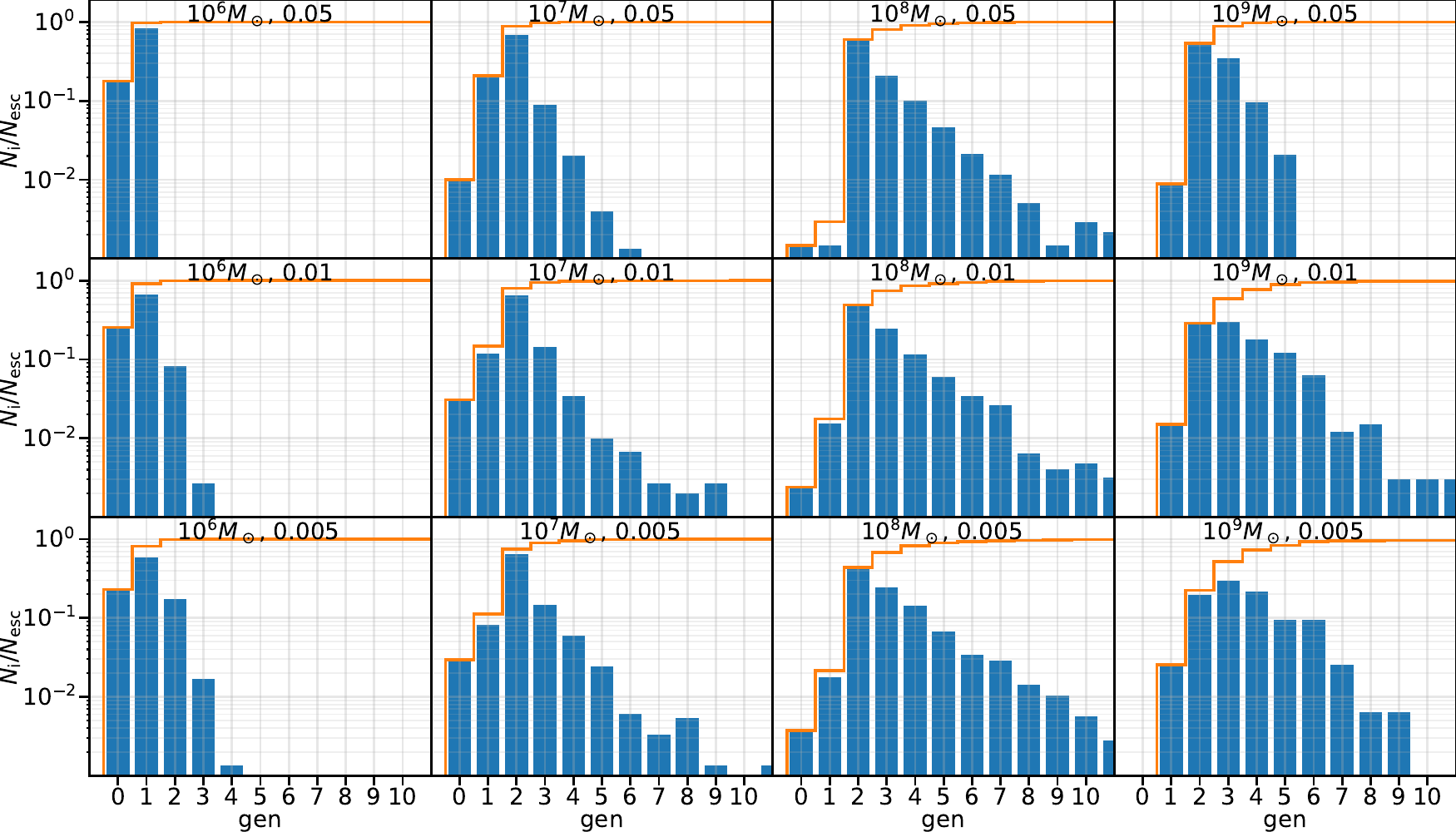}{\\(b) $m_{\rm s0} = 100 \,M_\odot$}
\end{minipage}
\caption{Bar chart displaying the fraction of escaping seed BHs which escape following their $N^{\rm th}$ merger. The $0^{\rm th}$ generation refers to a seed which escaped without ever undergoing a merger, $1^{\rm st}$ generation refers to a seed escaping following one merger, etc. The orange line is the cumulative distribution which caps at unity when all escaping BHs have been considered.}
\label{fig:probescvsgen}
\end{figure*}

\begin{figure}
\centering
\includegraphics[width=.48\textwidth]{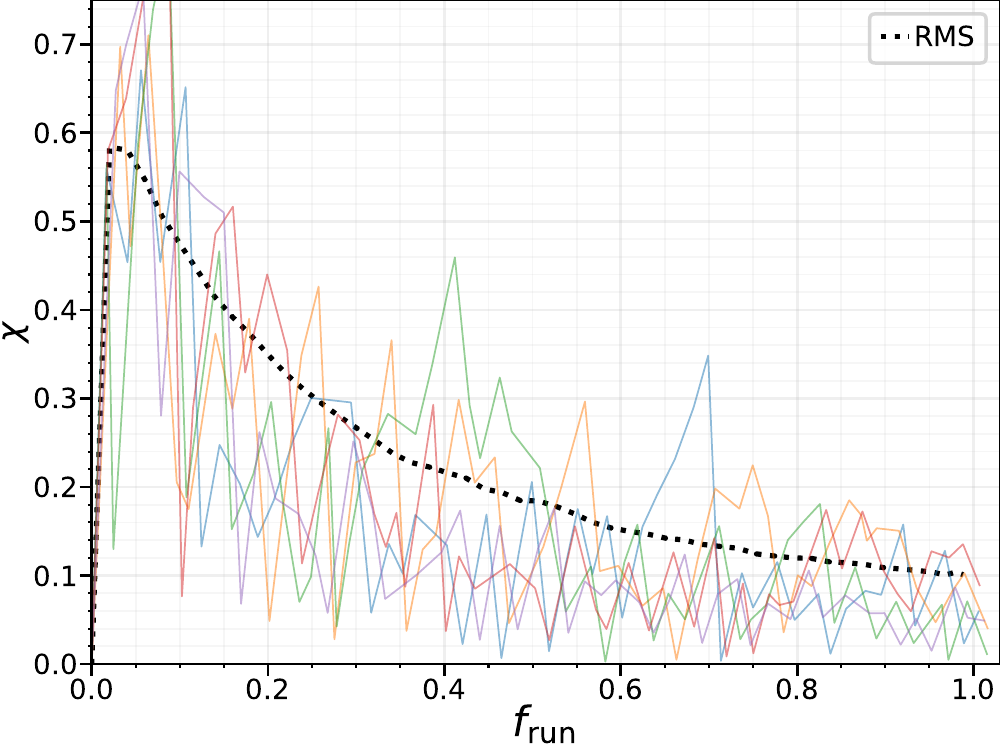}
\caption{Dimensionless spin $(\chi)$ vs \textit{effective runaway fraction} for five randomly selected runaways sourced from $50\, M_\odot$ seeds. The black dotted line represents the RMS spin following successive mergers across all ${m_{\rm s0} = 50\, M_\odot}$ runaways independent of cluster initial condition.}
\label{fig:mvx_50.pdf}
\end{figure}

Escaping BHs are most likely to leave their host clusters by their second merger across all models (Fig.~\ref{fig:probescvsgen}). This is due to the kick amplification caused by a spinning BH seed. Since the spin of the seed BH is usually highest after the first merger, with successive mergers tending to slowly reduce the spin (Fig.~\ref{fig:mvx_50.pdf}), it follows that the most powerful GW recoil kicks peak at $2^{\rm nd}$ generation mergers--this effect is well depicted in Fig.~\ref{fig:qvsvfk}. Since all BHs in our sequences are initiated non-spinning, the zero-spin curve (shown in blue) is prominent as it exclusively hosts first generation mergers; $2^{\rm nd}$ generation mergers run the gamut of the \{mass ratio\}/\{kick velocity\} parameter space, almost unanimously amplifying the strength of GW recoil kicks by ${\sim}2 - 10$ times the zero-spin curve for spins between $0.4-0.8$.

\begin{figure*}
\begin{minipage}[b]{\textwidth}
\centering
\includegraphics[width=\textwidth]{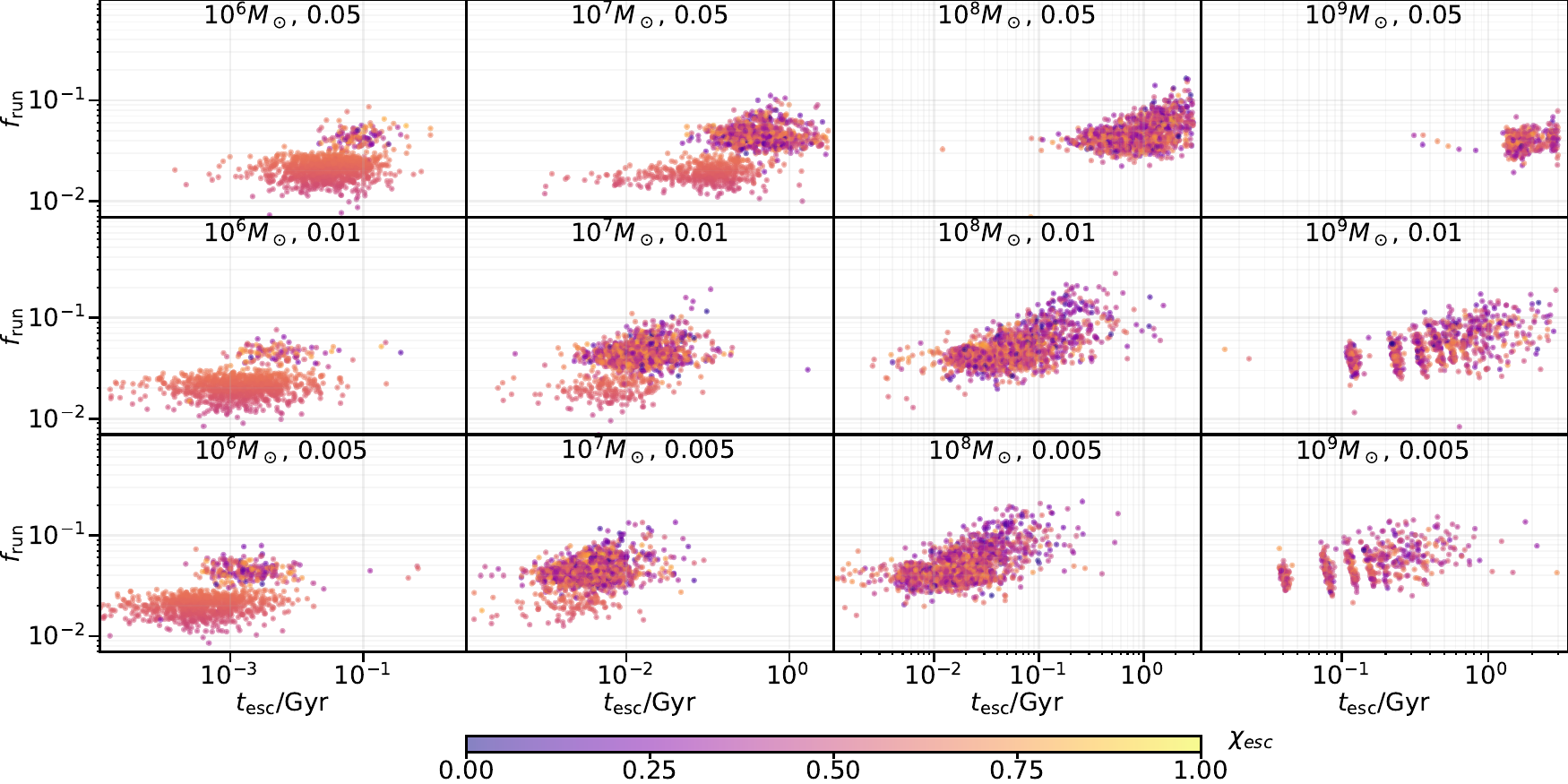}{\\(a) $m_{\rm s0} = 50 \,M_\odot$}
\end{minipage}
\par\bigskip
\begin{minipage}[b]{\textwidth}
\centering
\includegraphics[width=\textwidth]{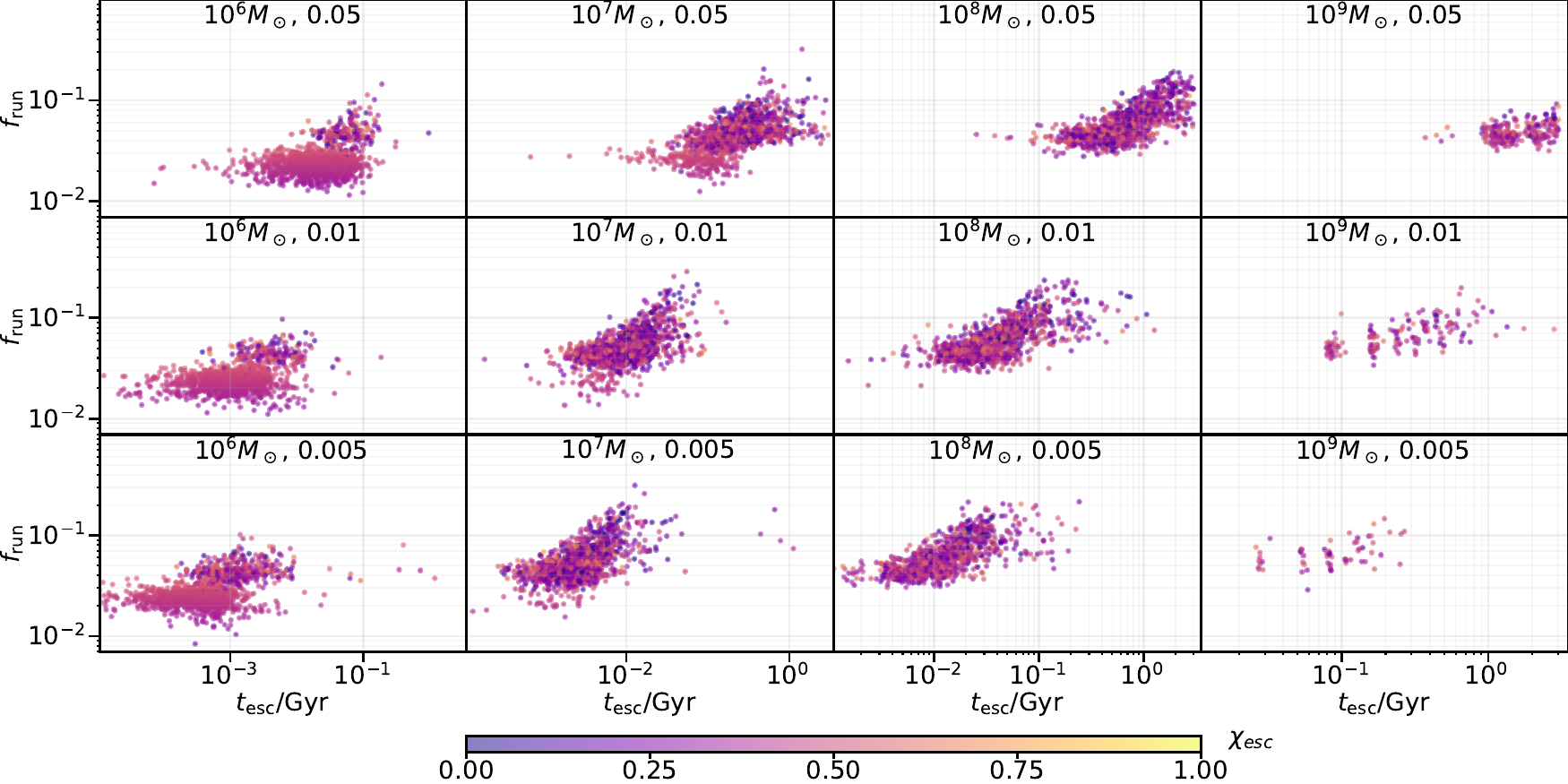}{\\(b) $m_{\rm s0} = 100 \,M_\odot$}
\end{minipage}
\caption{A grid of scatter plots displaying the time, $t_{\rm esc}$, when a seed mass escapes from its cluster; its mass expressed in terms of the \textit{effective runaway fraction}. The color indicates the spin magnitude at the time of escape. Note that, in many models, there are BH seeds escaping at their initial mass ($f_{\rm run} = 0)$; these are all escapee's due to dynamical, non-GW recoil kicks and are not displayed here.}
\label{fig:tvsmesc}
\end{figure*}

Across all models, except $\left(10^9 \, M_\odot, 0.05\right)$, if a seed is to escape its host cluster, it will most likely do so by 1~Gyr, escaping before 0.5 Gyr has elapsed in most models. The time of escape is directly correlated to the merger rate since GW recoils are the primary form of ejection. The merger rate decreases with cluster mass, as expressed in the movement and spread of escaping ``clumps" depicted in Fig.~\ref{fig:tvsmesc}. The reader may notice two clumps in many models. These clumps are a direct result of $1^{\rm st}$ generation GW recoil kicks which eject the seed from the BH sub-cluster, but fail to eject the seed from the macro-cluster. This prompts a time delay for the seed to decay back to the center of the cluster through dynamical friction, at which point it has another chance to merge with another projectile. This clumping tendency also appears in Fig.~\ref{fig:mfdist}. It should be noted that BH seeds with initial mass $50\, M_\odot$ are at a disadvantage to attain runaway status compared to $100\, M_\odot$ seeds. This is because the $m_{\rm s0} = 50\,  M_\odot$ BH will have a spin of about $0.5$ by the time it reaches $100 \, M_\odot$, dramatically increasing the average GW recoil magnitude at $m_{\rm s} = 100 \, M_\odot$ in comparison to a non-spinning, $m_{\rm s0} = 100 \, M_\odot$ BH seed.

\subsection{Varying Seed Mass in Low-Mass Clusters}

\begin{figure*}
\centering
\includegraphics[width=\textwidth]{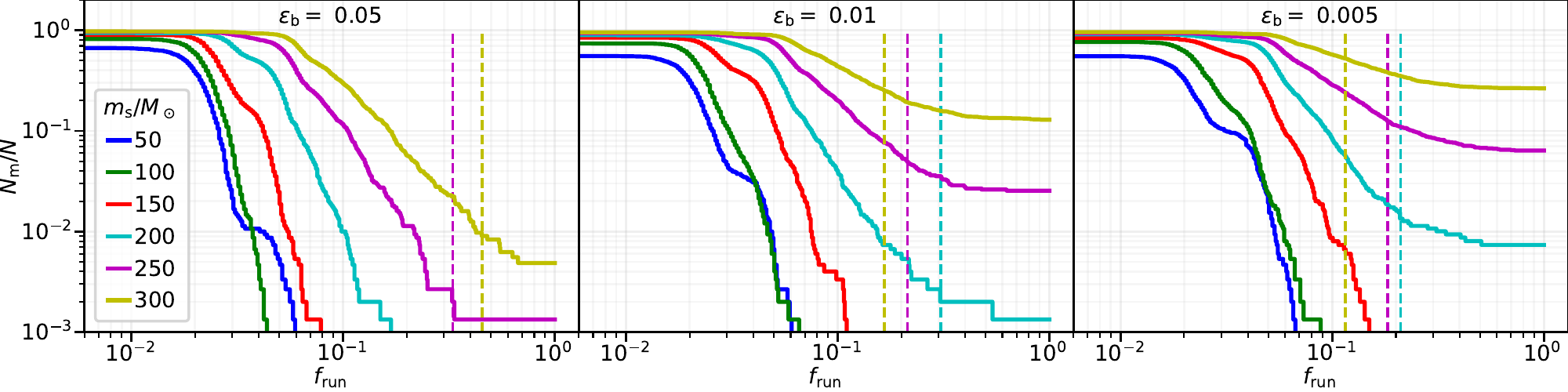}
\textbf{$M_{\rm cl} = 10^6 \, M_\odot$}
\caption{The fraction of BH seeds, $N_{\rm m}/N$, exceeding the \textit{effective runaway fraction}, $f_{\rm run}$ (see eq.~\ref{eq:effrunfrac}), in our low-mass $M_{\rm cl} = 10^6 \, M_\odot$ models. The colored, dashed lines mark $m_{\rm crit}$ (in terms of $f_{\rm run}$), defined as the point where 50\% of seeds will also reach $m_{\rm run}$ within the 3~Gyr evolution time. The $m_{\rm s}/M_\odot = \{250, 300\}$ seeds produce the only sequences with a finite probability of reaching $m_{\rm run}$ in an $\epsilon_{\rm b} = 0.05$ double-Plummer model across any set of initial conditions considered in this work.}
\label{fig:massbranchingratio1e+06}
\end{figure*}

Earlier works \citep{portegies_zwart_formation_2004, 2020ApJ...903...45K,gonzalez_intermediate-mass_2021,2021ApJ...907L..25W} found that low-mass clusters are inherently capable of forming massive BH seeds through collisions of massive stars at early times. While this is a promising development for populating the BH upper mass gap, the question of cluster retention requires further exploration. \citet[][]{Gonzalez_2022} explore retention through standard, cluster Monte Carlo models and finds that the vast majority of BH seeds with $m_{\rm s0}\leq 300 \, M_\odot$ are ejected from their host cluster, about $64\%$ through GW recoil; Martinez et al. (in prep) also explore ejection probabilities using small-$N$ scattering experiments and comes to a similar conclusion. BH seed retention in low-mass clusters is also explored in this work in addition to runaway tendency using \textsc{CuspBuilding}. For GC type masses, we find that the critical mass to undergo runaway is $\gtrsim$$350 M_\odot$ and will likely be ejected otherwise (Fig.~\ref{fig:massbranchingratio1e+06}), in agreement with \citet{Gonzalez_2022}.

\begin{figure*}
\centering
\includegraphics[width=\textwidth]{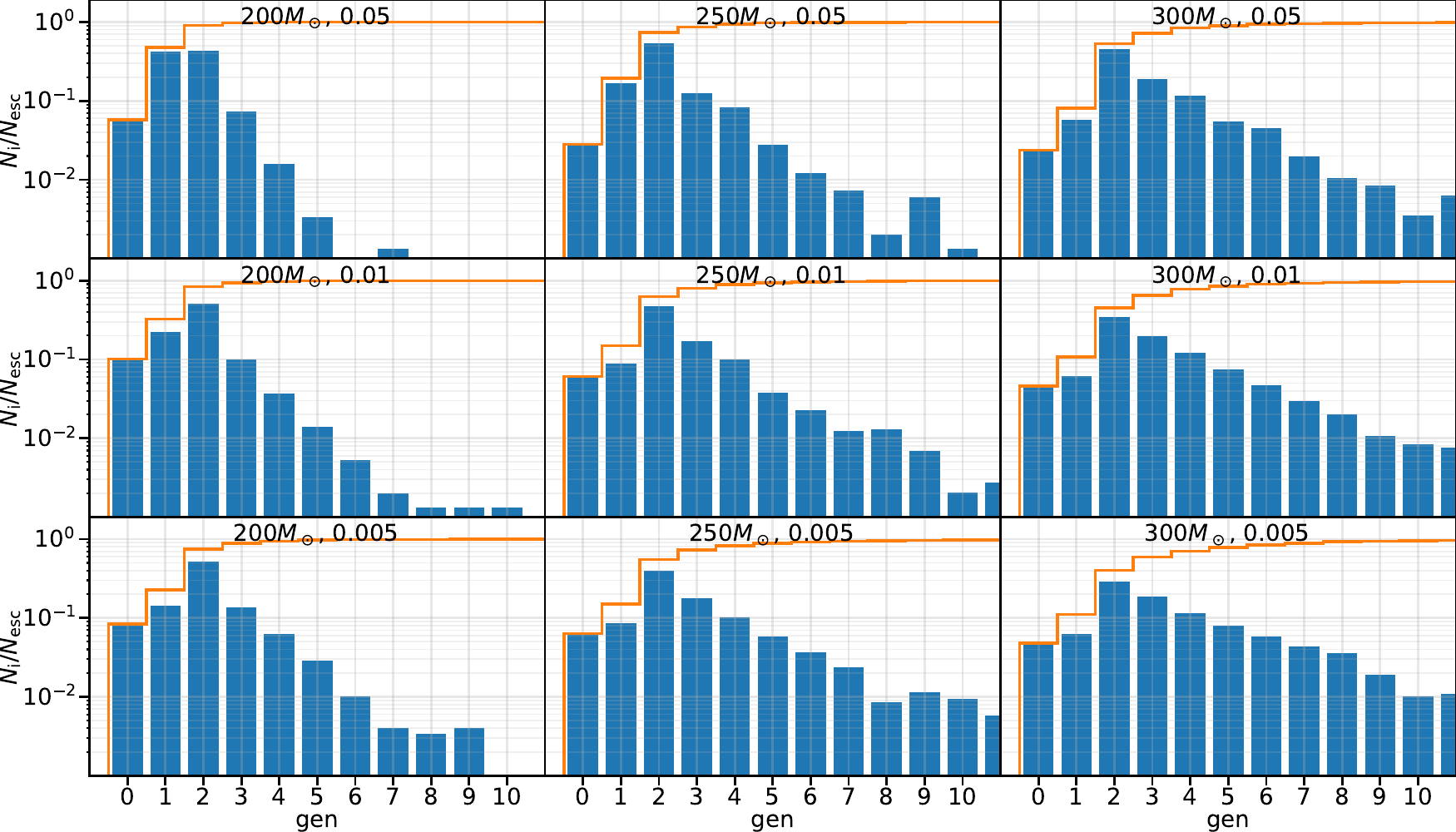}
\textbf{\\$M_{\rm cl} = 10^6 \,M_\odot$}

\caption{Bar chart displaying the fraction of escaping seed BHs which escape following their $N^{\rm th}$ merger in our ${10^6 \,M_\odot}$ double-Plummer models. The $0^{\rm th}$ generation refers to a seed which escaped without ever undergoing a merger, $1^{\rm st}$ generation refers to a seed escaping following one merger, etc. The orange line is the cumulative distribution, capping at unity when all escaping BHs have been considered.}
\label{fig:probescvsgen1e+06}
\end{figure*}

\begin{figure*}
\centering
\includegraphics[width=.9\textwidth]{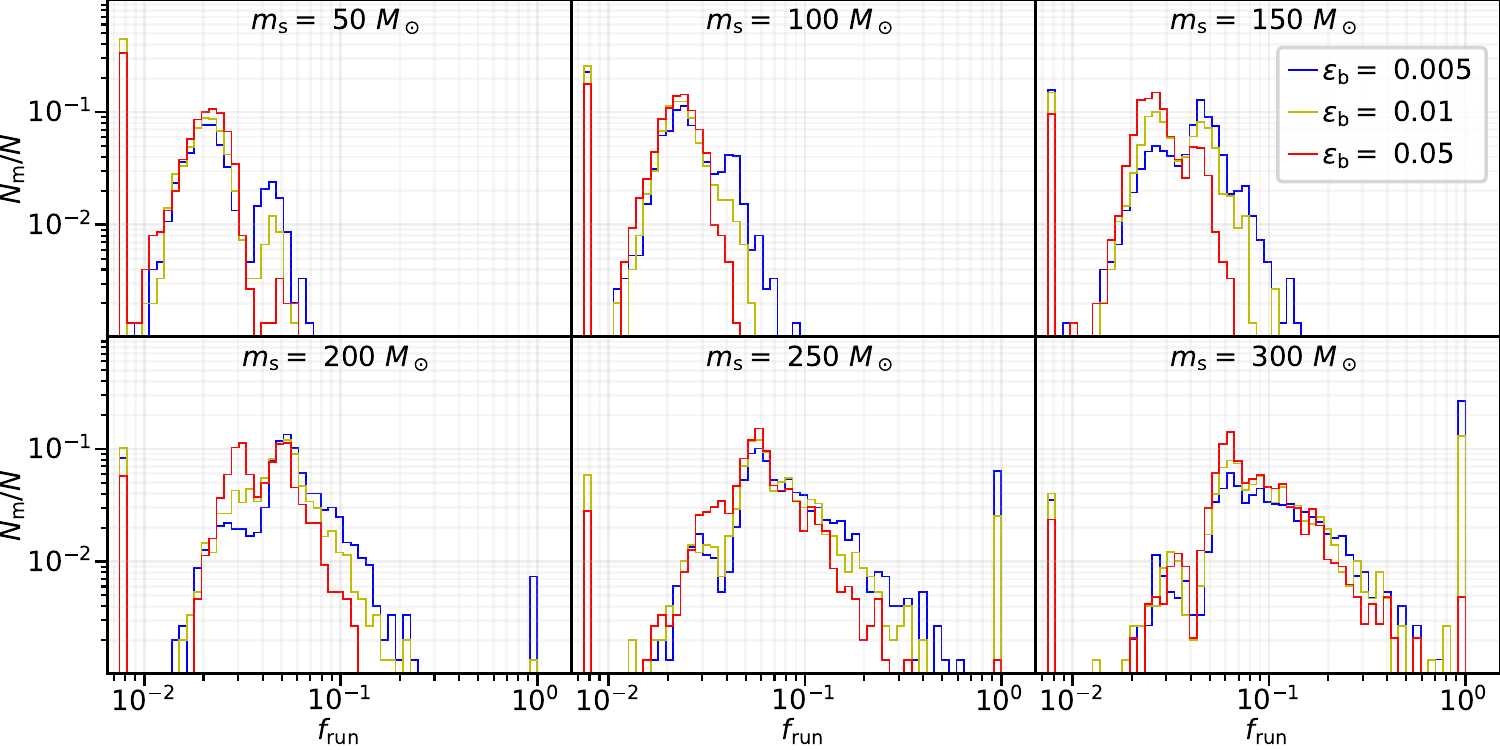}
\textbf{\\$M_{\rm cl} = 10^6 \,M_\odot$}
\caption{Distribution of final seed masses expressed in terms of the \textit{effective runaway fraction}, $f_{\rm run}$ (see eq.~\ref{eq:effrunfrac}), within our low-mass cluster models. Each grid member tracks the final seed mass distribution from an initial seed mass, $m_{\rm s}$. BH seeds which do not undergo a merger during a sequence are deposited into the leftmost bin of each grid ($f_{\rm run} = 8\times10^{-3}$).}
\label{fig:mfdist_1e06}
\end{figure*}

\begin{figure*}
\centering
\includegraphics[width=\textwidth]{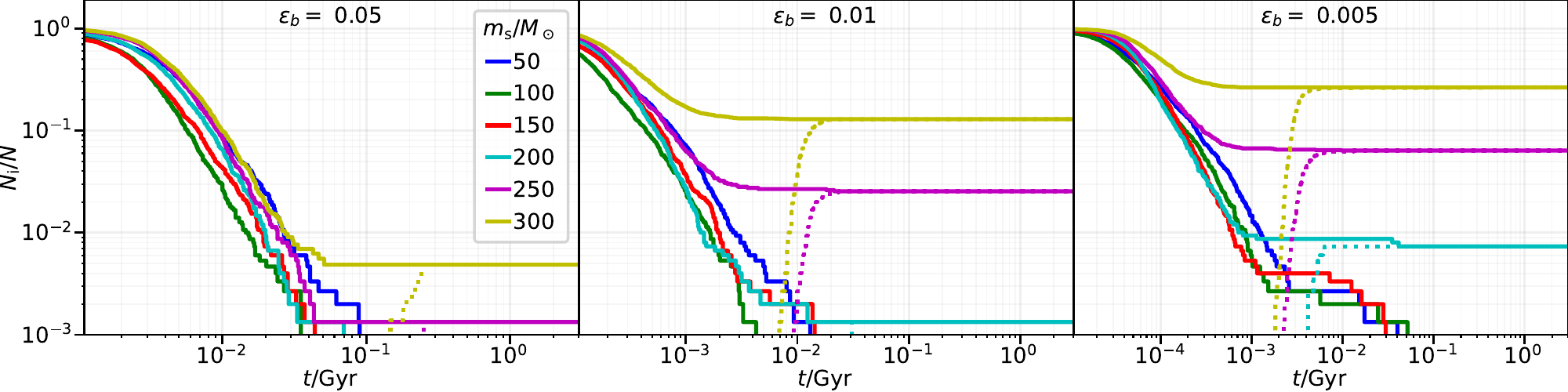}
\textbf{$M_{\rm cl} = 10^6 \,M_\odot$}
\caption{A grid depicting the fraction of BH seeds of initial mass, $m_{\rm s}$, retained in ${M_{\rm cl}=10^6\, M_\odot}$ clusters with density scaling parameter, $\epsilon_{\rm b}$. The dotted lines represent the fraction of seed BHs reaching our runaway threshold, $m_{\rm run} = 1000 M_\odot$, by time, $t$.}
\label{fig:remrunvst_1e06}
\end{figure*}

Following the $m_{\rm s0}/M_\odot = \{50,100\}$ models discussed previously, GW recoil kicks remain the primary source of ejection--most likely ejecting a BH seed following it's $2^{\rm nd}$ merger due to the kick amplification a spinning seed provides (Figs.~\ref{fig:probescvsgen1e+06} and \ref{fig:mfdist_1e06}). We find a $100\%$ ejection probability for {$50-150 \, M_\odot$} BH seeds. The most likely end-state for $m_{\rm s0}\geq200 \,M_\odot$ BH seeds is GW recoil ejection from the host cluster, but runaway is almost guaranteed for the few BH seeds which are not ejected by their $10^{\rm th}$ merger (Figs.~\ref{fig:probescvsgen1e+06}, \ref{fig:remrunvst_1e06} and Table~\ref{table:cusptable2}). These seeds constitute $0.1-0.7\%$, $2.5-6.4\%$, and $12.9-26.5\%$ of 200, 250, and $300 \,M_\odot$ BH seeds in $\epsilon_{\rm b} = \{0.01, 0.005\}$ models, respectively (Table~\ref{table:cusptable2}). In addition, the only $\epsilon_{\rm b} = 0.05$ models which host a runaway are sourced from $m_{\rm s0}/M_\odot = \{250, 300\}$ sequences. The probability of runaways here is small (${<1\%}$), with ${>99\%}$ of BH seeds escaping. This suggests clusters with $v_{\rm e,cl}\approx80-100$~km/s prohibit multi-generation seed growth and retention if $m_{\rm s0}\lesssim 250 \,M_\odot$.

\begin{figure*}
\centering
\includegraphics[width=\textwidth]{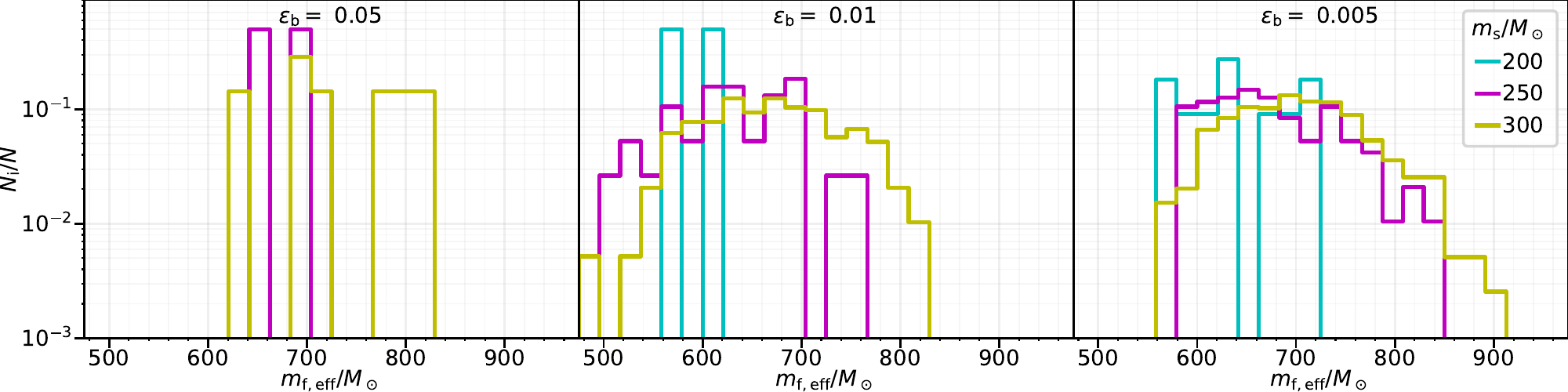}
\textbf{$M_{\rm cl} = 10^6 \,M_\odot$}\par\medskip

\caption{Distribution of runaway BH mass in our low-mass cluster expressed at the point at which $(>90\%)$ of the BH reservoir was ejected during the sequence of encounters. Alternatively stated, encounters $P_{\rm t}$ has with $\{ P_{\rm p} \}$ have a finite chance of dynamically ejecting a BH to infinity; if we assume ${\sim}1\%$ of our $10^6 \, M_\odot$ cluster mass is in BHs and the average BH mass is about $20 \, M_\odot$, then there are $500$ BHs in our static clusters. Therefore, $m_{\rm f,eff}$ is the seed mass at the point when $\approx450$ (non-seed) BHs had escaped to infinity during the sequence of encounters.}
\label{fig:mfeffplot_1e+06.pdf}
\end{figure*}

Despite runaway being a common end-state for $\gtrsim200 \,M_\odot$ seeds, $P_{\rm t}$ ejects $\gtrsim 90\%$ of the BH reservoir ($N_{\rm bh}$; see eq.~\ref{eq:N_Ns_Nb}) during every runaway sequence through dynamical kicks. Background statistics (see Table.~\ref{table:IC}) are fixed independent of what happens to reservoir BHs allowing sequences to continue uninterrupted, but the validity of a static reservoir is heavily challenged in low-mass clusters. Fig.~\ref{fig:mfeffplot_1e+06.pdf} displays the ending mass of a BH seed if a sequence was abruptly halted at the point $0.9 N_{\rm bh}$ BHs were ejected from the double-Plummer cluster. Unless it is assumed that the BH sub-cluster is replenished at a rate of about $40 \, \rm{BHs} /\rm{Myr}$, the average BH escape rate across all sequences, then the maximal mass an $m_{\rm s0} = 200-300 \, M_\odot$ runaway may reach in our $M_{\rm cl} = 10^6 \, M_\odot$ cluster model is between $500-800 \, M_\odot$.

\subsection{Varying Binary Fraction in High-Mass Clusters}

\begin{figure*}
\centering
\includegraphics[width=\textwidth]{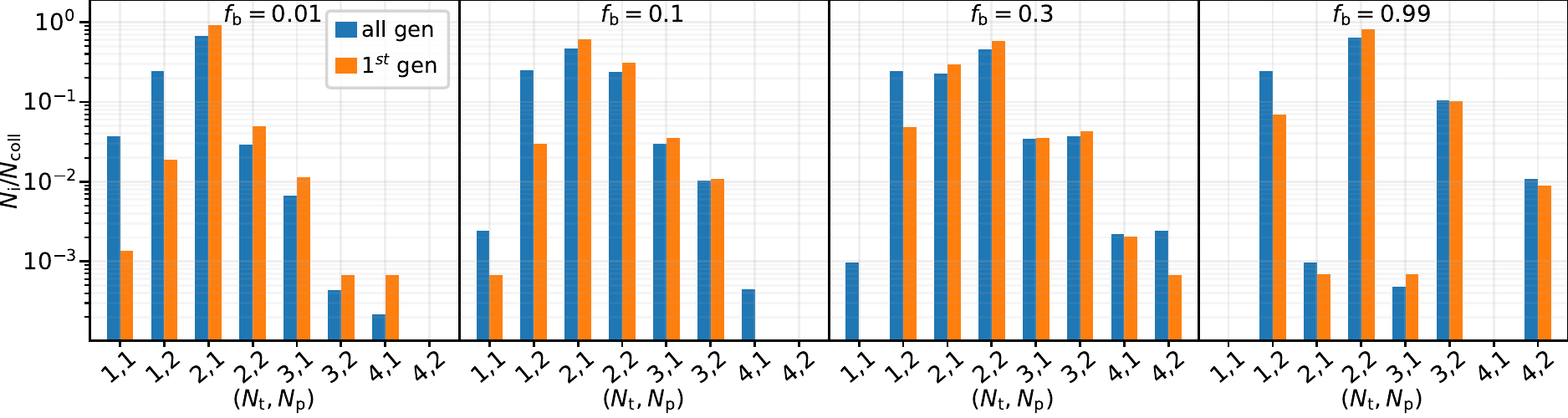}
\par\medskip
\includegraphics[width=\textwidth]{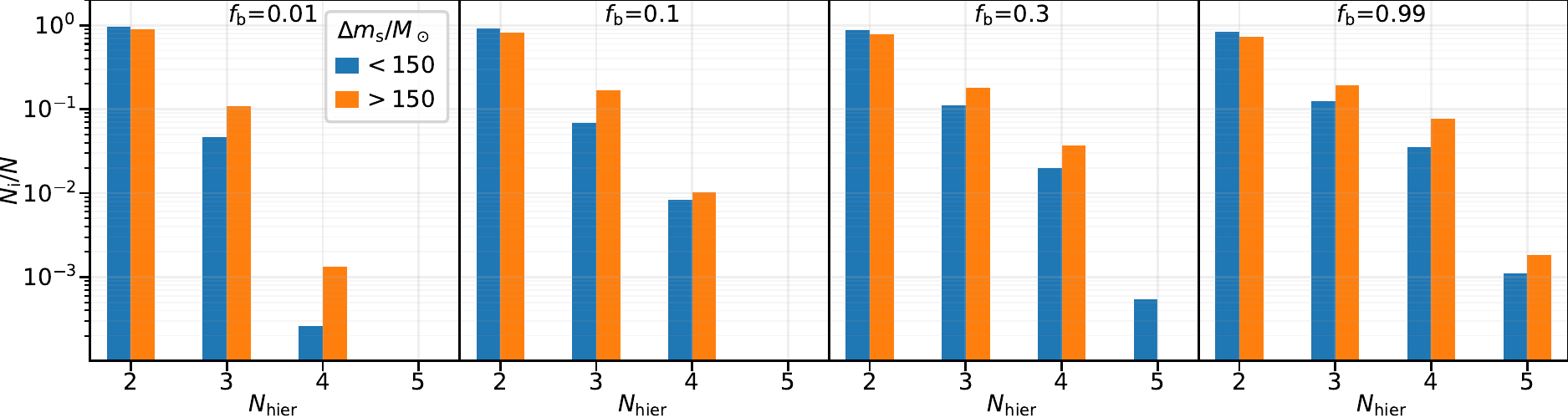}
\textbf{$(M_{\rm cl}, \epsilon_{\rm b}) = (10^8 \,M_\odot, 0.01)$}\par\medskip

\caption{From left to right, each panel corresponds to a different binary fraction, $f_{\rm b}$. Top: Fraction of $(N_{\rm t}, N_{\rm p})$ scatterings in which the seed undergoes a merger, akin to Fig.~\ref{fig:scatcollhist}. Bottom: Fraction of seeds undergoing merger within a certain hierarchy rank, akin to Fig.~\ref{fig:collhierhist}. As expected, the rate of larger hierarchy interactions increases as $f_{\rm b}$ increases.}
\label{fig:hierfbplots}
\end{figure*}

\begin{figure}
\centering
\includegraphics[width=.99\columnwidth]{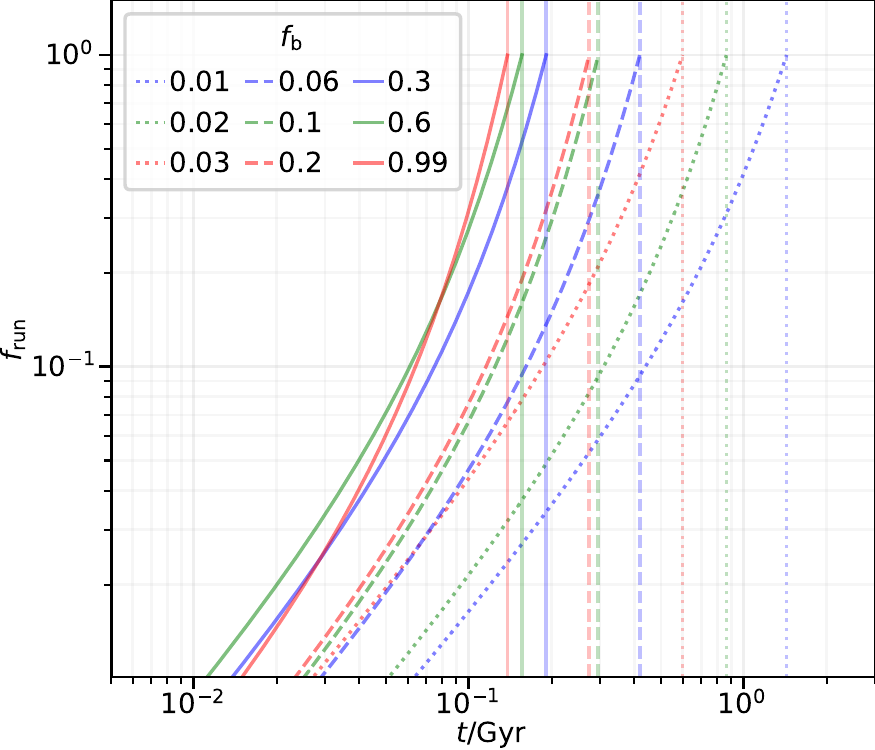}

\textbf{$(M_{\rm cl}, \epsilon_{\rm b}) = (10^8 \,M_\odot, 0.01)$}\par\medskip

\caption{Akin to Fig.~\ref{fig:masshistrem}, the solid lines are simple exponential fits to the median effective runaway fraction $f_{\rm run}$ across all runaway sequences for each binary fraction, $f_{\rm b}$. The dashed vertical line denotes the time the median growth curve reaches our runaway threshold, here set to $m_{\rm run} = 800 \, M_\odot$ for computational efficiency. As $f_{\rm b}$ increases, the time to cross the runaway threshold decreases; this is expected since the encounter rate increases with $f_{\rm b}$.}
\label{fig:masshistoryrem_avg50_fb.pdf}
\end{figure}

Throughout this work, we have held $f_{\rm b}$ constant. In reality, the evolving binary fraction is likely dependent on the evolving cluster properties and vice versa \citep{2003gmbp.book.....H,theoretical_uncertainties_chatterjee2017}. We reserve for future work a more detailed treatment of the binary fraction, especially as it relates to three-body-binary formation within the cores of dense stellar clusters \citep{Heggie_1993}. For now, to explore the possible role of binary fraction on our results, we present a small subset of simulations with varying binary fraction. We vary $f_{\rm b}$ in an $(M_{\rm cl}, \epsilon_{\rm b}) = \{10^8 M_\odot, 0.01\}$ model seeded with a $50 \, M_\odot$ BH.

As expected, the rate of large hierarchy ($N>2$) mergers increases with $f_{\rm b}$ as well as the probability of $N_{\rm t} + N_{\rm p} > 3$ interactions immediately preceding a merger (Fig.~\ref{fig:hierfbplots}). Despite this, and following from the trends observed in Figs.~\ref{fig:scatcollhist} and \ref{fig:collhierhist}, mergers within $N>2$ hierarchies remain less than $20\%$ of all merger scenarios, even for a $99\%$ binary fraction. While substantial, it is clear that the high-velocity BH cores of high-mass clusters are not suitable environments for maintaining $3+$ hierarchies. In addition, we find that the rate of runaways is largely unchanged across $f_{\rm b}$, ranging between $1\%$ and $2\%$ of all sequences for all $f_{\rm b}$ without a (presently) resolvable trend.

The most substantial effect increasing $f_{\rm b}$ has on our BH sequences is to increase the encounter rate and, in turn, reduce the time to runaway, $t_{\rm run}$, (Fig.~\ref{fig:masshistoryrem_avg50_fb.pdf}). This is expected, since BBHs have a much larger cross section than stellar mass BH singles. The time to runaway spans an order of magnitude, between $1.5$~Gyr to $0.5$~Gyr for ${f_{\rm b} = 1\% \text{ and } 99\%}$, respectively.

\section{Discussion \& Future Considerations}
\label{sec:disc}
The primary methods explored for producing MBHs from smaller BH seeds include (i) accretion of matter from local gaseous environments \citep[e.g.,][]{2014Sci...345.1330A, 2015ApJ...810...51M, 2016MNRAS.458.3047P, 2017ApJ...850L..42P, 2020MNRAS.493.3732D}, (ii) mergers of galaxies hosting sub-massive BHs, (iii) collapse from repeat stellar collisions \citep[e.g.,][]{gonzalez_intermediate-mass_2021}, and (iv) repeat mergers with other BHs \citep[e.g.,][]{2019arXiv190500902A, 2020ApJ...903...45K, 2021ApJ...907L..25W, 2022MNRAS.511.2631A}. Unfortunately, accretion growth is limited as a source for the rapid growth of seed BHs due to the fragile conditions necessary to maintain (super-)Eddington accretion rates, with numerous self-feedback and environmental mechanisms limiting its viability \citep{2020ARA&A..58...27I}. Galaxy-galaxy mergers tend to occur on timescales on the order of a Hubble Time, and are categorically left out of consideration for rapid BH growth as a consequence. 

There are several considerations which would further increase the plausibility of our calculations. The basis of our models is to consider static, double-Plummer clusters. The evolution of an NSC over 3~Gyr is likely extremely complex and may not be faithfully represented with a static double-Plummer system, even for 1~Gyr of evolution time. Evolving the global and local properties in time--such as the half-mass radii, total cluster mass, BH mass and spin distributions, binary fraction, and velocity dispersion--using up to date analytic or semi-analytic prescriptions of stellar cluster evolution will be implemented in Paper II. A BH reservoir with initially spinning BHs would likely reduce the fraction of BH seeds reaching runaway for any initial BH seed mass by further amplifying the effect GW recoil kicks have on retention and runaway probabilities \citep[e.g.,][]{2019_Rodriguez_review}. 

%

Repeat BH-BH mergers in \textsc{Cuspbuilding} were performed within the context of dense and gas-poor NSCs. A gas-rich environment could increase gas drag on massive bodies \citep{2022ApJ...931..149R},  further condensing a central BH population and likely increasing merger rates. The possibility of a BH seed encountering an object other than a BH has also been ignored, but could provide another avenue for mass growth \citep[e.g.,][]{giersz_mocca_2015,Rose2022} and potentially increase the relevance of gas-drag should the seed retain bound stellar material after a stellar collision.

In a single interaction, we only considered the possibility of an individual binary or single BH system interacting with $P_{\rm t}$. Though rare, three-hierarchy interactions are likely to occur and could further increase the availability of rich dynamical interactions to probe, especially in the densest cluster models. A self-consistent method for three separate hierarchies to interact simultaneously will be considered in future work. 

The tendency of an individual BH seed to be retained and runaway has been the focus of this work, but NSCs may contain many potential seeds over several Gyrs. This suggests that a small runaway probability for an individual BH seed may translate to a large runaway probability in an NSC. The possibility for two seeds to simultaneously and independently runaway has also been ignored, but would, instead, threaten the sustained growth of massive BH seeds. The probability of two seeds merging at the center of a BH sub-cluster would be large and resultant GW recoil kicks would be powerful, pushing our predicted runaway regimes to higher critical masses.

As shown in Fig.~\ref{fig:mfeffplot_1e+06.pdf}, assuming a large and constant BH reservoir is not self-consistent for lower mass (${\sim}10^6 M_\odot$) NSCs in isolation. GCs tend to migrate to the central regions of galaxies through dynamical friction effects and inevitably merge with central NSCs \citep{2014MNRAS.444.3738A, 2022A&A...658A.172F, Fragione2022nsc}. If a reasonable fraction of the mass of migrating GCs is composed of stellar mass BHs, BH losses from mergers and/or recoil kicks in the central NSC may be replenished. Thus, it may not be unreasonable to assume a roughly static reservoir of BHs under some contexts. Given a reservoir which tends to replenish BHs lost to recoil kicks and mergers, the ultimate numerical destiny of a seed on the runaway track is to become an SMBH ($>$$10^5\, M_\odot$); in some cases, within 1~Gyr. Given the context that our models implement the fits of \citet{10.1093/mnras/stw093}, where many individual NSC's are denser than the results of the ``late-type'' numerical fit we selected, it may not be unreasonable to consider these calculations a conservative estimate on the likelihood of BH runaway in the early universe. We note that the fits we use are relative to present time (small~$z$) and do not consider the contributions of a pre-existing SMBH to density profiles.



\section{Conclusion}\label{sec:conclusion}

The evolution of ${50-300 \,M_\odot}$ seed BHs in our NSC models provides novel insight into the repeat merger prospects of BH seeds within various host cluster environments. On timescales of {${\sim}0.01 - 1 \, \rm{Gyr}$} following their formation (Figs.~\ref{fig:masshistrem} and \ref{fig:tvsmesc}), BH seeds born in dense clusters will accumulate mergers rapidly enough to either escape their host cluster through (primarily GW) recoil kicks or quickly grow too heavy to be ejected. This suggests that BHs between $50-300 \,M_\odot$ should be very rare in dense massive clusters older than ${\sim}1\,{\rm Gyr}$. In a sense, rapid mergers enforce a \textit{dynamically-mediated mass gap} between ${50-300 \, M_\odot}$ in an NSC. Under the context of runaway timescales that we predict in our persistently dense BH reservoirs, this mass gap may extend far beyond $300 \, M_\odot$.

 The most likely end-state of an individual BH seed is to escape its host cluster except in the heaviest (${\sim}10^9 \, M_\odot$) cluster models. GW recoil kicks with at least one spinning member subject the BH merger product to violent kicks often surpassing the clusters central escape velocity. Since our seeds are initiated with zero spin, it is $2^{\rm nd}$ and $3^{\rm rd}$ generation mergers with spun-up seeds which act as the primary obstacle to runaway growth. Upon surpassing this generational threshold, the mass of the BH seed becomes large compared to the typical mass of the surrounding BH reservoir, suppressing GW recoil kicks of merger remnants. The likelihood of runaway is nearly guaranteed across all models if the seed does not escape following the $7^{\rm th}$ merger.

Mass vs spin distributions of our seeds agree with previous findings (see Sec.~\ref{sec:intro}). First and second generation mergers fall between ${\chi_{\rm eff} = 0.5 - 0.75}$. As runaway occurs, back-to-back mergers tend to spin down the seed to dimensionless spin magnitudes of {${\sim}10^{-2} - 10^{-1}$} (Fig.~\ref{fig:mvx_50.pdf}). 

Two dominant factors determine the seed sequence end-state (i.e. runaway, stall, or escape): the cluster's core escape velocity and dynamical friction timescale. The former controls the ease of BH ejection from the core, which either stalls continued runaway growth or brings it to a halt if the BH escapes the cluster entirely. The former controls how quickly a kicked BH returns to the core, and thereby how long such stalls last. As intuitively expected, the frequency of runaway growth increases in proportion to $M_{\rm cl}$, $\rho_{\rm c,bh}$, and $m_{\rm s}$, primarily affecting $v_{\rm e}$, $n_{\rm c,bh}$, and $v_{\rm kick}$, respectively. Independent of runaway status, both the mean and variance of the final mass of each seed increases with cluster mass. 

Low-mass (${\leq10^7 \,M_\odot}$) clusters struggle to retain BHs with runaway growth except in cases where a very large seed is already present---e.g., our clusters with $\{m_{\rm s}, M_{\rm cl}\}={\{100 \,M_\odot, 10^7\, M_\odot\}}$, or $\{m_{\rm s}, M_{\rm cl}\} = {\{\geq200 \,M_\odot, 10^6\, M_\odot\}}$. In the most challenging initial conditions for runaway growth, our models predict a $50 M_\odot$ seed has a $1.0-3.7\%$ chance of runaway within a $10^8 \,M_\odot$ cluster, a $100\, M_\odot$ seed has a $0.5\%$ chance of runaway within a $10^7 \, M_\odot$ cluster, and a $200 \,M_\odot$ seed has a $0.1-0.7\%$ chance of runaway within a $10^6 \, M_\odot$ cluster with probabilities of runaway growth dramatically increasing with $m_{\rm s}$, $M_{\rm cl}$, and $\rho_{\rm c,bh}$. Notably, a $100 \, M_\odot$ BH has a $76-89\%$ chance of achieving runaway within our $10^9 \,M_\odot$ cluster environment, ensuring a near guarantee of runaway growth within such a densely populated stellar community if even a handful of $50-100 \, M_\odot$ BHs form in these environments (Tables~\ref{table:cusptable}, \ref{table:cusptable2}; Figs.~\ref{fig:fvsmf}, \ref{fig:mfdist}, \ref{fig:mfdist_1e06}, \ref{fig:remrunvst_1e06}). 

Lower-mass clusters (${M_{\rm cl} = 10^6-10^7 \, M_\odot}$) with low escape speeds feature encounter rates amplified by stronger gravitational focusing. The enhanced encounter rate ensures a speedy conclusion to a seed's dynamical history ($<100 \, \rm{Myr}$ to escape or runaway), with seeds being ejected in ${(>99\%)}$ of sequences. Though expanding onto a runaway growth ``track'' in a ${\sim}10^6\, M_\odot$ cluster is possible for high-mass BH seeds of ${200-300 \,M_\odot}$, frequent dynamical ejections of stellar-mass BH projectiles produce the primary bottleneck to runaway growth (Fig.~\ref{fig:mfeffplot_1e+06.pdf}). In these runaways, the seed hierarchies dynamically eject $>90\%$ of the number of BHs in the sub-cluster before a BH seed can exceed about $500-800\,M_\odot$. Depletion of the BH reservoir ceases to be a serious concern in our simulations for ${M_{\rm cl} \geq 10^7\, M_\odot}$ clusters within our 3~Gyr evolution time and ${m_{\rm run} = 1000\, M_\odot}$ runaway threshold.


For computational efficiency, the formation and dynamics of triples and higher-order hierarchies is typically neglected in cluster modeling. Our calculations help to inform the impact these higher-order hierarchies have on mergers, merger dynamics, and runaways. The survivability of multi-component hierarchies in clusters is constrained by two limiting factors: (i) the average inter-particle distance limiting the maximum size a hierarchy may expand into and (ii) the GW merger timescale. If the latter is comparable to or smaller than the encounter timescale, the innermost orbit will decay before another interaction may occur, reducing or completely splitting the hierarchy. Given these limitations, multi-component hierarchical assembly still plays a significant part in the merger history of BH seeds. We find that mergers occurring within large, 3+ hierarchies total about $45\%$ and $11\%$ ($30\%$ for runaway track BHs) of all mergers in clusters of mass $10^6 \,M_\odot$ and $10^8 M_\odot$, respectively (Fig.~\ref{fig:collhierhist}). We also find that mergers during an interaction involving a seed hierarchy of rank 3+ contribute roughly $30\%$ and $5\%$ of all mergers in clusters of mass $10^6\, M_\odot$ and $10^8\, M_\odot$, respectively (Fig.~\ref{fig:scatcollhist}). Unsurprisingly, the contribution to the overall rate from 3+ hierarchy mergers in the $10^9\, M_\odot$ clusters is very low (about $8-10\%$ for runaway track BHs), due to the small inter-particle distance that limits the size of stable hierarchies in such massive clusters.

In accordance with our results, the most likely environment for rapid and unrestrained runaway BH growth are young and dense NSCs as they are likely to provide a population rich in stellar mass BHs and a gravitational potential well deep enough to retain BH merger products. We predict that a runaway track BH of ${m_{\rm bh}\geq10^3 \, M_\odot}$ should form within any ${\sim}10^8 \, M_\odot$ NSC roughly $100-300 \,$Myr following the assembly of a BH sub-cluster. This finding is subject to significant uncertainty in how well our assumed static double-Plummer model describes an NSC that does not yet host an IMBH/SMBH. High redshift observations from JWST will help assess the validity of our young NSC models and inform future developments.

\begin{table*}
\caption{Table depicting average values relevant to the net conglomerate of BH lifetime sequences simulated for each of 12 cluster models, with $m_{\rm s0}/M_\odot = \{50, 100\}$, varying cluster mass, $M_{\rm cl}$. The first 6 columns detail terms describing the physical characteristics of the cluster models held constant over a sequence while the following columns list statistics of interest. From left to right: average final seed mass, $\langle m_{\rm f} \rangle$, percentage of BH seeds which escape during the sequence (escape), percentage of BH seeds which merge $N$ times (generation), and the percentage of sequences which hosted a $P_{\rm t}$ reaching a maximal hierarchy rank $(3,4,5)$ at some point during the sequence.}
\raggedright
\fontsize{7.8pt}{10pt}\selectfont
\addtolength{\tabcolsep}{-2.2pt} 
\begin{tabular}{c || l l r r r r | r r | r r r r r r r r r | r r r ||}
\toprule
{} &  $M_{\rm cl}$ &          $\epsilon_{\rm b}$ & $r_{\rm h}$ & $v_{\rm e,cl}$ & $ \rho_{\rm c,bh}$ & $ \sigma_{\rm c,bh}$ & $\langle m_{\rm f}\rangle$ & escape & \multicolumn{9}{l}{generation $(\%)$} & \multicolumn{3}{l}{max rank $(\%)$} \\
{} & $(M_{\odot})$ & $\frac{r_{\rm h,bh}}{r_{\rm h}}$ &      $(pc)$ &           (km/s) &           $(M_\odot \, pc^{-3})$ &                             (km/s) &                $(M_\odot)$ & $(\%)$ &                 1 &     2 &     3 &     4 &     5 &     6 &     7 &     8 &   $\infty$ &                    3 &     4 &    5 \\
\midrule
0 & $10^{6}$ & 0.05   & 2.1  & 79   & ${3.3} \times 10^{6}$ &  9 &   64 & 100.0 & 67.4  & 1.4 &  0.1 &  0.0 &  0.0 &  0.0 &  0.0 &  0.0 &  0.0 & 91.1 & 25.2 & 0.4 \\
1 & $10^{6}$ & 0.01   & 2.1  & 103  & ${4.1} \times 10^{8}$ &  20 &  62 & 100.0 & 55.9  & 4.3 &  0.1 &  0.0 &  0.0 &  0.0 &  0.0 &  0.0 &  0.0 & 77.1 &  8.7 & 0.0 \\
2 & $10^{6}$ & 0.005  & 2.1  & 126  & ${3.3} \times 10^{9}$ &  29 &  63 & 100.0 & 55.6 & 10.7 &  0.7 &  0.0 &  0.0 &  0.0 &  0.0 &  0.0 &  0.0 & 71.8 &  7.3 & 0.1 \\
\hline
3 & $10^{7}$ & 0.05   & 4.5  & 174  & ${3.6} \times 10^{6}$ &  20 &  81 & 98.9  & 94.8 & 56.3 &  4.1 &  0.5 &  0.1 &  0.0 &  0.0 &  0.0 &  0.0 & 85.3 & 15.7 & 0.5 \\
4 & $10^{7}$ & 0.01   & 4.5  & 224  & ${4.5} \times 10^{8}$ &  44 &  83 & 100.0 & 90.5 & 71.9 &  8.9 &  0.7 &  0.1 &  0.0 &  0.0 &  0.0 &  0.0 & 64.3 &  5.8 & 0.1 \\
5 & $10^{7}$ & 0.005  & 4.5  & 274  & ${3.6} \times 10^{9}$ &  63 &  85 & 99.9  & 90.6 & 73.2 & 12.7 &  2.5 &  0.6 &  0.3 &  0.2 &  0.1 &  0.0 & 52.7 &  2.9 & 0.1 \\
\hline
6 & $10^{8}$ & 0.05   & 9.4  & 379  & ${3.9} \times 10^{6}$ &  43 &  96 & 96.3  & 99.4 & 97.8 & 27.1 &  9.0 &  3.1 &  1.0 &  0.7 &  0.5 &  0.0 & 49.9 &  2.6 & 0.0 \\
7 & $10^{8}$ & 0.01   & 9.4  & 490  & ${4.9} \times 10^{8}$ &  97 & 106 & 98.9  & 98.8 & 95.3 & 32.7 & 13.7 &  6.3 &  3.3 &  2.3 &  1.9 &  1.0 & 20.3 &  1.3 & 0.0 \\
8 & $10^{8}$ & 0.005  & 9.4  & 600  & ${3.9} \times 10^{9}$ & 137 & 133 & 95.9  & 98.8 & 94.0 & 41.4 & 19.8 & 11.7 &  8.3 &  6.4 &  5.4 &  3.7 & 16.0 &  1.3 & 0.0 \\
\hline
9 & $10^{9}$ & 0.05   & 19.6 & 829  & ${4.2} \times 10^{6}$ &  95 &  86 & 38.8  & 98.9 & 79.9 & 16.9 &  2.0 &  0.1 &  0.0 &  0.0 &  0.0 &  0.0 &  4.3 &  0.1 & 0.0 \\
10 & $10^{9}$ & 0.01  & 19.6 & 1070 & ${5.3} \times 10^{8}$ & 212 & 335 & 70.4  & 99.9 & 98.6 & 67.4 & 48.4 & 36.4 & 31.7 & 28.9 & 27.5 & 24.2 & 20.0 &  1.7 & 0.1 \\
11 & $10^{9}$ & 0.005 & 19.6 & 1310 & ${4.2} \times 10^{9}$ & 300 & 528 & 52.8  & 99.9 & 98.2 & 77.7 & 62.5 & 54.2 & 50.8 & 48.5 & 47.4 & 46.5 & 28.4 &  1.6 & 0.0 \\
\bottomrule
\end{tabular}
\centering
{\\(a) $m_{\rm s0} = 50 M_\odot$}
\par\bigskip
\begin{tabular}{c || l l r r r r | r r | r r r r r r r r r | r r r ||}
\toprule
{} &  $M_{\rm cl}$ &          $\epsilon_{\rm b}$ & $r_{\rm h}$ & $v_{\rm e,cl}$ & $ \rho_{\rm c,bh}$ & $ \sigma_{\rm c,bh}$ & $\langle m_{\rm f}\rangle$ & escape & \multicolumn{9}{l}{generation $(\%)$} & \multicolumn{3}{l}{max rank $(\%)$} \\
{} & $(M_{\odot})$ & $\frac{r_{\rm h,bh}}{r_{\rm h}}$ &      $(pc)$ &           (km/s) &           $(M_\odot \, pc^{-3})$ &                             (km/s) &                $(M_\odot)$ & $(\%)$ &                 1 &      2 &     3 &     4 &     5 &     6 &     7 &     8 &   $\infty$ &                    3 &     4 &    5 \\
\midrule
0 & $10^{6}$ & 0.05 & 2.10 & 79 & ${3.3} \times 10^{6}$ & 9 & 117        & 100.0  & 82.3  &  2.1 &   0. &   0. &   0.0 & 0.0  &  0.0    & 0.0  & 0.0  & 95.3 & 37.3 & 1.3  \\
1 & $10^{6}$ & 0.01 & 2.10 & 103 & ${4.1} \times 10^{8}$ & 20 & 117      & 100.0  & 74.4  &  8.5 &  0.3 &  0.1 &   0.0 & 0.0  &  0.0    & 0.0  & 0.0  & 86.8 & 17.3 & 0.5 \\
2 & $10^{6}$ & 0.005 & 2.10 & 126 & ${3.3} \times 10^{9}$ & 29 & 119     &  99.9  & 77.1  & 19.2 &  1.8 &  0.1 &   0.1 & 0.0  &  0.0    & 0.0  & 0.0  & 82.5 & 15.0 & 0.3 \\
\hline
3 & $10^{7}$ & 0.05 & 4.50 & 174 & ${3.6} \times 10^{6}$ & 20 & 139      &  99.7  & 99.0  & 78.9 & 11.4 &  2.5 &   0.5 & 0.1  &  0.1    & 0.0  & 0.0  & 93.8 & 26.7 & 0.9 \\
4 & $10^{7}$ & 0.01 & 4.50 & 224 & ${4.5} \times 10^{8}$ & 44 & 142      & 100.0  & 96.9  & 85.2 & 20.2 &  5.8 &   2.4 & 1.4  &  0.7    & 0.5  & 0.0  & 77.4  & 8.8 & 0.2 \\
5 & $10^{7}$ & 0.005 & 4.50 & 274 & ${3.6} \times 10^{9}$ & 63 & 149     &  99.5  & 97.1  & 88.8 & 25.4 & 10.7 &   4.8 & 2.4  &  1.8    & 1.5  & 0.5  & 66.4  & 5.7 & 0.1 \\
\hline
6 & $10^{8}$ & 0.05 & 9.40 & 379 & ${3.9} \times 10^{6}$ & 43 & 183      &  91.3  & 99.9  & 99.7 & 43.8 & 23.7 &  14.1 & 9.6  &  7.3    & 6.2  & 1.0  & 63.3  & 8.5 & 0.4 \\
7 & $10^{8}$ & 0.01 & 9.40 & 490 & ${4.9} \times 10^{8}$ & 97 & 296      &  83.7  & 99.8  & 98.5 & 58.6 & 37.9 &  28.1 & 23.0 & 20.1    & 17.9 & 16.0 & 42.1  & 9.5 & 0.4 \\
8 & $10^{8}$ & 0.005 & 9.40 & 600 & ${3.9} \times 10^{9}$ & 137 & 405    &  71.4  & 99.7  & 98.5 & 68.8 & 51.4 &  41.2 & 36.4 & 33.9    & 31.8 & 28.4 & 43.4  & 8.1 & 0.5 \\
\hline
9 & $10^{9}$ & 0.05 & 19.60 & 829 & ${4.2} \times 10^{6}$ & 95 & 153     &  22.5  & 100.0 & 97.1 & 57.9 & 24.6 &   7.1 & 1.5  &  0.1    & 0.0  & 0.0  & 10.5  & 0.1 & 0.0 \\
10 & $10^{9}$ & 0.01 & 19.60 & 1070 & ${5.3} \times 10^{8}$ & 212 & 814  &  22.4  & 100.0 & 99.7 & 93.3 & 86.2 &  82.3 & 79.5 & 78.0    & 77.7 & 76.4 & 61.8  & 4.2 & 0.3 \\
11 & $10^{9}$ & 0.005 & 19.60 & 1310 & ${4.2} \times 10^{9}$ & 300 & 922 &  10.6  & 100.0 & 99.7 & 97.6 & 94.5 &  92.2 & 91.2 & 90.1    & 89.9 & 89.3 & 48.5  & 2.4 & 0.1 \\
\bottomrule
\end{tabular}
\centering
{\\(b) $m_{\rm s0} = 100 M_\odot$}\\

\label{table:cusptable}
\end{table*}

\begin{table*}%
\caption{Table depicting average values relevant to BH sequences simulated for each of 18 cluster models varying initial seed mass, $m_{\rm{s0}}$, within cluster mass, $M_{\rm{cl}}=10^6 M_\odot$. It is organized identically to Table~\ref{table:cusptable}.}
\raggedright
\fontsize{7.9pt}{10pt}\selectfont
\centering
{$M_{\rm cl} = 10^6 M_\odot$}
\addtolength{\tabcolsep}{-2.2pt} 
\begin{tabular}{c || r l r r r r | r r | r r r r r r r r r | r r r r ||}
\toprule
{} &   $m_{\rm s0}$ &          $\epsilon_{\rm b}$ & $r_{\rm h}$ & $v_{\rm e,cl}$ & $\rho_{\rm c,bh}$ & $\sigma_{\rm c,bh}$ & $\langle m_{\rm f}\rangle$ & escape & \multicolumn{9}{l}{generation $(\%)$} & \multicolumn{3}{l}{max rank $(\%)$} \\
{} & $(M_{\odot})$ & $\frac{r_{\rm h,bh}}{r_{\rm h}}$ &      (pc) &           (km/s) &           $(M_\odot \, pc^{-3})$ &                             (km/s) &                $(M_\odot)$ & $(\%)$ &                 1 &     2 &     3 &     4 &     5 &     6 &     7 &     8 &   $\infty$ &                    3 &     4 &     5 \\
\midrule
0 & 50 & 0.05 & 2.1 & 79 & ${3.3} \times 10^{6}$ & 9 & 64 & 100.0 & 67.4 & 1.4 & 0.1 & 0.0 & 0.0 & 0.0 & 0.0 & 0.0 & 0.0 & 91.1 & 25.2 & 0.4 \\
1 & 100 & 0.05 & 2.1 & 79 & ${3.3} \times 10^{6}$ & 9 & 117 & 100.0 & 82.3 & 2.1 & 0.0 & 0.0 & 0.0 & 0.0 & 0.0 & 0.0 & 0.0 & 95.3 & 37.3 & 1.3 \\
2 & 150 & 0.05 & 2.1 & 79 & ${3.3} \times 10^{6}$ & 9 & 172 & 100.0 & 90.4 & 18.1 & 1.1 & 0.1 & 0.1 & 0.0 & 0.0 & 0.0 & 0.0 & 96.9 & 49.7 & 3.4 \\
3 & 200 & 0.05 & 2.1 & 79 & ${3.3} \times 10^{6}$ & 9 & 233 & 100.0 & 94.3 & 52.3 & 9.5 & 2.1 & 0.5 & 0.2 & 0.1 & 0.1 & 0.0 & 98.7 & 70.5 & 6.7 \\
4 & 250 & 0.05 & 2.1 & 79 & ${3.3} \times 10^{6}$ & 9 & 299 & 99.9 & 97.2 & 80.6 & 26.7 & 14.2 & 5.9 & 3.1 & 1.9 & 1.2 & 0.1 & 99.8 & 85.9 & 14.8 \\
5 & 300 & 0.05 & 2.1 & 79 & ${3.3} \times 10^{6}$ & 9 & 369 & 99.5 & 97.6 & 91.9 & 46.9 & 28.1 & 16.7 & 11.2 & 6.8 & 4.9 & 0.5 & 100.0 & 91.0 & 25.1 \\
\hline
6 & 50 & 0.01 & 2.1 & 103 & ${4.1} \times 10^{8}$ & 20 & 62 & 100.0 & 55.9 & 4.3 & 0.1 & 0.0 & 0.0 & 0.0 & 0.0 & 0.0 & 0.0 & 77.1 & 8.7 & 0.0 \\
7 & 100 & 0.01 & 2.1 & 103 & ${4.1} \times 10^{8}$ & 20 & 117 & 100.0 & 74.4 & 8.5 & 0.3 & 0.1 & 0.0 & 0.0 & 0.0 & 0.0 & 0.0 & 86.8 & 17.3 & 0.5 \\
8 & 150 & 0.01 & 2.1 & 103 & ${4.1} \times 10^{8}$ & 20 & 176 & 100.0 & 85.1 & 38.8 & 5.2 & 0.9 & 0.2 & 0.1 & 0.1 & 0.1 & 0.0 & 94.0 & 34.0 & 1.5 \\
9 & 200 & 0.01 & 2.1 & 103 & ${4.1} \times 10^{8}$ & 20 & 239 & 99.9 & 89.9 & 67.8 & 16.4 & 6.5 & 2.8 & 1.4 & 0.9 & 0.7 & 0.1 & 97.9 & 49.8 & 4.2 \\
10 & 250 & 0.01 & 2.1 & 103 & ${4.1} \times 10^{8}$ & 20 & 323 & 97.5 & 94.1 & 85.5 & 39.2 & 22.6 & 12.9 & 9.2 & 7.0 & 5.8 & 2.5 & 99.4 & 67.3 & 8.8 \\
11 & 300 & 0.01 & 2.1 & 103 & ${4.1} \times 10^{8}$ & 20 & 456 & 87.1 & 96.0 & 90.7 & 60.5 & 43.2 & 32.6 & 26.1 & 22.0 & 19.4 & 12.9 & 99.8 & 79.6 & 22.1 \\
\hline
12 & 50 & 0.005 & 2.1 & 126 & ${3.3} \times 10^{9}$ & 29 & 63 & 100.0 & 55.6 & 10.7 & 0.7 & 0.0 & 0.0 & 0.0 & 0.0 & 0.0 & 0.0 & 71.8 & 7.3 & 0.1 \\
13 & 100 & 0.005 & 2.1 & 126 & ${3.3} \times 10^{9}$ & 29 & 119 & 99.9 & 77.1 & 19.2 & 1.8 & 0.1 & 0.1 & 0.0 & 0.0 & 0.0 & 0.0 & 82.5 & 15.0 & 0.3 \\
14 & 150 & 0.005 & 2.1 & 126 & ${3.3} \times 10^{9}$ & 29 & 181 & 100.0 & 84.5 & 56.9 & 10.7 & 2.7 & 0.8 & 0.5 & 0.1 & 0.1 & 0.0 & 93.5 & 25.9 & 1.1 \\
15 & 200 & 0.005 & 2.1 & 126 & ${3.3} \times 10^{9}$ & 29 & 250 & 99.3 & 91.6 & 77.4 & 26.0 & 12.4 & 6.2 & 3.3 & 2.3 & 1.9 & 0.7 & 97.5 & 41.1 & 2.2 \\
16 & 250 & 0.005 & 2.1 & 126 & ${3.3} \times 10^{9}$ & 29 & 358 & 93.6 & 94.1 & 86.0 & 48.7 & 32.2 & 22.7 & 17.2 & 13.8 & 11.6 & 6.4 & 98.9 & 60.0 & 8.9 \\
17 & 300 & 0.005 & 2.1 & 126 & ${3.3} \times 10^{9}$ & 29 & 548 & 73.5 & 96.5 & 91.8 & 70.4 & 56.5 & 48.0 & 42.1 & 37.9 & 34.7 & 26.5 & 99.0 & 78.1 & 26.5 \\

\bottomrule
\end{tabular}
\centering

\label{table:cusptable2}
\end{table*}

\section*{Acknowledgements}
This work was supported by NASA Grant~80NSSC21K1722 and NSF Grant~AST-2108624 at Northwestern University.
We thank Jeremy Rath for helpful discussions on secular evolution of orbits. This work was supported through the computational resources and staff contributions provided for the Quest high performance computing facility at Northwestern University. Quest is jointly supported by the Office of the Provost, the Office for Research, and Northwestern University Information Technology. This work also used computing resources at CIERA funded by NSF Grant~PHY-1726951. AAT is supported by JSPS Grants-in-Aid for Scientific Research~19K03907 and~21K13914. DVA acknowledges support from the NSF Graduate Research Fellowship Program under Grant~DGE-1842165. KK is supported by an NSF Astronomy and Astrophysics Postdoctoral Fellowship under award~AST-2001751.

\section{Data Availability}
The data underlying this article will be shared on reasonable request to the corresponding author.

\clearpage
\clearpage

\bibliographystyle{mnras}
\bibliography{Atallah_2022} 

\begin{thebibliography}{}
\makeatletter
\relax
\def\mn@urlcharsother{\let\do\@makeother \do\$\do\&\do\#\do\^\do\_\do\%\do\~}
\def\mn@doi{\begingroup\mn@urlcharsother \@ifnextchar [ {\mn@doi@}
  {\mn@doi@[]}}
\def\mn@doi@[#1]#2{\def\@tempa{#1}\ifx\@tempa\@empty \href
  {http://dx.doi.org/#2} {doi:#2}\else \href {http://dx.doi.org/#2} {#1}\fi
  \endgroup}
\def\mn@eprint#1#2{\mn@eprint@#1:#2::\@nil}
\def\mn@eprint@arXiv#1{\href {http://arxiv.org/abs/#1} {{\tt arXiv:#1}}}
\def\mn@eprint@dblp#1{\href {http://dblp.uni-trier.de/rec/bibtex/#1.xml}
  {dblp:#1}}
\def\mn@eprint@#1:#2:#3:#4\@nil{\def\@tempa {#1}\def\@tempb {#2}\def\@tempc
  {#3}\ifx \@tempc \@empty \let \@tempc \@tempb \let \@tempb \@tempa \fi \ifx
  \@tempb \@empty \def\@tempb {arXiv}\fi \@ifundefined
  {mn@eprint@\@tempb}{\@tempb:\@tempc}{\expandafter \expandafter \csname
  mn@eprint@\@tempb\endcsname \expandafter{\@tempc}}}

\bibitem[\protect\citeauthoryear{{Aarseth}}{{Aarseth}}{1999}]{Aarseth_1999}
{Aarseth} S.~J.,  1999, \mn@doi [\pasp] {10.1086/316455}, \href
  {https://ui.adsabs.harvard.edu/abs/1999PASP..111.1333A} {111, 1333}

\bibitem[\protect\citeauthoryear{{Aarseth}}{{Aarseth}}{2003}]{aarseth-nbody}
{Aarseth} S.~J.,  2003, {Gravitational N-Body Simulations}

\bibitem[\protect\citeauthoryear{{Abbott} et~al.,}{{Abbott}
  et~al.}{2019}]{LIGO_2019_population}
{Abbott} B.~P.,  et~al., 2019, \mn@doi [\apjl] {10.3847/2041-8213/ab3800},
  \href {https://ui.adsabs.harvard.edu/abs/2019ApJ...882L..24A} {882, L24}

\bibitem[\protect\citeauthoryear{{Alexander} \& {Natarajan}}{{Alexander} \&
  {Natarajan}}{2014}]{2014Sci...345.1330A}
{Alexander} T.,  {Natarajan} P.,  2014, \mn@doi [Science]
  {10.1126/science.1251053}, \href
  {https://ui.adsabs.harvard.edu/abs/2014Sci...345.1330A} {345, 1330}

\bibitem[\protect\citeauthoryear{Antognini \& Thompson}{Antognini \&
  Thompson}{2016a}]{antognini_dynamical_2016}
Antognini J. M.~O.,  Thompson T.~A.,  2016a, \mn@doi [Monthly Notices of the
  Royal Astronomical Society] {10.1093/mnras/stv2938}, 456, 4219

\bibitem[\protect\citeauthoryear{{Antognini} \& {Thompson}}{{Antognini} \&
  {Thompson}}{2016b}]{2016MNRAS.456.4219A}
{Antognini} J. M.~O.,  {Thompson} T.~A.,  2016b, \mn@doi [\mnras]
  {10.1093/mnras/stv2938}, \href
  {https://ui.adsabs.harvard.edu/abs/2016MNRAS.456.4219A} {456, 4219}

\bibitem[\protect\citeauthoryear{Antonini \& Rasio}{Antonini \&
  Rasio}{2016a}]{antonini_merging_2016}
Antonini F.,  Rasio F.~A.,  2016a, \mn@doi [The Astrophysical Journal]
  {10.3847/0004-637X/831/2/187}, 831, 187

\bibitem[\protect\citeauthoryear{{Antonini} \& {Rasio}}{{Antonini} \&
  {Rasio}}{2016b}]{Antonini2016}
{Antonini} F.,  {Rasio} F.~A.,  2016b, \mn@doi [\apj]
  {10.3847/0004-637X/831/2/187}, \href
  {https://ui.adsabs.harvard.edu/abs/2016ApJ...831..187A} {831, 187}

\bibitem[\protect\citeauthoryear{{Antonini}, {Gieles}  \&
  {Gualandris}}{{Antonini} et~al.}{2019}]{antonini_bhgrowth_2019}
{Antonini} F.,  {Gieles} M.,   {Gualandris} A.,  2019, \mn@doi [\mnras]
  {10.1093/mnras/stz1149}, \href
  {https://ui.adsabs.harvard.edu/abs/2019MNRAS.486.5008A} {486, 5008}

\bibitem[\protect\citeauthoryear{{Arca-Sedda} \&
  {Capuzzo-Dolcetta}}{{Arca-Sedda} \&
  {Capuzzo-Dolcetta}}{2014}]{2014MNRAS.444.3738A}
{Arca-Sedda} M.,  {Capuzzo-Dolcetta} R.,  2014, \mn@doi [\mnras]
  {10.1093/mnras/stu1683}, \href
  {https://ui.adsabs.harvard.edu/abs/2014MNRAS.444.3738A} {444, 3738}

\bibitem[\protect\citeauthoryear{{Arca Sedda}, {Askar}  \& {Giersz}}{{Arca
  Sedda} et~al.}{2019}]{2019arXiv190500902A}
{Arca Sedda} M.,  {Askar} A.,   {Giersz} M.,  2019, arXiv e-prints, \href
  {https://ui.adsabs.harvard.edu/abs/2019arXiv190500902A} {p. arXiv:1905.00902}

\bibitem[\protect\citeauthoryear{{Askar}, {Davies}  \& {Church}}{{Askar}
  et~al.}{2022}]{2022MNRAS.511.2631A}
{Askar} A.,  {Davies} M.~B.,   {Church} R.~P.,  2022, \mn@doi [\mnras]
  {10.1093/mnras/stab3741}, \href
  {https://ui.adsabs.harvard.edu/abs/2022MNRAS.511.2631A} {511, 2631}

\bibitem[\protect\citeauthoryear{Banerjee}{Banerjee}{2017}]{banerjee_stellar-mass_2017}
Banerjee S.,  2017, \mn@doi [Monthly Notices of the Royal Astronomical Society]
  {10.1093/mnras/stw3392}, 467, 524

\bibitem[\protect\citeauthoryear{{Banerjee} \& {Kroupa}}{{Banerjee} \&
  {Kroupa}}{2011}]{2011ApJ...741L..12B}
{Banerjee} S.,  {Kroupa} P.,  2011, \mn@doi [\apjl]
  {10.1088/2041-8205/741/1/L12}, \href
  {https://ui.adsabs.harvard.edu/abs/2011ApJ...741L..12B} {741, L12}

\bibitem[\protect\citeauthoryear{{Binney} \& {Tremaine}}{{Binney} \&
  {Tremaine}}{2008}]{2008gady.book.....B}
{Binney} J.,  {Tremaine} S.,  2008, {Galactic Dynamics: Second Edition}.
Princeton University Press

\bibitem[\protect\citeauthoryear{{Blanchet}}{{Blanchet}}{2014}]{2014LRR....17....2B}
{Blanchet} L.,  2014, \mn@doi [Living Reviews in Relativity]
  {10.12942/lrr-2014-2}, \href
  {https://ui.adsabs.harvard.edu/abs/2014LRR....17....2B} {17, 2}

\bibitem[\protect\citeauthoryear{{Breen} \& {Heggie}}{{Breen} \&
  {Heggie}}{2013}]{2013MNRAS.432.2779B}
{Breen} P.~G.,  {Heggie} D.~C.,  2013, \mn@doi [\mnras] {10.1093/mnras/stt628},
  \href {https://ui.adsabs.harvard.edu/abs/2013MNRAS.432.2779B} {432, 2779}

\bibitem[\protect\citeauthoryear{Campanelli, Lousto, Zlochower  \&
  Merritt}{Campanelli et~al.}{2007}]{campanelli_maximum_2007}
Campanelli M.,  Lousto C.~O.,  Zlochower Y.,   Merritt D.,  2007, \mn@doi
  [Physical Review Letters] {10.1103/PhysRevLett.98.231102}, 98, 231102

\bibitem[\protect\citeauthoryear{{Castellano} et~al.,}{{Castellano}
  et~al.}{2022}]{2022_JWST_earlyhighgalaxies}
{Castellano} M.,  et~al., 2022, arXiv e-prints, \href
  {https://ui.adsabs.harvard.edu/abs/2022arXiv220709436C} {p. arXiv:2207.09436}

\bibitem[\protect\citeauthoryear{{Chatterjee}, {Rodriguez}  \&
  {Rasio}}{{Chatterjee}
  et~al.}{2017}]{theoretical_uncertainties_chatterjee2017}
{Chatterjee} S.,  {Rodriguez} C.~L.,   {Rasio} F.~A.,  2017, \mn@doi [\apj]
  {10.3847/1538-4357/834/1/68}, \href
  {https://ui.adsabs.harvard.edu/abs/2017ApJ...834...68C} {834, 68}

\bibitem[\protect\citeauthoryear{{Dittmann} \& {Miller}}{{Dittmann} \&
  {Miller}}{2020}]{2020MNRAS.493.3732D}
{Dittmann} A.~J.,  {Miller} M.~C.,  2020, \mn@doi [\mnras]
  {10.1093/mnras/staa463}, \href
  {https://ui.adsabs.harvard.edu/abs/2020MNRAS.493.3732D} {493, 3732}

\bibitem[\protect\citeauthoryear{Downing, Benacquista, Giersz  \&
  Spurzem}{Downing et~al.}{2010}]{downing_compact_2010}
Downing J. M.~B.,  Benacquista M.~J.,  Giersz M.,   Spurzem R.,  2010, \mn@doi
  [Monthly Notices of the Royal Astronomical Society]
  {10.1111/j.1365-2966.2010.17040.x}, 407, 1946

\bibitem[\protect\citeauthoryear{{Fahrion}, {Leaman}, {Lyubenova}  \& {van de
  Ven}}{{Fahrion} et~al.}{2022}]{2022A&A...658A.172F}
{Fahrion} K.,  {Leaman} R.,  {Lyubenova} M.,   {van de Ven} G.,  2022, \mn@doi
  [\aap] {10.1051/0004-6361/202039778}, \href
  {https://ui.adsabs.harvard.edu/abs/2022A&A...658A.172F} {658, A172}

\bibitem[\protect\citeauthoryear{{Farmer}, {Renzo}, {de Mink}, {Marchant}  \&
  {Justham}}{{Farmer} et~al.}{2019}]{2019ApJ...887...53F}
{Farmer} R.,  {Renzo} M.,  {de Mink} S.~E.,  {Marchant} P.,   {Justham} S.,
  2019, \mn@doi [\apj] {10.3847/1538-4357/ab518b}, \href
  {https://ui.adsabs.harvard.edu/abs/2019ApJ...887...53F} {887, 53}

\bibitem[\protect\citeauthoryear{{Feldmeier-Krause}, {Zhu}, {Neumayer}, {van de
  Ven}, {de Zeeuw}  \& {Sch{\"o}del}}{{Feldmeier-Krause}
  et~al.}{2017}]{MW_NSC_SMBH_correlation}
{Feldmeier-Krause} A.,  {Zhu} L.,  {Neumayer} N.,  {van de Ven} G.,  {de Zeeuw}
  P.~T.,   {Sch{\"o}del} R.,  2017, \mn@doi [\mnras] {10.1093/mnras/stw3377},
  \href {https://ui.adsabs.harvard.edu/abs/2017MNRAS.466.4040F} {466, 4040}

\bibitem[\protect\citeauthoryear{{Ferrarese} et~al.,}{{Ferrarese}
  et~al.}{2006}]{2006_Ferrarese_SMBHNSC}
{Ferrarese} L.,  et~al., 2006, \mn@doi [\apjl] {10.1086/505388}, \href
  {https://ui.adsabs.harvard.edu/abs/2006ApJ...644L..21F} {644, L21}

\bibitem[\protect\citeauthoryear{{Fishbach}, {Holz}  \& {Farr}}{{Fishbach}
  et~al.}{2017}]{fischbach_are_2017}
{Fishbach} M.,  {Holz} D.~E.,   {Farr} B.,  2017, \mn@doi [\apjl]
  {10.3847/2041-8213/aa7045}, \href
  {https://ui.adsabs.harvard.edu/abs/2017ApJ...840L..24F} {840, L24}

\bibitem[\protect\citeauthoryear{{Fragione}}{{Fragione}}{2022}]{Fragione2022nsc}
{Fragione} G.,  2022, arXiv e-prints, \href
  {https://ui.adsabs.harvard.edu/abs/2022arXiv220205618F} {p. arXiv:2202.05618}

\bibitem[\protect\citeauthoryear{{Fragione} \& {Silk}}{{Fragione} \&
  {Silk}}{2020}]{fragione_repeat_2020}
{Fragione} G.,  {Silk} J.,  2020, \mn@doi [\mnras] {10.1093/mnras/staa2629},
  \href {https://ui.adsabs.harvard.edu/abs/2020MNRAS.498.4591F} {498, 4591}

\bibitem[\protect\citeauthoryear{Fragione, Ginsburg  \& Kocsis}{Fragione
  et~al.}{2018}]{fragione_gravitational_2018}
Fragione G.,  Ginsburg I.,   Kocsis B.,  2018, \mn@doi [The Astrophysical
  Journal] {10.3847/1538-4357/aab368}, 856, 92

\bibitem[\protect\citeauthoryear{{Fragione}, {Loeb}  \& {Rasio}}{{Fragione}
  et~al.}{2020}]{FragioneLoeb2020}
{Fragione} G.,  {Loeb} A.,   {Rasio} F.~A.,  2020, \mn@doi [\apjl]
  {10.3847/2041-8213/abbc0a}, \href
  {https://ui.adsabs.harvard.edu/abs/2020ApJ...902L..26F} {902, L26}

\bibitem[\protect\citeauthoryear{Fragione, Kocsis, Rasio  \& Silk}{Fragione
  et~al.}{2022a}]{fragione_repeated_2022}
Fragione G.,  Kocsis B.,  Rasio F.~A.,   Silk J.,  2022a, \mn@doi [The
  Astrophysical Journal] {10.3847/1538-4357/ac5026}, 927, 231

\bibitem[\protect\citeauthoryear{{Fragione}, {Loeb}, {Kocsis}  \&
  {Rasio}}{{Fragione} et~al.}{2022b}]{FragioneLoeb2022}
{Fragione} G.,  {Loeb} A.,  {Kocsis} B.,   {Rasio} F.~A.,  2022b, \mn@doi
  [\apj] {10.3847/1538-4357/ac75d0}, \href
  {https://ui.adsabs.harvard.edu/abs/2022ApJ...933..170F} {933, 170}

\bibitem[\protect\citeauthoryear{{Fregeau} \& {Rasio}}{{Fregeau} \&
  {Rasio}}{2007}]{Fregeau_2007}
{Fregeau} J.~M.,  {Rasio} F.~A.,  2007, \mn@doi [\apj] {10.1086/511809}, \href
  {https://ui.adsabs.harvard.edu/abs/2007ApJ...658.1047F} {658, 1047}

\bibitem[\protect\citeauthoryear{Fregeau, Cheung, Portegies~Zwart  \&
  Rasio}{Fregeau et~al.}{2004a}]{fregeau_stellar_2004}
Fregeau J.~M.,  Cheung P.,  Portegies~Zwart S.~F.,   Rasio F.~A.,  2004a,
  \mn@doi [Monthly Notices of the Royal Astronomical Society]
  {10.1111/j.1365-2966.2004.07914.x}, 352, 1

\bibitem[\protect\citeauthoryear{{Fregeau}, {Cheung}, {Portegies Zwart}  \&
  {Rasio}}{{Fregeau} et~al.}{2004b}]{2004MNRAS.352....1F}
{Fregeau} J.~M.,  {Cheung} P.,  {Portegies Zwart} S.~F.,   {Rasio} F.~A.,
  2004b, \mn@doi [\mnras] {10.1111/j.1365-2966.2004.07914.x}, \href
  {https://ui.adsabs.harvard.edu/abs/2004MNRAS.352....1F} {352, 1}

\bibitem[\protect\citeauthoryear{Freitag, Gürkan  \& Rasio}{Freitag
  et~al.}{2006}]{freitag_runaway_2006}
Freitag M.,  Gürkan M.~A.,   Rasio F.~A.,  2006, \mn@doi [Monthly Notices of
  the Royal Astronomical Society] {10.1111/j.1365-2966.2006.10096.x}, 368, 141

\bibitem[\protect\citeauthoryear{Georgiev, Böker, Leigh, Lützgendorf  \&
  Neumayer}{Georgiev et~al.}{2016}]{10.1093/mnras/stw093}
Georgiev I.~Y.,  Böker T.,  Leigh N.,  Lützgendorf N.,   Neumayer N.,  2016,
  \mn@doi [Monthly Notices of the Royal Astronomical Society]
  {10.1093/mnras/stw093}, 457, 2122

\bibitem[\protect\citeauthoryear{Gerosa \& Berti}{Gerosa \&
  Berti}{2019}]{gerosa_escape_2019}
Gerosa D.,  Berti E.,  2019, \mn@doi [Physical Review D]
  {10.1103/PhysRevD.100.041301}, 100, 041301

\bibitem[\protect\citeauthoryear{{Gerosa} \& {Fishbach}}{{Gerosa} \&
  {Fishbach}}{2021}]{Gerosa_hierarchical_2021}
{Gerosa} D.,  {Fishbach} M.,  2021, \mn@doi [Nature Astronomy]
  {10.1038/s41550-021-01398-w}, \href
  {https://ui.adsabs.harvard.edu/abs/2021NatAs...5..749G} {5, 749}

\bibitem[\protect\citeauthoryear{{Gerssen}, {van der Marel}, {Gebhardt},
  {Guhathakurta}, {Peterson}  \& {Pryor}}{{Gerssen}
  et~al.}{2002}]{2002AJ....124.3270G}
{Gerssen} J.,  {van der Marel} R.~P.,  {Gebhardt} K.,  {Guhathakurta} P.,
  {Peterson} R.~C.,   {Pryor} C.,  2002, \mn@doi [\aj] {10.1086/344584}, \href
  {https://ui.adsabs.harvard.edu/abs/2002AJ....124.3270G} {124, 3270}

\bibitem[\protect\citeauthoryear{{Gieles}, {Balbinot}, {Yaaqib},
  {H{\'e}nault-Brunet}, {Zocchi}, {Peuten}  \& {Jonker}}{{Gieles}
  et~al.}{2018}]{2018MNRAS.473.4832G}
{Gieles} M.,  {Balbinot} E.,  {Yaaqib} R. I.~S.~M.,  {H{\'e}nault-Brunet} V.,
  {Zocchi} A.,  {Peuten} M.,   {Jonker} P.~G.,  2018, \mn@doi [\mnras]
  {10.1093/mnras/stx2694}, \href
  {https://ui.adsabs.harvard.edu/abs/2018MNRAS.473.4832G} {473, 4832}

\bibitem[\protect\citeauthoryear{{Giersz}, {Heggie}, {Hurley}  \&
  {Hypki}}{{Giersz} et~al.}{2013}]{2013MNRAS.431.2184G}
{Giersz} M.,  {Heggie} D.~C.,  {Hurley} J.~R.,   {Hypki} A.,  2013, \mn@doi
  [\mnras] {10.1093/mnras/stt307}, \href
  {https://ui.adsabs.harvard.edu/abs/2013MNRAS.431.2184G} {431, 2184}

\bibitem[\protect\citeauthoryear{Giersz, Leigh, Hypki, Lützgendorf  \&
  Askar}{Giersz et~al.}{2015}]{giersz_mocca_2015}
Giersz M.,  Leigh N.,  Hypki A.,  Lützgendorf N.,   Askar A.,  2015, \mn@doi
  [Monthly Notices of the Royal Astronomical Society] {10.1093/mnras/stv2162},
  454, 3150

\bibitem[\protect\citeauthoryear{{Gonz{\'a}lez Prieto}, {Kremer}, {Fragione},
  {Martinez}, {Weatherford}, {Zevin}  \& {Rasio}}{{Gonz{\'a}lez Prieto}
  et~al.}{2022}]{Gonzalez_2022}
{Gonz{\'a}lez Prieto} E.,  {Kremer} K.,  {Fragione} G.,  {Martinez} M. A.~S.,
  {Weatherford} N.~C.,  {Zevin} M.,   {Rasio} F.~A.,  2022, arXiv e-prints,
  \href {https://ui.adsabs.harvard.edu/abs/2022arXiv220807881G} {p.
  arXiv:2208.07881}

\bibitem[\protect\citeauthoryear{González, Kremer, Chatterjee, Fragione,
  Rodriguez, Weatherford, Ye  \& Rasio}{González
  et~al.}{2021}]{gonzalez_intermediate-mass_2021}
González E.,  Kremer K.,  Chatterjee S.,  Fragione G.,  Rodriguez C.~L.,
  Weatherford N.~C.,  Ye C.~S.,   Rasio F.~A.,  2021, \mn@doi [The
  Astrophysical Journal] {10.3847/2041-8213/abdf5b}, 908, L29

\bibitem[\protect\citeauthoryear{{G{\"u}ltekin}, {Miller}  \&
  {Hamilton}}{{G{\"u}ltekin} et~al.}{2004}]{Gultekin2004}
{G{\"u}ltekin} K.,  {Miller} M.~C.,   {Hamilton} D.~P.,  2004, \mn@doi [\apj]
  {10.1086/424809}, \href
  {https://ui.adsabs.harvard.edu/abs/2004ApJ...616..221G} {616, 221}

\bibitem[\protect\citeauthoryear{{G{\"u}rkan}, {Freitag}  \&
  {Rasio}}{{G{\"u}rkan} et~al.}{2004}]{2004ApJ...604..632G}
{G{\"u}rkan} M.~A.,  {Freitag} M.,   {Rasio} F.~A.,  2004, \mn@doi [\apj]
  {10.1086/381968}, \href
  {https://ui.adsabs.harvard.edu/abs/2004ApJ...604..632G} {604, 632}

\bibitem[\protect\citeauthoryear{Gürkan, Fregeau  \& Rasio}{Gürkan
  et~al.}{2006}]{gurkan_massive_2006}
Gürkan M.~A.,  Fregeau J.~M.,   Rasio F.~A.,  2006, \mn@doi [The Astrophysical
  Journal] {10.1086/503295}, 640, L39

\bibitem[\protect\citeauthoryear{{Haehnelt}, {Natarajan}  \& {Rees}}{{Haehnelt}
  et~al.}{1998}]{Haehnelt_1998_AGNmass}
{Haehnelt} M.~G.,  {Natarajan} P.,   {Rees} M.~J.,  1998, \mn@doi [\mnras]
  {10.1046/j.1365-8711.1998.01951.x}, \href
  {https://ui.adsabs.harvard.edu/abs/1998MNRAS.300..817H} {300, 817}

\bibitem[\protect\citeauthoryear{{Hayashi}, {Trani}  \& {Suto}}{{Hayashi}
  et~al.}{2022}]{hayashi2022}
{Hayashi} T.,  {Trani} A.~A.,   {Suto} Y.,  2022, \mn@doi [\apj]
  {10.3847/1538-4357/ac8f48}, \href
  {https://ui.adsabs.harvard.edu/abs/2022ApJ...939...81H} {939, 81}

\bibitem[\protect\citeauthoryear{{Healy} \& {Lousto}}{{Healy} \&
  {Lousto}}{2018}]{2018PhRvD..97h4002H}
{Healy} J.,  {Lousto} C.~O.,  2018, \mn@doi [\prd]
  {10.1103/PhysRevD.97.084002}, \href
  {https://ui.adsabs.harvard.edu/abs/2018PhRvD..97h4002H} {97, 084002}

\bibitem[\protect\citeauthoryear{{Hees} et~al.,}{{Hees}
  et~al.}{2017}]{Ghez_review_SMBH_dyn_meas}
{Hees} A.,  et~al., 2017, \mn@doi [\prl] {10.1103/PhysRevLett.118.211101},
  \href {https://ui.adsabs.harvard.edu/abs/2017PhRvL.118u1101H} {118, 211101}

\bibitem[\protect\citeauthoryear{{Heger} \& {Woosley}}{{Heger} \&
  {Woosley}}{2002}]{Heger_Woosley_2002}
{Heger} A.,  {Woosley} S.~E.,  2002, \mn@doi [\apj] {10.1086/338487}, \href
  {https://ui.adsabs.harvard.edu/abs/2002ApJ...567..532H} {567, 532}

\bibitem[\protect\citeauthoryear{{Heggie} \& {Hut}}{{Heggie} \&
  {Hut}}{1993}]{Heggie_1993}
{Heggie} D.~C.,  {Hut} P.,  1993, \mn@doi [\apjs] {10.1086/191768}, \href
  {https://ui.adsabs.harvard.edu/abs/1993ApJS...85..347H} {85, 347}

\bibitem[\protect\citeauthoryear{{Heggie} \& {Hut}}{{Heggie} \&
  {Hut}}{2003}]{2003gmbp.book.....H}
{Heggie} D.,  {Hut} P.,  2003, {The Gravitational Million-Body Problem: A
  Multidisciplinary Approach to Star Cluster Dynamics}

\bibitem[\protect\citeauthoryear{{Hellstr{\"o}m}, {Askar}, {Trani}, {Giersz},
  {Church}  \& {Samsing}}{{Hellstr{\"o}m} et~al.}{2022}]{hellstrom2022}
{Hellstr{\"o}m} L.,  {Askar} A.,  {Trani} A.~A.,  {Giersz} M.,  {Church} R.~P.,
    {Samsing} J.,  2022, \mn@doi [\mnras] {10.1093/mnras/stac2808}, \href
  {https://ui.adsabs.harvard.edu/abs/2022MNRAS.517.1695H} {517, 1695}

\bibitem[\protect\citeauthoryear{{Holley-Bockelmann}, {G{\"u}ltekin},
  {Shoemaker}  \& {Yunes}}{{Holley-Bockelmann} et~al.}{2008}]{HB2008}
{Holley-Bockelmann} K.,  {G{\"u}ltekin} K.,  {Shoemaker} D.,   {Yunes} N.,
  2008, \mn@doi [\apj] {10.1086/591218}, \href
  {https://ui.adsabs.harvard.edu/abs/2008ApJ...686..829H} {686, 829}

\bibitem[\protect\citeauthoryear{{Hut}}{{Hut}}{1981}]{hut81}
{Hut} P.,  1981, \aap, \href
  {https://ui.adsabs.harvard.edu/abs/1981A&A....99..126H} {99, 126}

\bibitem[\protect\citeauthoryear{{Hypki} \& {Giersz}}{{Hypki} \&
  {Giersz}}{2013}]{2013MNRAS.429.1221H}
{Hypki} A.,  {Giersz} M.,  2013, \mn@doi [\mnras] {10.1093/mnras/sts415}, \href
  {https://ui.adsabs.harvard.edu/abs/2013MNRAS.429.1221H} {429, 1221}

\bibitem[\protect\citeauthoryear{{Inayoshi}, {Visbal}  \& {Haiman}}{{Inayoshi}
  et~al.}{2020}]{2020ARA&A..58...27I}
{Inayoshi} K.,  {Visbal} E.,   {Haiman} Z.,  2020, \mn@doi [\araa]
  {10.1146/annurev-astro-120419-014455}, \href
  {https://ui.adsabs.harvard.edu/abs/2020ARA&A..58...27I} {58, 27}

\bibitem[\protect\citeauthoryear{{Jeans}}{{Jeans}}{1919}]{jeans1919}
{Jeans} J.~H.,  1919, \mn@doi [\mnras] {10.1093/mnras/79.6.408}, \href
  {https://ui.adsabs.harvard.edu/abs/1919MNRAS..79..408J} {79, 408}

\bibitem[\protect\citeauthoryear{{Kamann} et~al.,}{{Kamann}
  et~al.}{2016}]{2016A&A...588A.149K}
{Kamann} S.,  et~al., 2016, \mn@doi [\aap] {10.1051/0004-6361/201527065}, \href
  {https://ui.adsabs.harvard.edu/abs/2016A&A...588A.149K} {588, A149}

\bibitem[\protect\citeauthoryear{{Kirsten} \& {Vlemmings}}{{Kirsten} \&
  {Vlemmings}}{2012}]{2012A&A...542A..44K}
{Kirsten} F.,  {Vlemmings} W.~H.~T.,  2012, \mn@doi [\aap]
  {10.1051/0004-6361/201218928}, \href
  {https://ui.adsabs.harvard.edu/abs/2012A&A...542A..44K} {542, A44}

\bibitem[\protect\citeauthoryear{{Kormendy} \& {Ho}}{{Kormendy} \&
  {Ho}}{2013}]{kormendy_2013_smbhgalaxyclusterreview}
{Kormendy} J.,  {Ho} L.~C.,  2013, \mn@doi [\araa]
  {10.1146/annurev-astro-082708-101811}, \href
  {https://ui.adsabs.harvard.edu/abs/2013ARA&A..51..511K} {51, 511}

\bibitem[\protect\citeauthoryear{Kovetz, Cholis, Kamionkowski  \& Silk}{Kovetz
  et~al.}{2018}]{kovetz_limits_2018}
Kovetz E.~D.,  Cholis I.,  Kamionkowski M.,   Silk J.,  2018, \mn@doi [Physical
  Review D] {10.1103/PhysRevD.97.123003}, 97, 123003

\bibitem[\protect\citeauthoryear{{Kremer}, {Chatterjee}, {Ye}, {Rodriguez}  \&
  {Rasio}}{{Kremer} et~al.}{2019}]{2019ApJ...871...38K}
{Kremer} K.,  {Chatterjee} S.,  {Ye} C.~S.,  {Rodriguez} C.~L.,   {Rasio}
  F.~A.,  2019, \mn@doi [\apj] {10.3847/1538-4357/aaf646}, \href
  {https://ui.adsabs.harvard.edu/abs/2019ApJ...871...38K} {871, 38}

\bibitem[\protect\citeauthoryear{{Kremer} et~al.,}{{Kremer}
  et~al.}{2020a}]{2020ApJ...903...45K}
{Kremer} K.,  et~al., 2020a, \mn@doi [\apj] {10.3847/1538-4357/abb945}, \href
  {https://ui.adsabs.harvard.edu/abs/2020ApJ...903...45K} {903, 45}

\bibitem[\protect\citeauthoryear{{Kremer} et~al.,}{{Kremer}
  et~al.}{2020b}]{Kremer_2020_IMBH}
{Kremer} K.,  et~al., 2020b, \mn@doi [\apj] {10.3847/1538-4357/abb945}, \href
  {https://ui.adsabs.harvard.edu/abs/2020ApJ...903...45K} {903, 45}

\bibitem[\protect\citeauthoryear{Lousto \& Zlochower}{Lousto \&
  Zlochower}{2013}]{PhysRevD.87.084027}
Lousto C.~O.,  Zlochower Y.,  2013, \mn@doi [Phys. Rev. D]
  {10.1103/PhysRevD.87.084027}, 87, 084027

\bibitem[\protect\citeauthoryear{{Manwadkar}, {Kol}, {Trani}  \&
  {Leigh}}{{Manwadkar} et~al.}{2021}]{manwadkar2021}
{Manwadkar} V.,  {Kol} B.,  {Trani} A.~A.,   {Leigh} N. W.~C.,  2021, \mn@doi
  [\mnras] {10.1093/mnras/stab1689}, \href
  {https://ui.adsabs.harvard.edu/abs/2021MNRAS.506..692M} {506, 692}

\bibitem[\protect\citeauthoryear{Mapelli}{Mapelli}{2016}]{mapelli_massive_2016}
Mapelli M.,  2016, \mn@doi [Monthly Notices of the Royal Astronomical Society]
  {10.1093/mnras/stw869}, 459, 3432

\bibitem[\protect\citeauthoryear{{Mapelli} et~al.,}{{Mapelli}
  et~al.}{2021}]{2021MNRAS.505..339M}
{Mapelli} M.,  et~al., 2021, \mn@doi [\mnras] {10.1093/mnras/stab1334}, \href
  {https://ui.adsabs.harvard.edu/abs/2021MNRAS.505..339M} {505, 339}

\bibitem[\protect\citeauthoryear{{Mayer}, {Fiacconi}, {Bonoli}, {Quinn},
  {Ro{\v{s}}kar}, {Shen}  \& {Wadsley}}{{Mayer}
  et~al.}{2015}]{2015ApJ...810...51M}
{Mayer} L.,  {Fiacconi} D.,  {Bonoli} S.,  {Quinn} T.,  {Ro{\v{s}}kar} R.,
  {Shen} S.,   {Wadsley} J.,  2015, \mn@doi [\apj]
  {10.1088/0004-637X/810/1/51}, \href
  {https://ui.adsabs.harvard.edu/abs/2015ApJ...810...51M} {810, 51}

\bibitem[\protect\citeauthoryear{{McNamara}, {Harrison}  \&
  {Anderson}}{{McNamara} et~al.}{2003}]{2003ApJ...595..187M}
{McNamara} B.~J.,  {Harrison} T.~E.,   {Anderson} J.,  2003, \mn@doi [\apj]
  {10.1086/377341}, \href
  {https://ui.adsabs.harvard.edu/abs/2003ApJ...595..187M} {595, 187}

\bibitem[\protect\citeauthoryear{{Mikkola} \& {Aarseth}}{{Mikkola} \&
  {Aarseth}}{1993}]{kschain}
{Mikkola} S.,  {Aarseth} S.~J.,  1993, \mn@doi [Celestial Mechanics and
  Dynamical Astronomy] {10.1007/BF00695714}, \href
  {https://ui.adsabs.harvard.edu/abs/1993CeMDA..57..439M} {57, 439}

\bibitem[\protect\citeauthoryear{{Mikkola} \& {Merritt}}{{Mikkola} \&
  {Merritt}}{2006}]{2006MNRAS.372..219M}
{Mikkola} S.,  {Merritt} D.,  2006, \mn@doi [\mnras]
  {10.1111/j.1365-2966.2006.10854.x}, \href
  {https://ui.adsabs.harvard.edu/abs/2006MNRAS.372..219M} {372, 219}

\bibitem[\protect\citeauthoryear{{Mikkola} \& {Merritt}}{{Mikkola} \&
  {Merritt}}{2008}]{2008AJ....135.2398M}
{Mikkola} S.,  {Merritt} D.,  2008, \mn@doi [\aj]
  {10.1088/0004-6256/135/6/2398}, \href
  {https://ui.adsabs.harvard.edu/abs/2008AJ....135.2398M} {135, 2398}

\bibitem[\protect\citeauthoryear{{Mikkola} \& {Tanikawa}}{{Mikkola} \&
  {Tanikawa}}{1999a}]{mikkola1999a}
{Mikkola} S.,  {Tanikawa} K.,  1999a, \mn@doi [Celestial Mechanics and
  Dynamical Astronomy] {10.1023/A:1008368322547}, \href
  {https://ui.adsabs.harvard.edu/abs/1999CeMDA..74..287M} {74, 287}

\bibitem[\protect\citeauthoryear{{Mikkola} \& {Tanikawa}}{{Mikkola} \&
  {Tanikawa}}{1999b}]{mikkola1999b}
{Mikkola} S.,  {Tanikawa} K.,  1999b, \mn@doi [\mnras]
  {10.1046/j.1365-8711.1999.02982.x}, \href
  {https://ui.adsabs.harvard.edu/abs/1999MNRAS.310..745M} {310, 745}

\bibitem[\protect\citeauthoryear{Miller \& Hamilton}{Miller \&
  Hamilton}{2002}]{miller_production_2002}
Miller M.~C.,  Hamilton D.~P.,  2002, \mn@doi [Monthly Notices of the Royal
  Astronomical Society] {10.1046/j.1365-8711.2002.05112.x}, 330, 232

\bibitem[\protect\citeauthoryear{{Miller} \& {Lauburg}}{{Miller} \&
  {Lauburg}}{2009}]{Miller2009}
{Miller} M.~C.,  {Lauburg} V.~M.,  2009, \mn@doi [\apj]
  {10.1088/0004-637X/692/1/917}, \href
  {https://ui.adsabs.harvard.edu/abs/2009ApJ...692..917M} {692, 917}

\bibitem[\protect\citeauthoryear{Mouri \& Taniguchi}{Mouri \&
  Taniguchi}{2002}]{mouri_runaway_2002}
Mouri H.,  Taniguchi Y.,  2002, \mn@doi [The Astrophysical Journal]
  {10.1086/339472}, 566, L17

\bibitem[\protect\citeauthoryear{{Murphy}, {Cohn}  \& {Lugger}}{{Murphy}
  et~al.}{2011}]{2011ApJ...732...67M}
{Murphy} B.~W.,  {Cohn} H.~N.,   {Lugger} P.~M.,  2011, \mn@doi [\apj]
  {10.1088/0004-637X/732/2/67}, \href
  {https://ui.adsabs.harvard.edu/abs/2011ApJ...732...67M} {732, 67}

\bibitem[\protect\citeauthoryear{{Myll{\"a}ri}, {Valtonen}, {Pasechnik}  \&
  {Mikkola}}{{Myll{\"a}ri} et~al.}{2018}]{2018MNRAS.476..830M}
{Myll{\"a}ri} A.,  {Valtonen} M.,  {Pasechnik} A.,   {Mikkola} S.,  2018,
  \mn@doi [\mnras] {10.1093/mnras/sty237}, \href
  {https://ui.adsabs.harvard.edu/abs/2018MNRAS.476..830M} {476, 830}

\bibitem[\protect\citeauthoryear{{Neumayer}, {Walcher}, {Andersen},
  {S{\'a}nchez}, {B{\"o}ker}  \& {Rix}}{{Neumayer}
  et~al.}{2011}]{Neumayer_2011_NSCloc}
{Neumayer} N.,  {Walcher} C.~J.,  {Andersen} D.,  {S{\'a}nchez} S.~F.,
  {B{\"o}ker} T.,   {Rix} H.-W.,  2011, \mn@doi [\mnras]
  {10.1111/j.1365-2966.2011.18266.x}, \href
  {https://ui.adsabs.harvard.edu/abs/2011MNRAS.413.1875N} {413, 1875}

\bibitem[\protect\citeauthoryear{{Neumayer}, {Seth}  \& {B{\"o}ker}}{{Neumayer}
  et~al.}{2020}]{NSC2020review}
{Neumayer} N.,  {Seth} A.,   {B{\"o}ker} T.,  2020, \mn@doi [\aapr]
  {10.1007/s00159-020-00125-0}, \href
  {https://ui.adsabs.harvard.edu/abs/2020A&ARv..28....4N} {28, 4}

\bibitem[\protect\citeauthoryear{O’Leary, Rasio, Fregeau, Ivanova  \&
  O’Shaughnessy}{O’Leary et~al.}{2006}]{oleary_binary_2006}
O’Leary R.~M.,  Rasio F.~A.,  Fregeau J.~M.,  Ivanova N.,   O’Shaughnessy
  R.,  2006, \mn@doi [The Astrophysical Journal] {10.1086/498446}, 637, 937

\bibitem[\protect\citeauthoryear{{Pacucci}, {Natarajan}, {Volonteri},
  {Cappelluti}  \& {Urry}}{{Pacucci} et~al.}{2017}]{2017ApJ...850L..42P}
{Pacucci} F.,  {Natarajan} P.,  {Volonteri} M.,  {Cappelluti} N.,   {Urry}
  C.~M.,  2017, \mn@doi [\apjl] {10.3847/2041-8213/aa9aea}, \href
  {https://ui.adsabs.harvard.edu/abs/2017ApJ...850L..42P} {850, L42}

\bibitem[\protect\citeauthoryear{{Padovani} et~al.,}{{Padovani}
  et~al.}{2017}]{agn_review_2017}
{Padovani} P.,  et~al., 2017, \mn@doi [\aapr] {10.1007/s00159-017-0102-9},
  \href {https://ui.adsabs.harvard.edu/abs/2017A&ARv..25....2P} {25, 2}

\bibitem[\protect\citeauthoryear{{Perera} et~al.,}{{Perera}
  et~al.}{2017}]{2017MNRAS.468.2114P}
{Perera} B.~B.~P.,  et~al., 2017, \mn@doi [\mnras] {10.1093/mnras/stx501},
  \href {https://ui.adsabs.harvard.edu/abs/2017MNRAS.468.2114P} {468, 2114}

\bibitem[\protect\citeauthoryear{Peters}{Peters}{1964}]{PhysRev.136.B1224}
Peters P.~C.,  1964, \mn@doi [Phys. Rev.] {10.1103/PhysRev.136.B1224}, 136,
  B1224

\bibitem[\protect\citeauthoryear{{Pezzulli}, {Valiante}  \&
  {Schneider}}{{Pezzulli} et~al.}{2016}]{2016MNRAS.458.3047P}
{Pezzulli} E.,  {Valiante} R.,   {Schneider} R.,  2016, \mn@doi [\mnras]
  {10.1093/mnras/stw505}, \href
  {https://ui.adsabs.harvard.edu/abs/2016MNRAS.458.3047P} {458, 3047}

\bibitem[\protect\citeauthoryear{Portegies~Zwart \& McMillan}{Portegies~Zwart
  \& McMillan}{2002}]{zwart_runaway_2002}
Portegies~Zwart S.~F.,  McMillan S. L.~W.,  2002, \mn@doi [The Astrophysical
  Journal] {10.1086/341798}, 576, 899

\bibitem[\protect\citeauthoryear{Portegies~Zwart, Makino, McMillan  \&
  Hut}{Portegies~Zwart et~al.}{1999}]{portegies_zwart_star_1999}
Portegies~Zwart S.~F.,  Makino J.,  McMillan S. L.~W.,   Hut P.,  1999,
  Astronomy and Astrophysics, 348, 117

\bibitem[\protect\citeauthoryear{Portegies~Zwart, Baumgardt, Hut, Makino  \&
  McMillan}{Portegies~Zwart et~al.}{2004}]{portegies_zwart_formation_2004}
Portegies~Zwart S.~F.,  Baumgardt H.,  Hut P.,  Makino J.,   McMillan S. L.~W.,
   2004, \mn@doi [Nature] {10.1038/nature02448}, 428, 724

\bibitem[\protect\citeauthoryear{Quinlan \& Shapiro}{Quinlan \&
  Shapiro}{1989}]{quinlan_dynamical_1989}
Quinlan G.~D.,  Shapiro S.~L.,  1989, \mn@doi [The Astrophysical Journal]
  {10.1086/167745}, 343, 725

\bibitem[\protect\citeauthoryear{{Rees}}{{Rees}}{1984}]{Rees1984}
{Rees} M.~J.,  1984, \mn@doi [\araa] {10.1146/annurev.aa.22.090184.002351},
  \href {https://ui.adsabs.harvard.edu/abs/1984ARA&A..22..471R} {22, 471}

\bibitem[\protect\citeauthoryear{{Ricarte} \& {Natarajan}}{{Ricarte} \&
  {Natarajan}}{2018}]{Ricarte_2018_bhseedaccretion}
{Ricarte} A.,  {Natarajan} P.,  2018, \mn@doi [\mnras] {10.1093/mnras/sty2448},
  \href {https://ui.adsabs.harvard.edu/abs/2018MNRAS.481.3278R} {481, 3278}

\bibitem[\protect\citeauthoryear{{Richardson}, {Simpson}, {Polimera},
  {Kannappan}, {Bellovary}, {Greene}  \& {Jenkins}}{{Richardson}
  et~al.}{2022}]{JWST_IMBH_2022}
{Richardson} C.~T.,  {Simpson} C.,  {Polimera} M.~S.,  {Kannappan} S.~J.,
  {Bellovary} J.~M.,  {Greene} C.,   {Jenkins} S.,  2022, \mn@doi [\apj]
  {10.3847/1538-4357/ac510c}, \href
  {https://ui.adsabs.harvard.edu/abs/2022ApJ...927..165R} {927, 165}

\bibitem[\protect\citeauthoryear{Rizzuto et~al.,}{Rizzuto
  et~al.}{2021}]{rizzuto_intermediate_2021}
Rizzuto F.~P.,  et~al., 2021, \mn@doi [Monthly Notices of the Royal
  Astronomical Society] {10.1093/mnras/staa3634}, 501, 5257

\bibitem[\protect\citeauthoryear{Rodriguez, Haster, Chatterjee, Kalogera  \&
  Rasio}{Rodriguez et~al.}{2016}]{rodriguez_dynamical_2016}
Rodriguez C.~L.,  Haster C.-J.,  Chatterjee S.,  Kalogera V.,   Rasio F.~A.,
  2016, \mn@doi [The Astrophysical Journal] {10.3847/2041-8205/824/1/L8}, 824,
  L8

\bibitem[\protect\citeauthoryear{{Rodriguez}, {Zevin}, {Amaro-Seoane},
  {Chatterjee}, {Kremer}, {Rasio}  \& {Ye}}{{Rodriguez}
  et~al.}{2019}]{2019_Rodriguez_review}
{Rodriguez} C.~L.,  {Zevin} M.,  {Amaro-Seoane} P.,  {Chatterjee} S.,  {Kremer}
  K.,  {Rasio} F.~A.,   {Ye} C.~S.,  2019, \mn@doi [\prd]
  {10.1103/PhysRevD.100.043027}, \href
  {https://ui.adsabs.harvard.edu/abs/2019PhRvD.100d3027R} {100, 043027}

\bibitem[\protect\citeauthoryear{{Rodriguez} et~al.,}{{Rodriguez}
  et~al.}{2022}]{CMCcodepaper}
{Rodriguez} C.~L.,  et~al., 2022, \mn@doi [\apjs] {10.3847/1538-4365/ac2edf},
  \href {https://ui.adsabs.harvard.edu/abs/2022ApJS..258...22R} {258, 22}

\bibitem[\protect\citeauthoryear{{Rose}, {Naoz}, {Sari}  \& {Linial}}{{Rose}
  et~al.}{2022}]{Rose2022}
{Rose} S.~C.,  {Naoz} S.,  {Sari} R.,   {Linial} I.,  2022, \mn@doi [\apjl]
  {10.3847/2041-8213/ac6426}, \href
  {https://ui.adsabs.harvard.edu/abs/2022ApJ...929L..22R} {929, L22}

\bibitem[\protect\citeauthoryear{{Rozner} \& {Perets}}{{Rozner} \&
  {Perets}}{2022}]{2022ApJ...931..149R}
{Rozner} M.,  {Perets} H.~B.,  2022, \mn@doi [\apj] {10.3847/1538-4357/ac6d55},
  \href {https://ui.adsabs.harvard.edu/abs/2022ApJ...931..149R} {931, 149}

\bibitem[\protect\citeauthoryear{{Rui}, {Weatherford}, {Kremer}, {Chatterjee},
  {Fragione}, {Rasio}, {Rodriguez}  \& {Ye}}{{Rui}
  et~al.}{2021}]{2021RNAAS...5...47R}
{Rui} N.~Z.,  {Weatherford} N.~C.,  {Kremer} K.,  {Chatterjee} S.,  {Fragione}
  G.,  {Rasio} F.~A.,  {Rodriguez} C.~L.,   {Ye} C.~S.,  2021, \mn@doi
  [Research Notes of the American Astronomical Society]
  {10.3847/2515-5172/abee77}, \href
  {https://ui.adsabs.harvard.edu/abs/2021RNAAS...5...47R} {5, 47}

\bibitem[\protect\citeauthoryear{{Samsing}, {Leigh}  \& {Trani}}{{Samsing}
  et~al.}{2018}]{samsing2018}
{Samsing} J.,  {Leigh} N. W.~C.,   {Trani} A.~A.,  2018, \mn@doi [\mnras]
  {10.1093/mnras/sty2247}, \href
  {https://ui.adsabs.harvard.edu/abs/2018MNRAS.481.5436S} {481, 5436}

\bibitem[\protect\citeauthoryear{{Seth}, {Ag{\"u}eros}, {Lee}  \&
  {Basu-Zych}}{{Seth} et~al.}{2008}]{Seth_2008_AGN_NSC}
{Seth} A.,  {Ag{\"u}eros} M.,  {Lee} D.,   {Basu-Zych} A.,  2008, \mn@doi
  [\apj] {10.1086/528955}, \href
  {https://ui.adsabs.harvard.edu/abs/2008ApJ...678..116S} {678, 116}

\bibitem[\protect\citeauthoryear{{Shi}, {Kremer}, {Grudi{\'c}},
  {Gerling-Dunsmore}  \& {Hopkins}}{{Shi} et~al.}{2022}]{ShiKremer2022}
{Shi} Y.,  {Kremer} K.,  {Grudi{\'c}} M.~Y.,  {Gerling-Dunsmore} H.~J.,
  {Hopkins} P.~F.,  2022, arXiv e-prints, \href
  {https://ui.adsabs.harvard.edu/abs/2022arXiv220805025S} {p. arXiv:2208.05025}

\bibitem[\protect\citeauthoryear{Spera \& Mapelli}{Spera \&
  Mapelli}{2017}]{10.1093/mnras/stx1576}
Spera M.,  Mapelli M.,  2017, \mn@doi [Monthly Notices of the Royal
  Astronomical Society] {10.1093/mnras/stx1576}, 470, 4739

\bibitem[\protect\citeauthoryear{{Stoer} \& {Bulirsch}}{{Stoer} \&
  {Bulirsch}}{1980}]{stoer1980}
{Stoer} J.,  {Bulirsch} R.,  1980, {Introduction to Numerical Analysis}.
{Springer-Verlag}, New York, \mn@doi{https://doi.org/10.1007/978-0-387-21738-3}

\bibitem[\protect\citeauthoryear{{Trani} \& {Spera}}{{Trani} \&
  {Spera}}{2022}]{tsunami2022}
{Trani} A.~A.,  {Spera} M.,  2022, arXiv e-prints, \href
  {https://ui.adsabs.harvard.edu/abs/2022arXiv220610583T} {p. arXiv:2206.10583}

\bibitem[\protect\citeauthoryear{{Trani}, {Fujii}  \& {Spera}}{{Trani}
  et~al.}{2019a}]{trani2019a}
{Trani} A.~A.,  {Fujii} M.~S.,   {Spera} M.,  2019a, \mn@doi [\apj]
  {10.3847/1538-4357/ab0e70}, \href
  {https://ui.adsabs.harvard.edu/abs/2019ApJ...875...42T} {875, 42}

\bibitem[\protect\citeauthoryear{{Trani}, {Spera}, {Leigh}  \& {Fujii}}{{Trani}
  et~al.}{2019b}]{trani2019b}
{Trani} A.~A.,  {Spera} M.,  {Leigh} N. W.~C.,   {Fujii} M.~S.,  2019b, \mn@doi
  [\apj] {10.3847/1538-4357/ab480a}, \href
  {https://ui.adsabs.harvard.edu/abs/2019ApJ...885..135T} {885, 135}

\bibitem[\protect\citeauthoryear{{Trani}, {Hamers}, {Geller}  \&
  {Spera}}{{Trani} et~al.}{2020}]{trani2020}
{Trani} A.~A.,  {Hamers} A.~S.,  {Geller} A.,   {Spera} M.,  2020, \mn@doi
  [\mnras] {10.1093/mnras/staa3098}, \href
  {https://ui.adsabs.harvard.edu/abs/2020MNRAS.499.4195T} {499, 4195}

\bibitem[\protect\citeauthoryear{{Tremou} et~al.,}{{Tremou}
  et~al.}{2018}]{2018ApJ...862...16T}
{Tremou} E.,  et~al., 2018, \mn@doi [\apj] {10.3847/1538-4357/aac9b9}, \href
  {https://ui.adsabs.harvard.edu/abs/2018ApJ...862...16T} {862, 16}

\bibitem[\protect\citeauthoryear{{Wang}, {Spurzem}, {Aarseth}, {Nitadori},
  {Berczik}, {Kouwenhoven}  \& {Naab}}{{Wang}
  et~al.}{2015}]{2015MNRAS.450.4070W}
{Wang} L.,  {Spurzem} R.,  {Aarseth} S.,  {Nitadori} K.,  {Berczik} P.,
  {Kouwenhoven} M.~B.~N.,   {Naab} T.,  2015, \mn@doi [\mnras]
  {10.1093/mnras/stv817}, \href
  {https://ui.adsabs.harvard.edu/abs/2015MNRAS.450.4070W} {450, 4070}

\bibitem[\protect\citeauthoryear{{Weatherford}, {Fragione}, {Kremer},
  {Chatterjee}, {Ye}, {Rodriguez}  \& {Rasio}}{{Weatherford}
  et~al.}{2021}]{2021ApJ...907L..25W}
{Weatherford} N.~C.,  {Fragione} G.,  {Kremer} K.,  {Chatterjee} S.,  {Ye}
  C.~S.,  {Rodriguez} C.~L.,   {Rasio} F.~A.,  2021, \mn@doi [\apjl]
  {10.3847/2041-8213/abd79c}, \href
  {https://ui.adsabs.harvard.edu/abs/2021ApJ...907L..25W} {907, L25}

\bibitem[\protect\citeauthoryear{{Wehner} \& {Harris}}{{Wehner} \&
  {Harris}}{2006}]{Wehner_2006_NSCSMBH}
{Wehner} E.~H.,  {Harris} W.~E.,  2006, \mn@doi [\apjl] {10.1086/505387}, \href
  {https://ui.adsabs.harvard.edu/abs/2006ApJ...644L..17W} {644, L17}

\bibitem[\protect\citeauthoryear{Zevin, Samsing, Rodriguez, Haster  \&
  Ramirez-Ruiz}{Zevin et~al.}{2019}]{zevin_eccentric_2019}
Zevin M.,  Samsing J.,  Rodriguez C.,  Haster C.-J.,   Ramirez-Ruiz E.,  2019,
  \mn@doi [The Astrophysical Journal] {10.3847/1538-4357/aaf6ec}, 871, 91

\bibitem[\protect\citeauthoryear{Ziosi, Mapelli, Branchesi  \& Tormen}{Ziosi
  et~al.}{2014}]{ziosi_dynamics_2014}
Ziosi B.~M.,  Mapelli M.,  Branchesi M.,   Tormen G.,  2014, \mn@doi [Monthly
  Notices of the Royal Astronomical Society] {10.1093/mnras/stu824}, 441, 3703

\makeatother
\end{thebibliography}

\bsp	
\label{lastpage}
\end{document}